\documentclass[twocolumn]{aastex7}

\usepackage{graphicx}	
\usepackage{amsmath}	

\DeclareRobustCommand{\ion}[2]{\textup{#1\,\textsc{\lowercase{#2}}}} 

\newcommand{\angstrom}{\text{\normalfont\AA}}
\def\he{\ion{He}{I}}

\def\hdu{HD118203}
\def\hd209{HD209458}
\def\hat12{HAT-P-12}
\def\hatu17{HAT-P-17}

\begin{document}


\title{ HET/HPF observations of Helium in warm, hot, and ultra-hot Jupiters}

\author[0000-0003-2066-8959]{Jaume Orell-Miquel}
\affiliation{Department of Astronomy, University of Texas at Austin, 2515 Speedway, Austin, TX 78712, USA}
\email[show]{jaume.miquel@austin.utexas.edu}

\author[]{Kyra Sampson} 
\affiliation{Department of Astronomy, University of Texas at Austin, 2515 Speedway, Austin, TX 78712, USA}
\email{sampson.kyra@gmail.com}

\author[0000-0002-4404-0456]{Caroline~V. Morley} 
\affiliation{Department of Astronomy, University of Texas at Austin, 2515 Speedway, Austin, TX 78712, USA}
\email{cmorley@utexas.edu}

\author[0000-0001-9662-3496]{William~D. Cochran}
\affiliation{McDonald Observatory, The University of Texas, Austin Texas USA}
\affiliation{Center for Planetary Systems Habitability, The University of Texas, Austin Texas USA}
\email{wdc@astro.as.utexas.edu}

\author[0000-0002-7119-2543]{Girish~M. Duvvuri}
\affiliation{Department of Physics and Astronomy, Vanderbilt University, Nashville, TN 37235, USA}
\email{girish.duvvuri@vanderbilt.edu}

\author[0000-0001-9626-0613]{Daniel~M. Krolikowski}
\affiliation{Steward Observatory, University of Arizona, 933 N. Cherry Ave, Tucson, AZ 85721, USA}
\email{krolikowski@arizona.edu}

\author[0000-0001-9596-7983]{Suvrath Mahadevan}
\affiliation{Department of Astronomy \& Astrophysics, The Pennsylvania State University, 525 Davey Laboratory, University Park, PA 16802, USA}
\affiliation{Center for Exoplanets and Habitable Worlds, The Pennsylvania State University, 525 Davey Laboratory, University Park, PA 16802, USA}
\email{suvrath@astro.psu.edu}

\author[0000-0001-6532-6755]{Quang~H. Tran}
\altaffiliation{51 Pegasi b Fellow.}
\affiliation{Department of Astronomy, Yale University, New Haven, CT 06511, USA}
\email{quang.tran@yale.edu}


\begin{abstract}
The near-infrared helium triplet line is a powerful tool for studying atmospheric escape processes of close-in exoplanets, especially irradiated gas giants. Line profile fitting provides direct insight into the mechanisms driving atmospheric mass loss of close-in, Jupiter-sized planets.
We present high-resolution transmission spectroscopy results for the helium triplet line of sixteen gas giants ($R_{\rm p} > 0.5 R_{\rm Jup}$). These observations are part of an extensive helium survey conducted using the Habitable Zone Planet Finder spectrograph on the 10\,m Hobby-Eberly Telescope.
For the first time, we provide constraints on the helium line for HAT-P-12\,b, HAT-P-17\,b, HD118203\,b, TrES-1\,b, and WASP-156\,b. Additionally, we are able to confirm previous robust or tentative detections for HD189733\,b, HD209458\,b, WASP-52\,b, WASP-69\,b, and WASP-76\,b, and non-detections for HAT-P-3\,b, WASP-11\,b, WASP-80\,b, WASP-127\,b, and WASP-177\,b. We do not confirm the previous helium narrow-band detection in HAT-P-26\,b using high-resolution observations.
To identify trends within the population of warm, hot, and ultra-hot Jupiters, we combined our results with available helium studies from the literature. As predicted by theory, we find that warm Jupiters with helium detections orbit K-type stars. However, the helium detections at equilibrium temperatures of $\sim$2000K are found in low-density planets orbiting F-type stars. We compiled a list of 46 irradiated gas giants, but more helium studies are needed to increase the sample and improve our understanding of atmospheric mass loss through helium observations.

\end{abstract}





\section{Introduction}  \label{sec:intro}





Gas giant exoplanets, particularly those on short‑period orbits, are the primary laboratories for atmospheric characterization because their large radii and extended atmospheres produce deep and readily detectable transit signals \citep{Seager2010}. Due to their proximity to the host star, they can reach high equilibrium temperatures ($T_{\rm eq}$), ranging from warm Jupiters (WJs, $T_{\rm eq}$ $\lesssim$ 1000\,K) through hot Jupiters (HJs, $T_{\rm eq}$ $\sim$1000--2000\,K) to ultra‑hot Jupiters (UHJs, $T_{\rm day}$ $\gtrsim$2150 K, \citealp{Parmentier2018}). During a transit, starlight filters through the planetary atmosphere, enabling transmission spectroscopy with high-resolution spectrographs, which probes the composition of the upper atmospheric layers. The combination of large scale heights, high irradiation, and short orbital periods makes warm, hot, and ultra-hot Jupiters particularly favorable for atmospheric studies \citep{Crossfield2015, Madhusudhan2019}. These planets have provided the first detections of exoplanetary atomic species, and escaping material \citep{HD209_Na2002, HD209_Lyalpha2003}.

The atmospheric escape of close-in planets is primarily hydrodynamic, driven by the thermal energy deposited into the exoplanet's atmosphere \citep{Pierrehumbert2010, Koskinen2014_Roche}. Photo-evaporation is the typical hydrodynamic escape process and is driven by stellar irradiation in the most energetic wavelengths, i.e., the X-ray and extreme ultraviolet (XUV\footnote{We define XUV as wavelengths up to 504\,\AA\ since this range is relevant to helium studies due to the ionization wavelength cutoff of the \ion{He}{i}. We note that other works may define the XUV range different, e.g. up to the H ionization cutoff at 912\AA.}) range \citep{Owen_Jackson_2012}. This energy input can inflate the upper atmosphere carrying light elements, such as hydrogen and helium, to higher altitudes.
Additionally, the outer layers of the atmosphere of extremely close-in planets can extend beyond the Roche lobe, resulting in a high atmospheric mass loss rate \citep{Owen2019_CloseIn}. Roche lobe overflow escape is independent of stellar flux and can act alongside other evaporation mechanisms \citep{Lecavelier2004_Roche}.
In extreme cases, these extended, escaping atmospheres can result in the formation of comet‑like tails that interact with the stellar wind, detectable via atmospheric escape tracers, such as the \ion{He}{i} triplet line (e.g., \citealp{HAT-P-32b_Zhang2023, HAT-P-67b_Gully2024}). Over time, these mass‑loss processes can sculpt planetary demographics, contributing to observed features such as the sub‑Jovian desert, and providing empirical constraints for models of photo-evaporation \citep{Owen_Jackson_2012,Owen_Lai_2018}. 

The \ion{He}{i} triplet at $\sim$10833\,\AA\ has become a key diagnostic for probing the upper atmospheres and escape processes of close‑in exoplanets, including irradiated gas giants \citep{Seager2000_He_inici, Oklopcic2018_He_inici,Lampon_2021_regimenes}. The absorption arises from the metastable 2$^3$S state of neutral helium, which is populated when stellar XUV radiation ionizes helium atoms, and subsequent recombination fills the triplet state. However, the helium metastable state has a lifetime of $\sim$2.2\,h \citep{Drake1971_He_lifetime}. Thus, a constant and intense irradiation is needed to sustain detectability of the \ion{He}{i} triplet line \citep{JorgeSanz2008}. Since its first detection in WJs (\citealp[WASP-107\,b]{Spake2018_WASP-107}; \citealp[WASP-69\,b]{Nortmann2018_WASP-69}; and \citealp[HAT-P-11\,b]{Allart2018_HAT-P-11b}), helium absorption has been reported in various types of exoplanets, providing a direct tracer of atmospheric escape accessible from ground-based telescopes. Moreover, ground-based high-resolution spectroscopy observations allow to retrieve several physical parameters from line profile fitting to study the hydrodynamical escape (e.g. \citealp{Lampon2023_molts_planetes}).

Helium observations provide a unique window into both the instantaneous escape dynamics and the long‑term sculpting of exoplanet populations. This spectral line has been extensively studied in a wide range of exoplanets using narrow-band photometry (e.g., \citealp{Vissapragada2022_narrowband}) and almost all ground-based near-infrared spectrographs (e.g., HPF, \citealp{GJ3470_Ninan2020}; NIRSPEC, \citealp{Kirk_2020}; CRIRES$^+$, \citealp{WASP-121_He}; CARMENES, \citealp{Nortmann2018_WASP-69, Allart2018_HAT-P-11b}; GIANO-B, \citealp{Guilluy_Edge_Neptune_Desert}; SPIRou, \citealp{SPIROU_He_survey_Masson2024, Allart_Helium_Survey}; NIRPS, \citealp{Allart_WASP_69b_2025}; and IRD, \citealp{IRD_helium}). Moreover, the convenience of observing the He line has been leveraged for large-scale surveys aimed at understanding the atmospheric retention, evolution, and population trends of exoplanets (e.g. \citealp{Vissapragada2022_narrowband, Guilluy_Edge_Neptune_Desert, Allart_Helium_Survey, MOPYS_Orell-Miquel2024}).

\begin{figure}[t!]
   \centering
   \includegraphics[width=\linewidth]{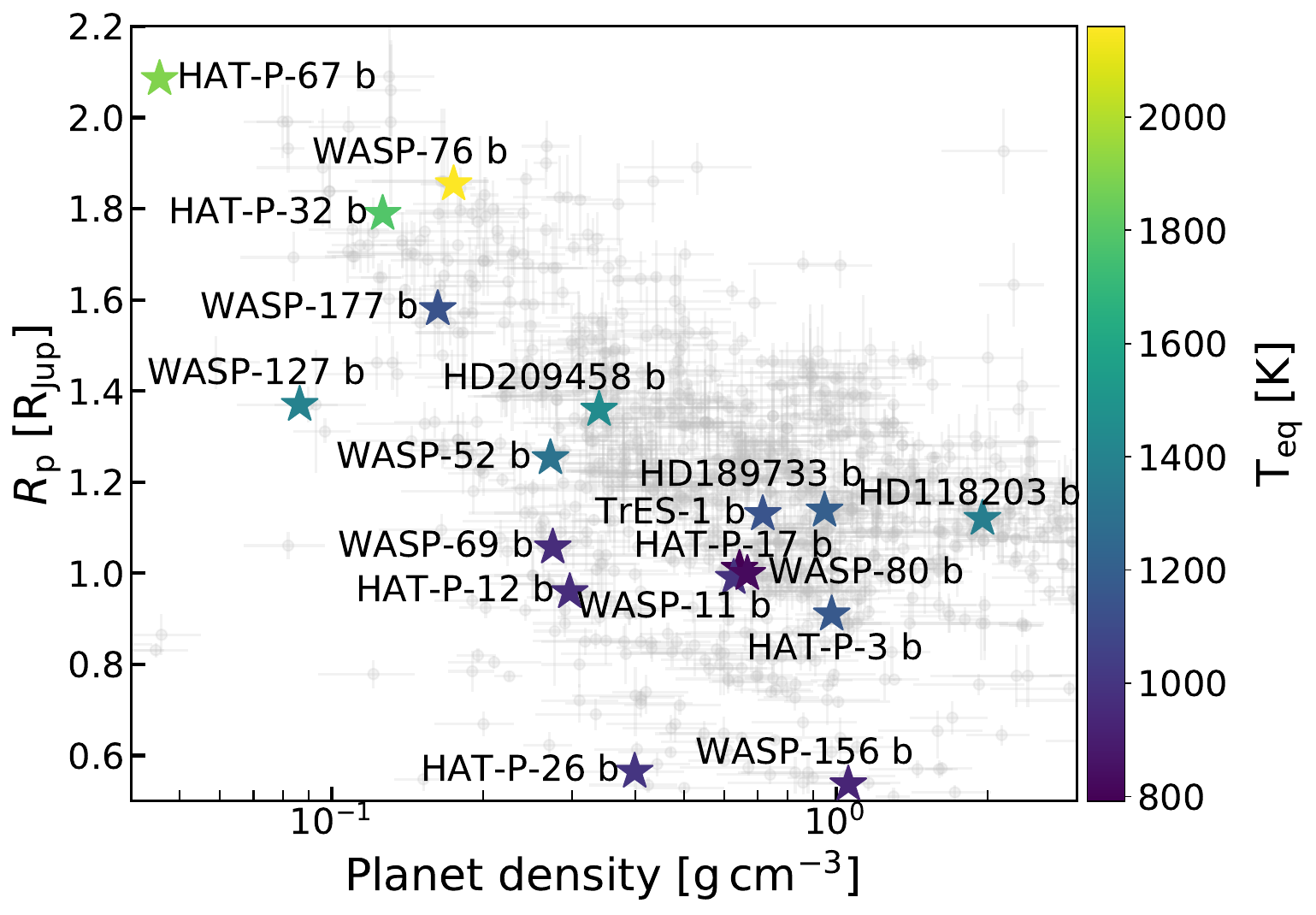}
   \caption{ \label{Fig: planet sample} Radius-density diagram for planets with $R_{\rm p}$ $>$ 0.5 R$_{\rm Jup}$. The planets observed by our HET/HPF Helium survey are labeled and color-coded according to their equilibrium temperature.
   The grey points represent all known planets with radius and density determined with precision better than 25\% and 33\%, respectively (data from NASA Exoplanet Archive as of 1 August 2025).
    }
\end{figure}

In this context, this work presents the results of an observational study aimed at characterizing atmospheric escape in gas giants. We have been carrying out a survey targeting the \ion{He}{i} triplet line for planets with $R_{\rm p}$ $>$ 0.5 R$_{\rm Jup}$ and $T_{\rm eq}$ ranging 800--2100\,K around F-, G-, and K-type host stars. Figure\,\ref{Fig: planet sample} presents the planets observed by our survey. The individual analyses and discussions for HAT-P-32\,b and HAT-P-67\,b are presented in \citet{HAT-P-32b_Zhang2023}, and \citet{HAT-P-67b_Gully2024}, respectively. Here, we only included those results, and not new analysis on these two systems in our helium detections discussion.

\section{Observations and data analysis}

\subsection{HPF spectroscopic observations}

We obtained spectroscopic observations of gas giant host stars with the Habitable Zone Planet Finder (HPF; \citealt{HPF_2012, HPF_2014, HPF_2019}) mounted on the 10\,m Hobby-Eberly Telescope (HET; \citealt{HET_1998, HET_queue_2007}). HPF spectrograph covers the near-infrared wavelength range from 0.81--1.38\,$\mu$m with a resolving power of $\mathcal{R}$\,$\simeq$\,55\,000. HPF data were reduced using the \texttt{Goldilocks}\footnote{\url{https://github.com/grzeimann/Goldilocks_Documentation}} pipeline \citep{Goldilocks_Ninan2018, Goldilocks_Kaplan2019}, obtaining the science, sky, and calibration. To avoid saturation or scattered light from the laser frequency comb, we did not use a simultaneous calibration source in the calibration fiber.

The HET telescope structure can rotate in azimuth but it is fixed at an elevation of 55 degrees. Target tracking is performed by a prime focus instrument package that moves across the focal plane to follow the target. Due to the design of the HET, we are limited to observations during ``tracks'' of typically $\sim$1 hour before (east track) and after (west track) target culmination, making it difficult observe a full transit, or to obtain in-transit and out-of-transit spectra on the same night. We organized HPF observations into visit campaigns around an exoplanet transit. We obtained in-transit spectra (IN) along with out-of-transit spectra (OUT) from surrounding nights. We also obtained random-phase spectra, referred to as P4 observations, to monitor long-term stellar variability. P4 observations are typically used to fill the HET queue or when the conditions are not ideal for higher-priority targets. Although P4 spectra are out-of-transit, they are not considered OUT because they are collected away from IN visits.

Table\,\ref{table - Log Table} provides the log of the observations for our targets, with the list of observed tracks, exposure time, and signal-to-noise ratio (S/N) per extracted pixel of the \'echelle order where the He triplet line is located. We provide a summary of the observations performed for each target:
\begin{itemize}
    
\item HAT-P-3\,b: We scheduled one visit with 16 spectra: 8 in-transit (1 February 2021) and 8 out-of-transit spectra. We additionally obtained 4 P4 spectra. 

\item HAT-P-12\,b: We collected 15 spectra on visit 1: 5 in-transit (31 March 2020) and 10 out-of-transit; and 12 spectra on visit 2: 8 in-transit (27 May 2025) and 4 out-of-transit. We also collected 2 P4 spectra. 

\item HAT-P-17\,b: We collected 23 spectra on visit 1: 11 in-transit (8 June 2020) and 12 out-of-transit; and 32 spectra on visit 2: 10 in-transit (30 July 2020) and 22 out-of-transit. We also collected 8 P4 spectra. 

\item HAT-P-26\,b: We observed one visit with 13 spectra: 5 in-transit (1 February 2021) and 8 out-of-transit spectra. We additionally obtained 2 P4 spectra.

\item HD118203\,b: We collected 26 spectra on visit 1: 18 in-transit (7 January 2021) and 8 out-of-transit; and 29 spectra on visit 2: 18 in-transit (19 February 2021) and 11 out-of-transit. We additionally obtained 2 P4 spectra. 

\item HD189733\,b: We obtained 9, 9, and 2 in-transit spectra for visits 1 (4 May 2019), 2 (13 June 2019), and 3 (1 August 2020), respectively. We also collected 11 spectra as out-of-transit in visit 4, complemented with 10 P4 spectra (two of them are later used as out-of-transit for visit 1).

\item HD209458\,b: We collected 6 spectra on visit 1: 5 in-transit (2 october 2019), and 1 out-of-transit; and 6 out-of-transit spectra on visit 2. 

\item TrES-1\,b: We obtained in total 33 spectra: 5 in-transit spectra on visit 1 (7 July 2020), 5 in-transit spectra on visit 2 (13 July 2020), and 23 out-of-transit spectra contemporaneous to both visits. We additionally obtained 9 P4 spectra. 

\item WASP-11\,b/HAT-P-10\,b: We collected 18 spectra on visit 1: 5 in-transit (7 December 2019) and 13 out-of-transit; and 12 spectra on visit 2: 5 in-transit (18 December 2019) and 5 out-of-transit. We also collected 8 P4 spectra. 

\item WASP-52\,b: We collected 9 spectra on visit 1: 4 in-transit (13 August 2019) and 5 out-of-transit; and 12 spectra on visit 2: 5 in-transit (10 October 2019) and 7 out-of-transit. 

\item WASP-69\,b: We collected a single in-transit spectrum on visit 1 (20 September 2019), and 23 spectra on visit 2: 7 in-transit (21 October 2019) and 16 out-of-transit. 

\item WASP-76\,b: We collected 22 spectra on visit 1: 10 in-transit (2 October 2020) and 12 out-of-transit; and 18 spectra on visit 2: 10 in-transit (4 October 2020) and 8 out-of-transit. 

\item WASP-80\,b: We observed one visit with 11 spectra: 8 in-transit (13 July 2019) and 3 out-of-transit spectra. We additionally obtained 16 P4 spectra. 

\item WASP-127\,b: We observed one visit with 35 spectra: 10 in-transit (13 March 2019) and 25 out-of-transit spectra. We additionally obtained 4 P4 spectra. 

\item WASP-156\,b: We collected 14 spectra on visit 1: 6 in-transit (24 October 2020) and 8 out-of-transit; and 18 spectra on visit 2: 5 in-transit (20 November 2020) and 13 out-of-transit. 

\item WASP-177\,b: We obtained in total 35 spectra: 4 in-transit spectra on visit 1 (20 August 2020), 4 in-transit spectra on visit 2 (23 August 2020), and 27 out-of-transit spectra contemporaneous to both visits. 
\end{itemize}


\subsection{Telluric correction}
\label{subsect: Telluric correction}

The main objective of these observations was the analysis of the planetary \ion{He}{I} triplet line, which is surrounded by H$_2$O absorption and OH emission lines from the Earth's atmosphere. The relative positions of the \ion{He}{I} triplet and telluric spectral lines change with the epoch due to the Doppler shift induced by the barycentric Earth radial velocity. Thus, we mainly scheduled the campaigns at epochs that minimize the overlap of the \ion{He}{I} planetary trace and the telluric lines \citep{GJ1214b_Orell-Miquel2022}. Moreover, we also correct the science spectra from telluric lines to minimize their impact on the planetary signals.
We use the \texttt{python} package \texttt{muler} \citep{muler_Gully2022} to correct the target spectra from the telluric emission lines. The \texttt{muler} code uses the simultaneous sky spectra and accounts for the different efficiencies between target and sky fibers \citep{HAT-P-32b_Zhang2023, HAT-P-67b_Gully2024}. We remove the telluric absorption lines using a telluric model computed from HPF/HET observations\footnote{The model is available at the \texttt{Goldilocks} pipeline documentation (telluric$\_$pca.fits).}, thus accounts for the instrumental profile and spectral resolution. For each individual spectra, we empirically fit the strongest H$_2$O lines to account for strength line differences between the model and the data. The observed spectrum is divided by the scaled model, removing the telluric absorption features (e.g., \citealp{Yan2015_telluricmodel, Nuria2017_telluric}).

\subsection{Transmission spectrum analysis}
\label{subsect: Telluric correction}

We analyzed the HPF spectroscopic time series using the single line transmission spectroscopy technique, which is a standard procedure that compares the in-transit spectra with the out-of-transit spectra (e.g., \citealp{Wyttenbach_2015, Nuria2017_telluric}). In particular, we adapted the methodology presented in \cite{GJ1214b_Orell-Miquel2022, Orell2023b_He_TOI1430, MOPYS_Orell-Miquel2024} to the particularities of the HET following \cite{HAT-P-32b_Zhang2023} and \cite{HAT-P-67b_Gully2024}. Table\,\ref{table - TRANSIT PARAMETERS} shows the planetary and stellar parameters needed to compute the transmission spectrum for our targets.

Once corrected from the telluric contributions, we normalized each spectrum by its own continuum. We used \texttt{muler} to fit a first order Chebyshev's polynomial to featureless spectral regions. Then, we shifted the normalized spectra into the stellar rest frame. We computed a high S/N stellar spectrum from the out-of-transit data (MasterOut). To minimize the impact of stellar variability, we analyzed each in-transit observation individually.
For each transit visit, we computed a different MasterOut spectrum using only the out-of-transit spectra that are closer to the in-transit track. We usually considered all the out-of-transit spectra from a particular visit to compute the MasterOut. This option provides the most optimal results: it takes into account the behavior of the He line in short time scales, and increases the number of spectra reducing the uncertainties. On the contrary to OUT spectra, P4 spectra are usually taken far in time from the visits and the stellar He line may have varied significantly for the required precision. Moreover, P4 spectra may not be taken at optimal epochs in terms of telluric contamination.
Then, we divided all the spectra (in-transit, out-of-transit, and P4) by the MasterOut to remove the stellar contribution and to obtain the residual map in the stellar rest frame. At this step, we computed the stellar light curve of the \he\ triplet integrating the counts within a 1.5\,\angstrom\ band pass centered at the doublet position. The light curves are shown in the Figures\,\ref{Fig: He LC plot 1} and \ref{Fig: He LC plot 2}, and provide information of the stellar variability and the presence of large planetary tails \citep{HAT-P-32b_Zhang2023, HAT-P-67b_Gully2024}.

We moved the residual map into the planet rest frame accounting for the planet movement around the star. We neglected the planets' eccentricity because it is usually poorly constrained and the impact is within the uncertainties derived for the signal shifts, except for HAT-P-17\,b, HAT-P-26\,b, and \hdu\,b. For these three planets, we used the \texttt{radvel} (\citealp{radvel}) package to compute the radial velocity contribution of the eccentricity during the transit.
We computed the transmission spectrum (TS) by averaging the in-transit spectra weighted by the inverse of the squared propagated errors. We also weighted the ingress and egress spectra to account for their true contribution into the TS average (e.g., \citealp{Allart_Helium_Survey, Allart_WASP_69b_2025}). In particular, we linearly interpolated from 0 at out-of-transit and 1 at full in-transit. If we obtained more than one in-transit visit, we computed the final TS from the combined residual maps. When the TS shows an absorption feature, we fitted the signal with a Gaussian profile using the Markov's chain Monte Carlo algorithm via its {\tt python} implementation \texttt{emcee} \citep{emcee}. If there were no clear signals in the TS, we derived conservative limits for the planetary absorption. We computed a 3$\sigma$ upper limit as three times the root-mean-squared value of a flat spectral region of the TS. We also estimated the upper limits in terms of equivalent width (EW) as EW\,=\,$\mid$\,$A$\,$\mid$\,$\sqrt{2 \pi \sigma^2}$, where $A$ is the 3$\sigma$ upper limit in units of normalized flux, and $\sigma$ is the average width of the He detections ($\sigma$\,=\,0.43\,m$\rm \AA$; \citealp{MOPYS_Orell-Miquel2024}).


\section{Transmission spectroscopy results}
\label{sect: Results}

\begin{table*}
\centering
\caption{
\label{table - Detections He}
Results from the Gaussian profile fitting to the \ion{He}{I} triplet detections. We fitted the depth, the width ($\sigma$), and the velocity shift from the doublet line ($\Delta v$). We computed the equivalent width (EW) and the full width at half maximum (FWHM).
}
\centering
\begin{tabular}{lccccc c }
\hline \hline 
\noalign{\smallskip} 

Planet & Depth & EW  & $\Delta v$ & $\sigma$  & FWHM & Singlet depth\vspace{0.05cm}\\
 &  (\%) & [m\AA] & [km\,s$^{-1}$] &  [\AA] & [\AA] & (\%) \vspace{0.05cm}\\
\hline
\noalign{\smallskip}

HD189733\,b &  0.45$\pm$0.08   &  2.20$\pm$0.40   &   $-$1.3$\pm$1.1  &  0.19$\pm$0.03   &  0.46$\pm$0.08  &  -- \\

\hd209\,b &  0.67$\pm$0.08   &  6.26$^{+0.75}_{-0.70}$   &   $-$6.5$\pm$1.5  &  0.37$^{+0.06}_{-0.05}$   &  0.88$^{+0.14}_{-0.12}$  &  -- \\

WASP-52\,b &  1.93$\pm$0.20   & 35.0$\pm$4.0   &  $-$3.1$\pm$2.5   &  0.73$^{+0.11}_{-0.10}$   &  1.71$\pm$0.25  &  -- \\

WASP-69\,b &  3.50$\pm$0.10   &  43.4$\pm$1.1   &  $-$5.6$\pm$0.4   &   0.460$\pm$0.017  &  1.08$\pm$0.04   &  0.68$\pm$0.12 \\

WASP-76\,b &  0.48$\pm$0.09   &  6.0$\pm$1.0   &   2.7$^{+2.3}_{-2.6}$  &  0.48$^{+0.13}_{-0.10}$   &  1.15$^{+0.30}_{-0.23}$  &  -- \\

\noalign{\smallskip}
\hline
\end{tabular}

\end{table*}

\begin{table}
\caption{
\label{table - Upper limits He}
Upper limits for the observations with no \ion{He}{I} detection. 
}
\centering
%

\begin{tabular}{ l c c }

\hline \hline 
\noalign{\smallskip} 

Planet  & Depth ($\%$) & EW [m\AA] \\

\noalign{\smallskip}
\hline
\noalign{\smallskip}

HAT-P-3\,b  &  $<$1.2 & $<$12.7 \\ 
\hline \noalign{\smallskip}

HAT-P-12\,b  &  $<$2 & $<$21.2 \\ 
\hline \noalign{\smallskip}

HAT-P-17\,b  &  $<$1.1 & $<$11.8 \\ 
\hline \noalign{\smallskip}

HAT-P-26\,b  &  $<$1.2 & $<$12.7 \\ 
\hline \noalign{\smallskip}

\hdu\,b  &  $<$0.2 & $<$2.1 \\ 
\hline \noalign{\smallskip}


TrES-1\,b  &  $<$1.3 & $<$13.8 \\ 
\hline \noalign{\smallskip}

WASP-11\,b  &  $<$1.2 & $<$12.7 \\ 
\hline \noalign{\smallskip}

WASP-80\,b  &  $<$0.8 & $<$8.5 \\ 
\hline \noalign{\smallskip}

WASP-127\,b  &  $<$0.6 & $<$6.4 \\ 
\hline \noalign{\smallskip}

WASP-156\,b  &  $<$0.9 & $<$9.5 \\ 
\hline \noalign{\smallskip}

WASP-177\,b  &  $<$1.1 & $<$11.6 \\ 

\hline

\end{tabular}


\end{table}

\begin{figure*}[ht!]
   \centering
   \includegraphics[width=\linewidth]{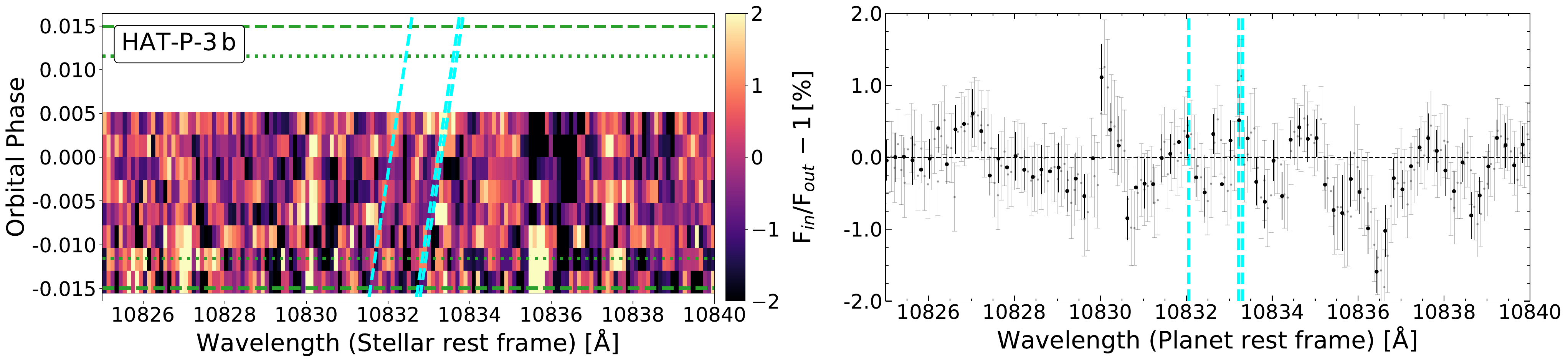}
   \includegraphics[width=\linewidth]{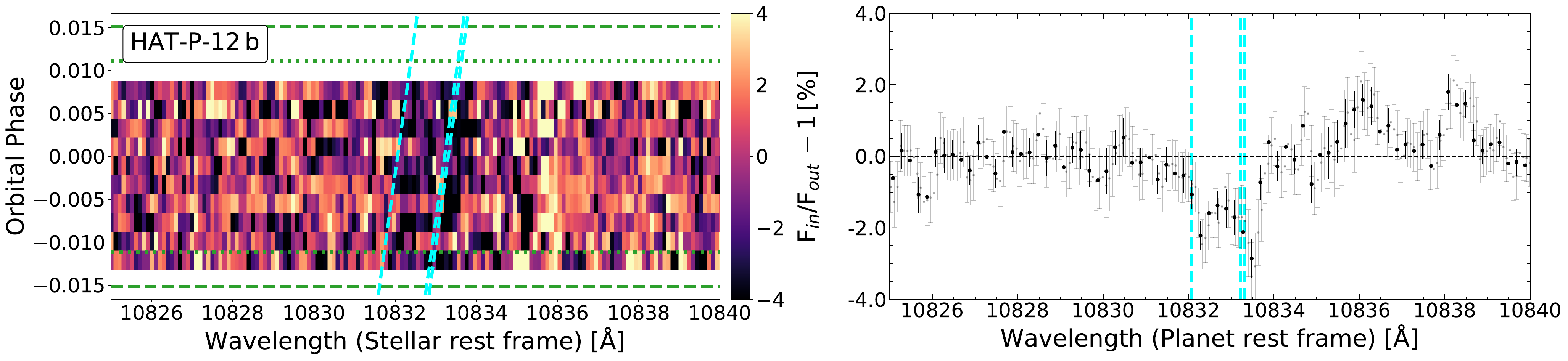}
   \includegraphics[width=\linewidth]{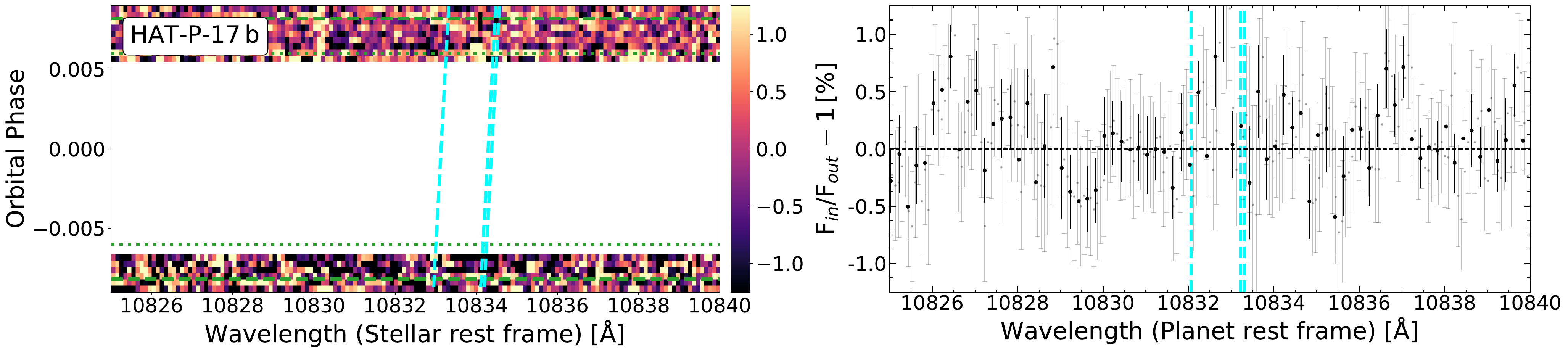}
   \includegraphics[width=\linewidth]{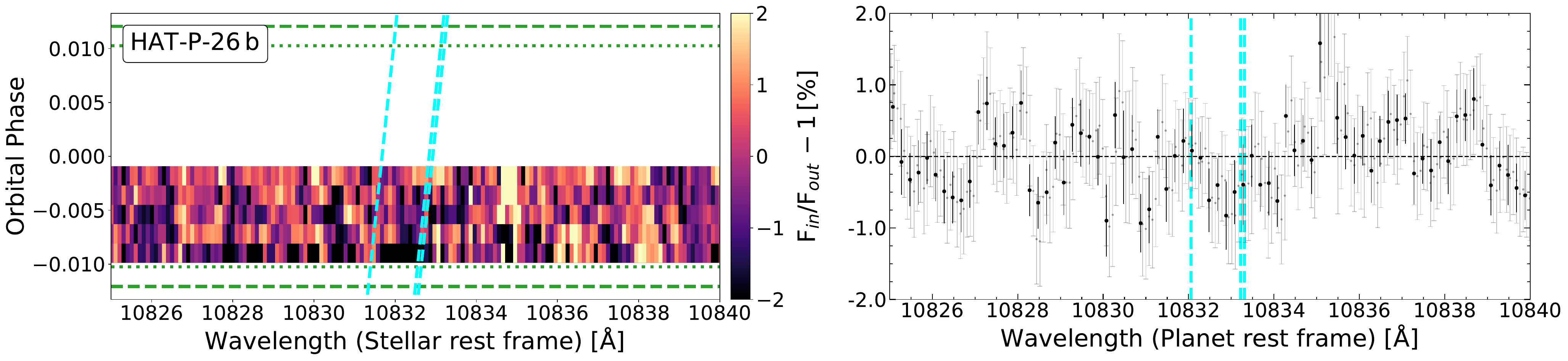}
   \caption{ \label{Fig: TS plot 1}
    Residual maps and transmission spectra around the \ion{He}{I} triplet lines for HAT-P-3\,b, HAT-P-12\,b, HAT-P-17\,b, and HAT-P-26\,b.
    \textit{Left panels}: Residual maps in the stellar rest frame as a function of wavelength and orbital phase. Relative absorption is color-coded. The dashed and dotted green horizontal lines indicate the different contacts during the transit. The dashed cyan tilted lines indicate the predicted trace of the planetary signal. The regions affected by strong OH residuals and telluric H$_2$O absorption are marked.
    \textit{Right panels}: Planet transmission spectra (TS) in the planet rest frame. We show the original data in light gray and the data binned by 0.2\,\AA\ in black. When an absorption signal is fitted, a red line and shaded region show the best fit model with its $1\sigma$ uncertainty. The dotted cyan vertical lines indicate the \ion{He}{I} triplet line position.
    }
\end{figure*}

\begin{figure*}[ht!]
   \centering
   \includegraphics[width=\linewidth]{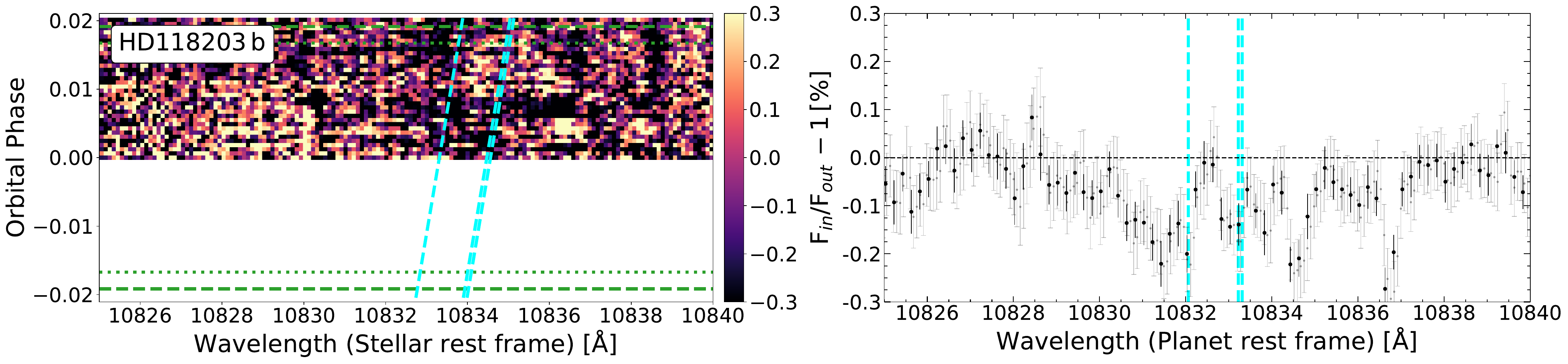}
   \includegraphics[width=\linewidth]{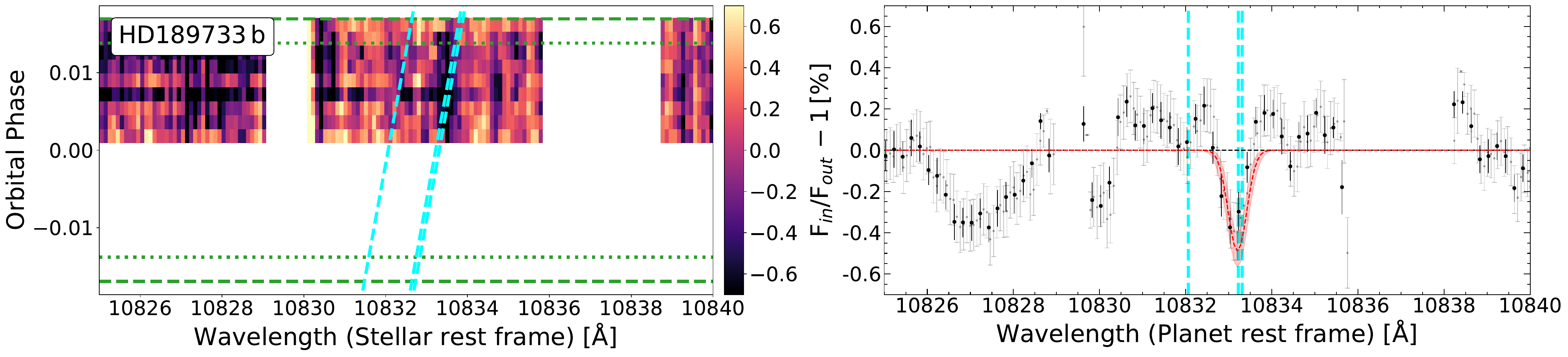}
   \includegraphics[width=\linewidth]{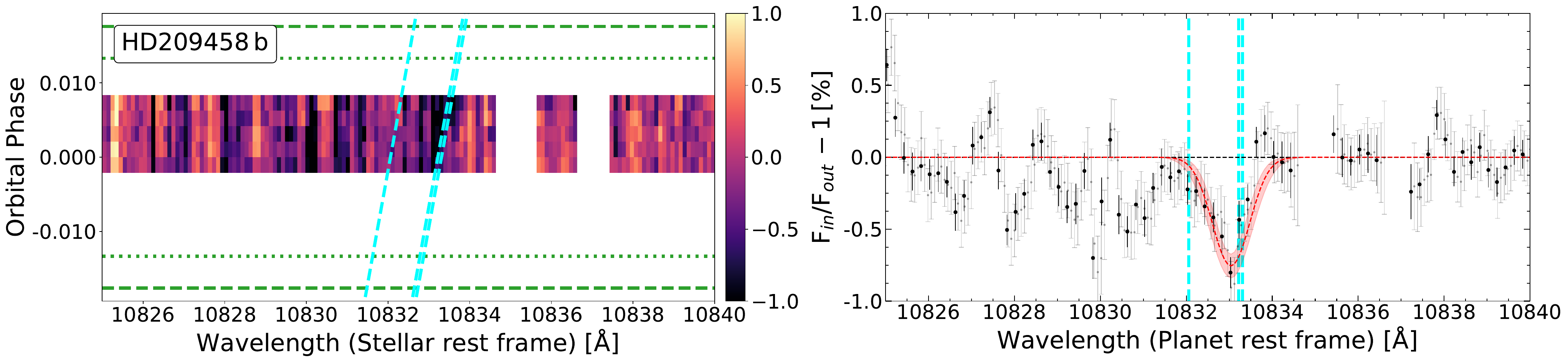}
   \includegraphics[width=\linewidth]{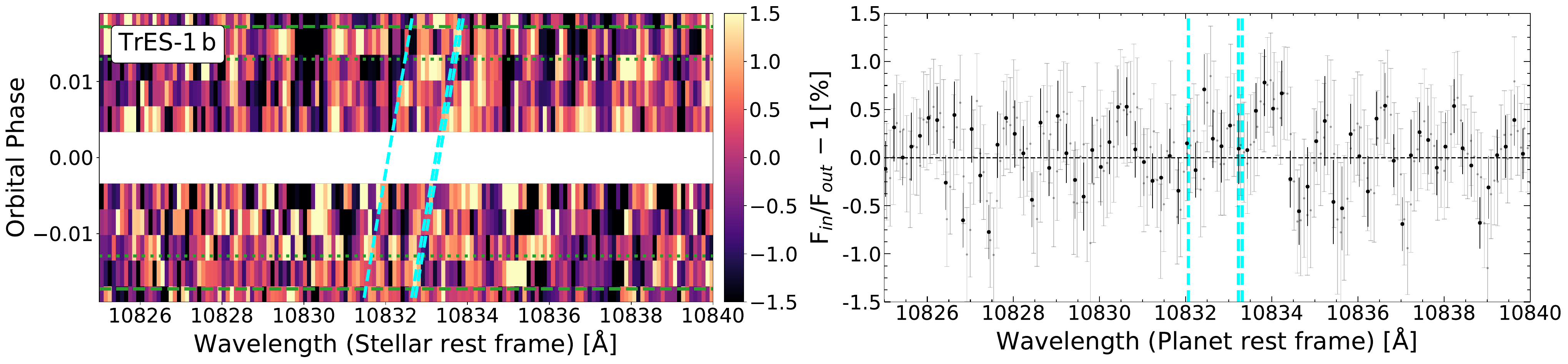}
   \caption{ \label{Fig: TS plot 2} Same as Figure\,\ref{Fig: TS plot 1} for HD118203\,b, HD189733\,b, \hd209\,b, TrES-1\,b.
    }
\end{figure*}

\begin{figure*}[ht!]
   \centering
   \includegraphics[width=\linewidth]{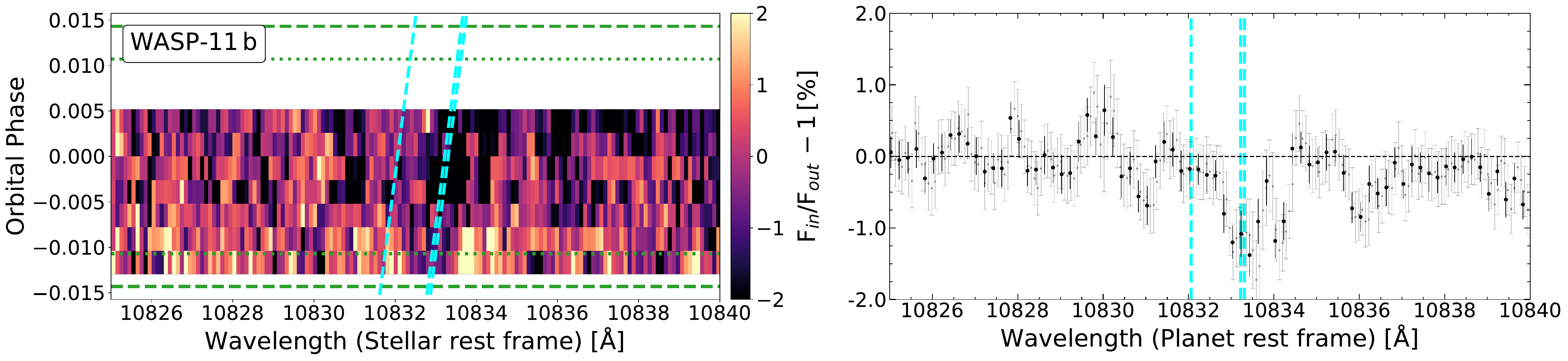}
   \includegraphics[width=\linewidth]{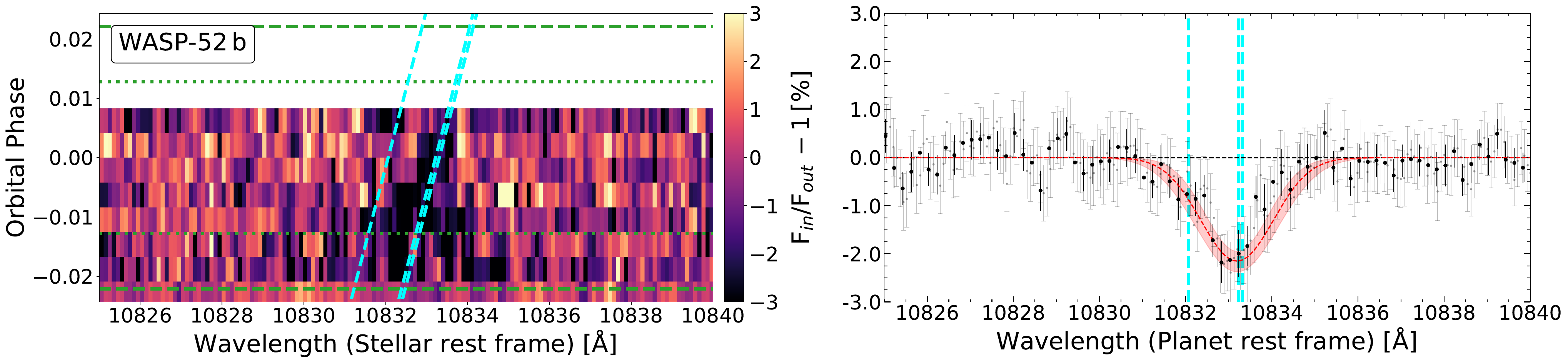}
   \includegraphics[width=\linewidth]{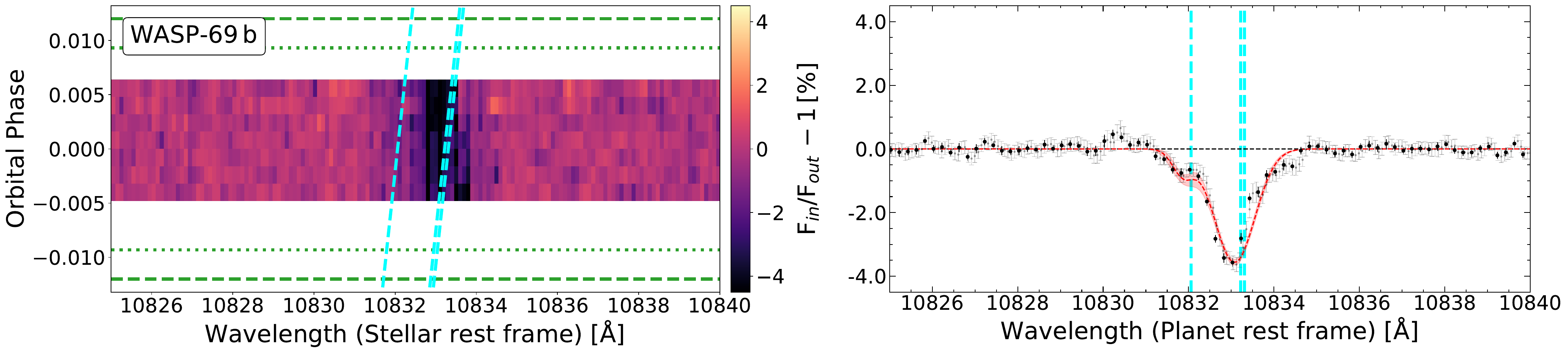}
   \includegraphics[width=\linewidth]{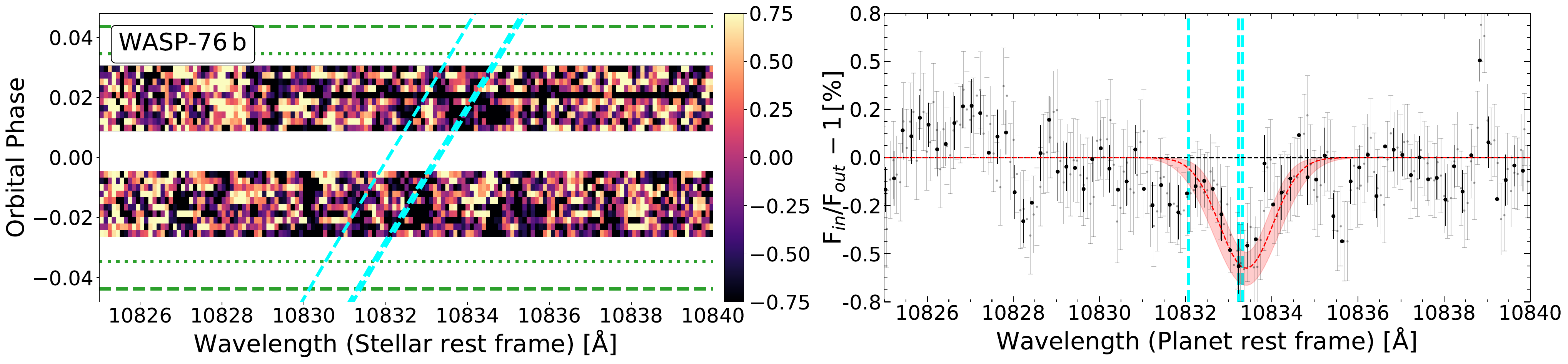}
   \caption{ \label{Fig: TS plot 3} Same as Figure\,\ref{Fig: TS plot 1} for WASP-11\,b (HAT-P-10\,b), WASP-52\,b, WASP-69\,b, WASP-76\,b.
    }
\end{figure*}

\begin{figure*}[ht!]
   \centering
   \includegraphics[width=\linewidth]{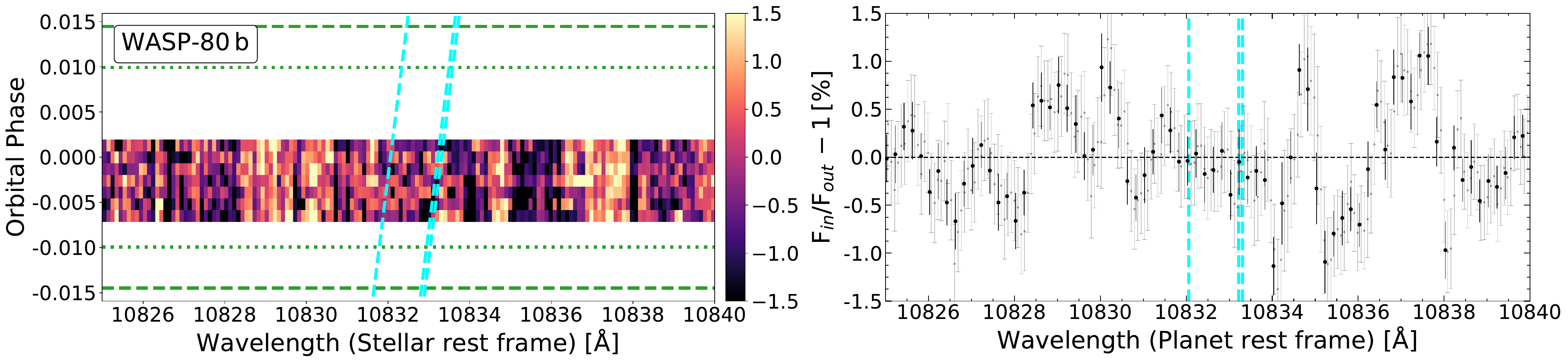}
   \includegraphics[width=\linewidth]{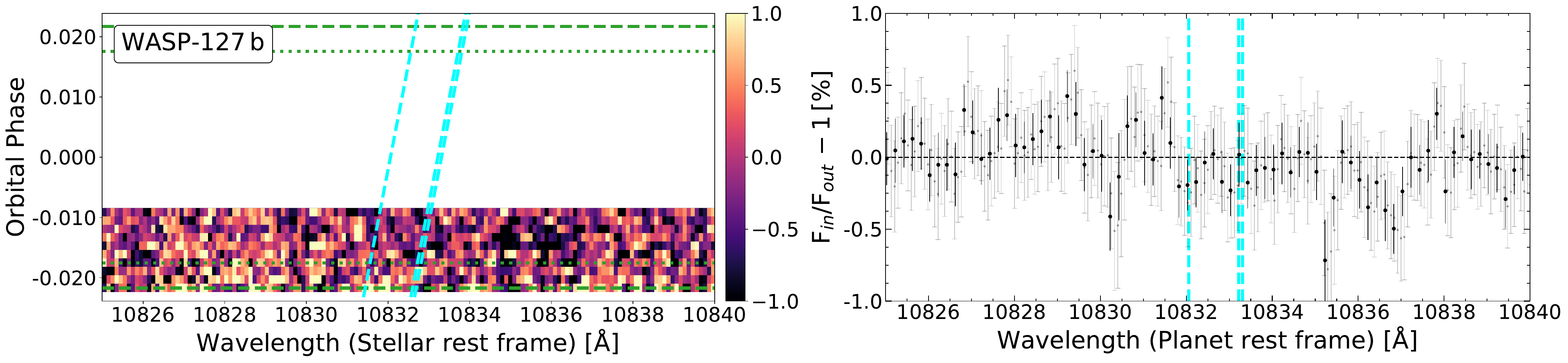}
   \includegraphics[width=\linewidth]{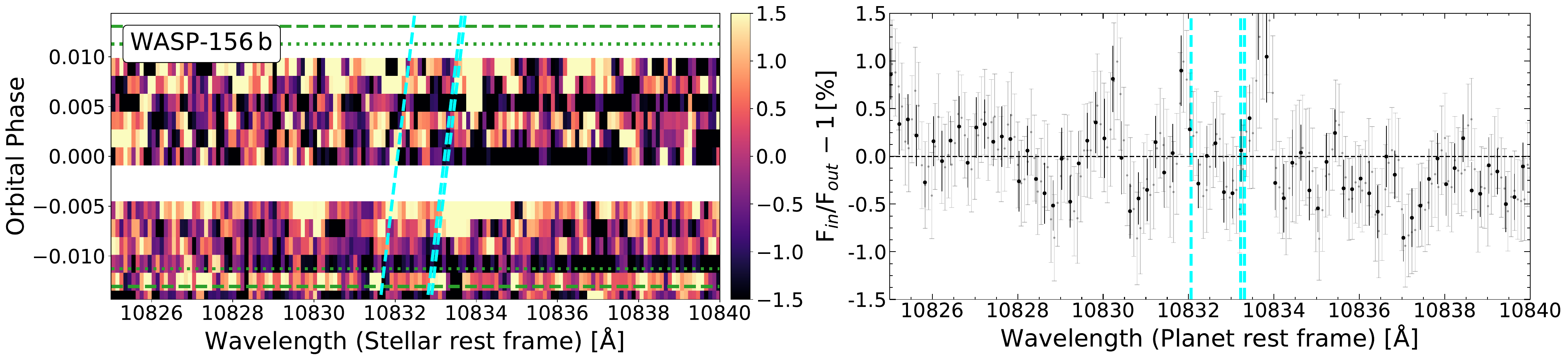}
   \includegraphics[width=\linewidth]{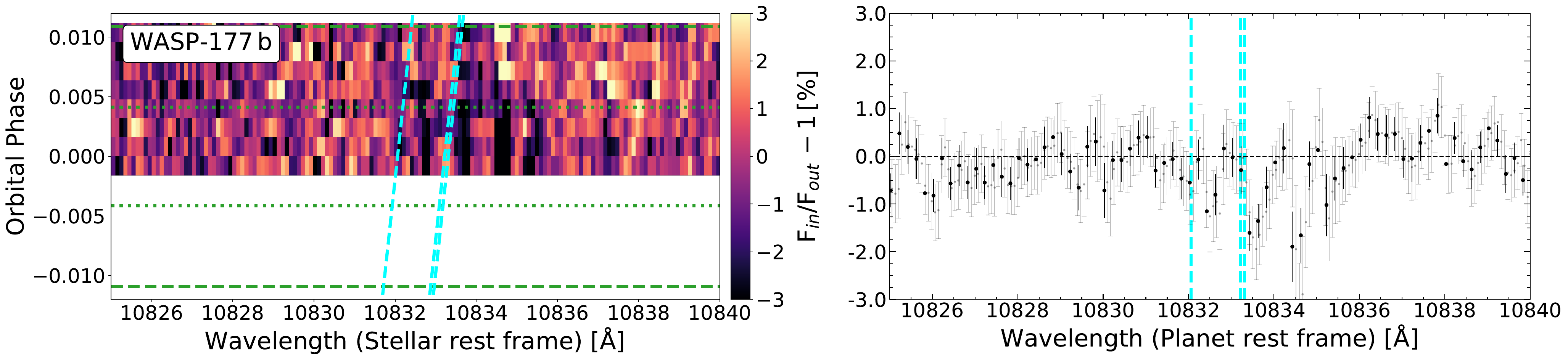}
   \caption{ \label{Fig: TS plot 4} Same as Figure\,\ref{Fig: TS plot 1} for WASP-80\,b, WASP-127\,b, WASP-156\,b, and WASP-177\,b.
    }
\end{figure*}

In this section, we present the \ion{He}{i} triplet results obtained by applying the methodology described in the previous section. The final residual maps and TS for each planet are shown in Figures\,\ref{Fig: TS plot 1}--\ref{Fig: TS plot 4}. Table\,\ref{table - Detections He} reports the results of our Gaussian profile fitting for the helium signals. In Table\,\ref{table - Upper limits He}, we report the derived upper limits for planetary \ion{He}{i} absorptions from our HPF TS.

Figures\,\ref{Fig: Visits plot 0}--\ref{Fig: Visits plot 3} show the performance of the telluric correction and the comparison between the MasterOut and the individual in-transit spectra. For multiple visits, we also present the individual TS. The \he\ triplet stellar light curves are shown in Figures\,\ref{Fig: He LC plot 1} and \ref{Fig: He LC plot 2}.

The main objective of this work is the study of the \ion{He}{i} triplet line, but other strong spectral features are accessible to HPF. In Appendix\,\ref{App: Ca IRT}, we provide a note on the study of the \ion{Ca}{ii} infrared triplet lines with HET/HPF.


\subsection{HAT-P-3\,b}

The \he\ triplet stellar line partially overlaps with the OH lines in the only visit to the target, but the telluric emission correction is well performed (Fig.\,\ref{Fig: Visits plot 0}). The residual map and TS do not indicate the presence of a planetary absorption in our data. We found no significant differences in the TS if we only use the pre- or post-transit spectra to compute the MO, compared to the fiducial results using both sets. The TS is slightly more noisy when we add the P4 spectra into the MasterOut computation. We derive a 3$\sigma$ upper limit to the \he\ excess absorption of 1.2\,\%.


\subsection{HAT-P-12\,b}

The planetary trace is overlapped by OH emission lines in the two visits (Fig.\,\ref{Fig: Visits plot 0}). While the TS from the first visit shows a broad $\sim$2\,\% excess absorption feature, the TS from the second visit is dominated by noise without a clear absorption signal. When combining both observations, final TS (Fig.\,\ref{Fig: TS plot 1}) looks similar to that of the first visit. The MasterOut from the visit 1 shows a less deep He stellar line, than that from visit 2. This could explain the prominent absorption feature in visit 1. Visit 1 has pre-transit observations during the same night, but those spectra have a relatively low S/N and we obtain a less noisy TS when combining all the out-of-transit spectra for visit 1. It is evident that both transits are impacted by some OH telluric residuals, which prevent us provide a clear results. We do not consider the P4 spectra because it is affected by He airglow, and also far from both visits.
Although, the S/N of the final TS would be sensitive to a \he\ triplet excess absorption of 1.3\,\%, we report a conservative upper limit of $\sim$2\,\% due to the telluric contamination.

\subsection{HAT-P-17\,b}

The stellar line does not overlap with the telluric lines, but the planetary trace is in between the OH emission line and the H$_2$O absorption line due to the RV shift caused by the planet's eccentricity (Fig.\,\ref{Fig: Visits plot 0}).
We obtained two visits which sampled the terminators of HAT-P-17\,b (visit 1: ingress, visit 2: egress). Although we did not get full in-transit spectra, these observations would still be able to detect evaporation tails.
Despite the telluric contamination, the TS are mainly flat without significant features (Fig.\,\ref{Fig: Visits plot 0}). However, visit 1 TS has larger uncertainties than that from visit 2, expected because we collected more spectra and with higher S/N in visit 2. There is no significant improvement when adding the P4 spectra for the MasterOut calculation. The combined TS (Fig.\,\ref{Fig: TS plot 1}) does not present absorption features from the planet atmosphere. We derive a 3$\sigma$ upper limit to the \he\ excess absorption of 1.1\,\%.

\subsection{HAT-P-26\,b}


The \he\ stellar line is free from tellurics, but the planetary trace falls over the blue doublet of the OH emission lines when accounting for the RV shift caused by the planet's eccentricity (Fig.\,\ref{Fig: Visits plot 0}).
Despite the telluric overlapping, the TS is mainly flat (Fig.\,\ref{Fig: TS plot 1}) and does not suggest \he\ excess absorption during the transit. There is no significant improvement if we consider the few P4 spectra.
The 3$\sigma$ upper limit to the \he\ excess absorption in our observations is of 1.2\,\%.

\subsection{HD118203\,b}

The \he\ stellar line is not contaminated by telluric lines (Fig.\,\ref{Fig: Visits plot 0}). However, the planetary signal falls close to the OH peak at 10835\,\AA\ when accounting for the RV shift caused by the planet's eccentricity. Visit 1 does not show evidence of planetary absorption, while visit 2 presents a broad $\sim$0.3\,\% absorption feature. However, this signal is similar to other features in both TS. The residual map (see Fig.\,\ref{Fig: TS plot 2}) does not show a clear planetary trace and the combined TS is mainly flat. Although the stellar LC is consistently negative at in-transit phases (Fig.\,\ref{Fig: He LC plot 1}), we do not consider that the signal is of a planetary origin.
If we combine the two visits, we can derive a 3$\sigma$ upper limit to the \he\ excess absorption of 0.2\,\%.

\subsection{HD189733\,b}

We obtained three in-transit visits that sample almost all the planetary transit (visits 1, 2, and 3 in Fig.\,\ref{Fig: He LC plot 1}). The planetary trace does not overlap with the telluric lines (Fig.\,\ref{Fig: Visits plot 1}). The only observations planned as out-of-transit baseline (visit 4 in Fig.\,\ref{Fig: He LC plot 1}) are more than one month apart from the in-transit visits and the resulting TS is dominated by \he\ stellar line variability and without planetary absorption (see Fig.\,\ref{Fig: Visits plot 1}). Fortunately, we took two spectra as P4 observations just a few days after visit 1 (those are the two spectra at phase $\sim$ $-$0.4 colored in blue, and included as visit 1 in Figure\,\ref{Fig: He LC plot 1}). We found that the MasterOut from visit 4 and from these two P4 spectra have a $\sim$1\,\% discrepancy at the He stellar line. This difference is similar in magnitude to the planetary signal reported in the literature. We computed the residual map and TS shown in Fig.\,\ref{Fig: TS plot 2} using those two originally P4 spectra as out-of-transit and only visit 1 as in-transit. We masked some strong telluric residuals as well. The TS show a significant absorption feature at the \ion{He}{i} triplet position (0.45$\pm$0.08\,\%). However, we emphasize that there are other features in the TS of the same amplitude.

\subsection{\hd209\,b}


We observed one in-transit visit, where the stellar line is free from tellurics (Fig.\,\ref{Fig: Visits plot 1}). Since we detect some stellar variability between visits (Fig.\,\ref{Fig: He LC plot 1}), we used the closer out-of-transit visit which has only one good S/N spectrum. Although there are some telluric residuals, the residual map and the TS present a tentative absorption feature at the planetary trace position. If we fit the signal, we derive a blue-shifted excess absorption of 0.67$\pm$0.08\,\%. However, we note that the amplitude of this signal is similar to that of the stellar variability (Fig.\,\ref{Fig: He LC plot 1}).

\subsection{TrES-1\,b}

The stellar line overlaps with the telluric blue OH doublet, but there are no significant telluric residuals in the individual TS (Fig.\,\ref{Fig: Visits plot 1}). The residual map and combined TS do not present evidences of planetary absorption (Fig.\,\ref{Fig: TS plot 2}) and we can only derive a 3$\sigma$ upper limit to the \he\ absorption of 1.1\,\%.

\subsection{WASP-11\,b / HAT-P-10\,b}

The stellar line overlaps with the OH line and some telluric residuals might be present in the individual TS of both visits (Fig.\,\ref{Fig: Visits plot 1}).
The first visit shows a feature which we fit as excess absorption of 1.0$\pm$0.2\,\% but with no clear origin. The second visit shows features which seem to have a clear origin in the telluric residuals. The combined residual map (Fig.\,\ref{Fig: TS plot 3}) presents a tentative absorption aligned with the planetary trace, and the TS shows an absorption feature similar to those detected in the individual TS. However, the combined transit is more noisy, influenced by the second transit data. We note that the stellar LC presents variability of similar amplitude to the observed feature (Fig.\,\ref{Fig: He LC plot 2}). Therefore, we consider the absorption from visit 1 as the upper limit of the He triplet absorption. The \he\ detection would need to be confirmed by completely uncontaminated transit observations.

\subsection{WASP-52\,b}

The first visit is slightly contaminated by OH emission, but the TS presents an absorption at the \he\ triplet position (Fig.\,\ref{Fig: Visits plot 1}). We fitted the feature obtaining a blue-shifted ($\Delta v$=$-$7$\pm$3\,km\,s$^{-1}$) excess absorption of 2.4$\pm$0.4\,\% (EW\,=\,33$\pm$5\,m\AA). The TS of the second visit shows a clear absorption feature as well. In the second case, we fitted an excess absorption of 1.9$\pm$0.3\,\% ($\Delta v$=$-$1$\pm$3\,km\,s$^{-1}$, EW\,=\,32$\pm$5\,m\AA). The larger absorption and shift detected in visit 1 may be due to the slightly telluric overlap. Figure\,\ref{Fig: TS plot 3} shows the residual map and the TS when combining both visits. Our observations provide a clear \he\ triplet detection with an excess absorption of 1.85$\pm$0.20\,\%.

\subsection{WASP-69\,b}

The stellar line from both visits slightly overlap with the OH emission line, but there are no obvious telluric residuals. Although the first visit only has one in-transit spectrum with a useful S/N, the He line is clearly deeper than the MasterOut spectrum (Fig.\,\ref{Fig: Visits plot 2}). From visit 1, we obtained a noisy TS showing a feature absorption of 5$\pm$2\,\%. The TS from visit 2 clearly shows an excess absorption signal from the planetary atmosphere. For this planet, we used two Gaussian profiles to fit simultaneously the doublet and singlet lines of the \he\ triplet.
We obtained an excess absorption of 3.50$\pm$0.10\,\% for the doublet and of 0.68$\pm$0.12\,\% for the singlet. The planetary signal presents a significant blue-shift ($\Delta v$=$-$5.6$\pm$0.4\,km\,s$^{-1}$). We report as our final results those from visit 2 alone (Fig.\,\ref{Fig: TS plot 3}).

\subsection{WASP-76\,b}

The \he\ stellar line does not overlap with telluric lines, although the planetary trace does due to the large velocity shift during the transit (Fig.\,\ref{Fig: Visits plot 2}). However, we do not detect residuals from the telluric correction. The TS of the first visit shows an absorption feature (0.55$\pm$0.15\,\%) coincident with the planetary position.
The TS of the second night is a bit more noisy with an absorption feature of 0.50$\pm$0.10\,\%.
When combining the spectra from both visits (Fig.\,\ref{Fig: TS plot 3}), the final TS shows a clear absorption feature with a significant absorption of 0.48$\pm$0.09\,\% ($\Delta v$=3$\pm$2\,km\,s$^{-1}$). The planetary absorption is also visible in the \he\ line curve (Fig.\,\ref{Fig: He LC plot 2}).




\subsection{WASP-80\,b}

The residual map and TS (Fig.\,\ref{Fig: TS plot 3}) do not show traces of planetary absorption. The telluric lines do not overlap with the planetary trace (Fig.\,\ref{Fig: Visits plot 2}), but there are some strong telluric residuals surrounding the \he\ triplet position.
We derive a 3$\sigma$ upper limit to the \he\ triplet excess absorption of 0.8\,\%.


\subsection{WASP-127\,b}

The planetary trace slightly overlaps with the blue OH doublet (Fig.\,\ref{Fig: Visits plot 2}), but the TS does not show residuals from the telluric correction. The residual map and TS are mainly flat without evidence of atmospheric absorption (Fig.\,\ref{Fig: TS plot 4}). We derive a 3$\sigma$ upper limit to the \he\ triplet excess absorption of 0.6\,\%.


\subsection{WASP-156\,b}

The stellar line partially overlaps with the telluric lines, mainly in the second visit (Fig.\,\ref{Fig: Visits plot 2}), thus the individual TS present some telluric residuals. The residual map and combined TS do not present evidence of planetary absorption (Fig.\,\ref{Fig: TS plot 4}). We derive a 3$\sigma$ upper limit to the \he\ absorption of 0.9\,\%. We obtain consistent results when considering different combinations of out-of-transit spectra for the MO.

\subsection{WASP-177\,b}

We note this planet has slightly grazing transits. We obtained two visits with in- and out-of-transit observations in the same night, as well as out-of-transit observations in the surrounding nights. In both visits, the planetary trace does not overlap with telluric lines (Fig.\,\ref{Fig: Visits plot 3}).
When analyzing visit 1 using the out-of-transit spectra from the surrounding nights, the TS shows a clear absorption-like feature. However, when using the out-of-transit spectra from the same night, this feature seems noisier. We decide to use this latter option (Fig.\,\ref{Fig: Visits plot 3}) because it minimizes any stellar variability impact.
For visit 2, we found no significant differences when using the out-of-transit spectra from the same night or from one night before and after. The TS from visit 2 is mainly flat, which increases the doubts about the planetary origin of the feature from visit 1.
The combined residual map and TS do not provide strong evidence for a planetary absorption. We report a 3$\sigma$ upper limit of 1.1\,\%.

\section{Discussion}

\subsection{Comparison of HPF results with literature measurements}
\label{sect: comparison results}

\begin{figure*}[t!]
   \centering
    \includegraphics[width=\linewidth]{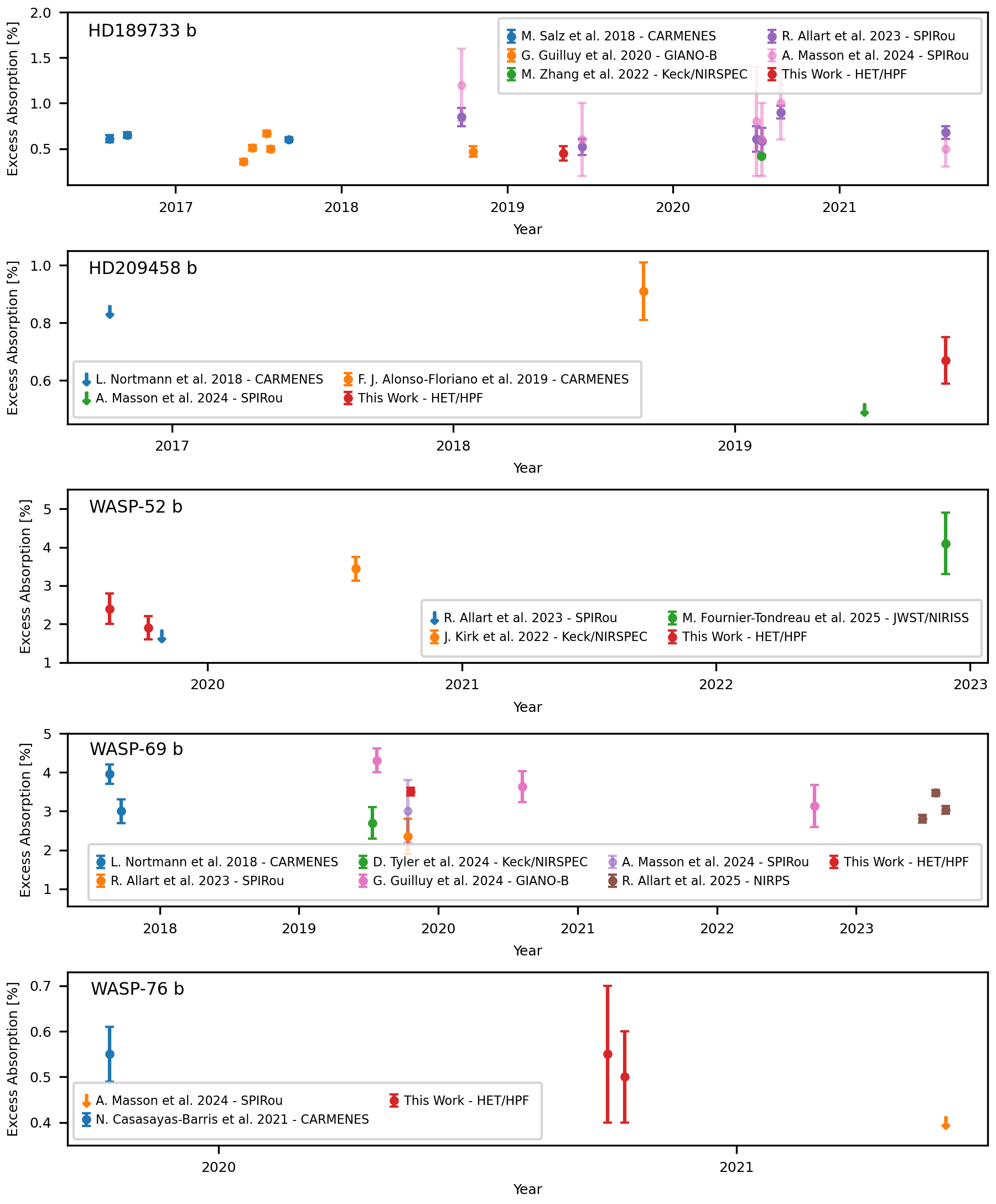}
   \caption{ \label{Fig: he_obs} Time evolution of the \ion{He}{i} signal for HD189733\,b, HD209458\,b, WASP-52\,b, WASP-69\,b, and WASP-76\,b (from top to bottom). This work's results are marked in red in all panels. When necessary, dates are averaged to correspond to the average absorption level over multiple visits. Note that \citet{Allart_Helium_Survey} and \citet{SPIROU_He_survey_Masson2024} analyzed the same datasets for HD189733\,b and WASP-69\,b reporting different results. Detections are denoted by dots with error bars, while upper limits are marked with arrows.
    }
\end{figure*}

We analyzed the He triplet line of five planets for the first time: HAT-P-12\,b, HAT-P-17\,b, HD118203\,b, TrES-1\,b, and WASP-156\,b. Although we do not firmly detect excess helium absorption, we provide the first constraints on excess absorption for these planets (see Table\,\ref{table - Upper limits He}).

We also report the first \ion{He}{i} triplet line spectroscopic results of HAT-P-26\,b, which has previously only been studied with helium narrow-band photometry by \citet{Vissapragada2022_narrowband}. 
We retrieve a flat TS that suggests no \ion{He}{i} absorption, and so we set our upper limit on excess absorption at 1.20\,\%.
However, \citet{Vissapragada2022_narrowband} found a significant \ion{He}{i} absorption signal with a mid-transit excess depth of 0.31$\pm$ 0.10\,\% using a 6.35\,\angstrom\ FWHM band-pass.
We computed the planet's helium light curve by binning the TS to match the FWHM. We measure a $\lesssim$0.10\,\% excess depth, which is not consistent with the narrow-band results.
The scatter present in \citet{Vissapragada2022_narrowband}'s light curves may help to explain the discrepancy between the two observations. This discrepancy could be due to temporal planetary variability, though more observations are needed to explore this hypothesis.

Additionally, we observed ten other planets, all of which have been the subject of previous spectroscopic observations. For HAT-P-3\,b, we set an upper limit on \ion{He}{i} excess absorption of 1.20\,\%, which improves the constraint of 1.90\,\% set by \citet{Guilluy_Edge_Neptune_Desert} using GIANO-B. Similarly, we found an upper limit of 1.20\,\% for WASP-11\,b. \citet{Allart_Helium_Survey} previously set the upper limit at 1.56\,\% with SPIRou for this planet. While we improve previous constraints, additional observations are needed to confirm the presence of metastable helium in either planet's atmosphere.

\citet{Salz_2016} first identified WASP-80\,b as a strong candidate for atmospheric escape, particularly due to the planet's high-energy irradiation from its host star \citep{King_2018}.
However, past observations by \citet[GIANO-B]{Fossati_2022}, \citet[SPIRou]{Allart_Helium_Survey}, and \citet[SPIRou]{SPIROU_He_survey_Masson2024} resulted in tentative detections with upper limits of 0.70\,\%, 1.24\,\%, and 2.40\,\%, respectively. Our upper limit agrees with past observations and suggests that WASP-80\,b does not have currently detectable helium outflow despite predictions. \citet{Fossati_2022} considered that a sub-solar He/H ratio in combination with a high stellar wind might contribute to these non-detections. These non-detections illustrate our limited understanding of the factors that drive the development of the \ion{He}{i} triplet.

WASP-127\,b has a large radius (1.31 R$_{\rm Jup}$), which indicates that the planet has an extended atmosphere and is a good candidate for detecting outflowing helium (\citealp{Lam_2017}). However, spectroscopic observations have only reported upper limits between 0.48--0.90\,\% (\citealp{dos_Santos_2020}, Gemini/Phoenix; \citealp{Allart_Helium_Survey}, SPIRou; \citealp{SPIROU_He_survey_Masson2024}, SPIRou). From our HET/HPF TS, we also put an excess absorption upper limit of 0.60\,\%, which is consistent with the previous results. As \citealp{dos_Santos_2020} suggested, the metastable He state may not form due to the host star's old age ($\sim$11\,Gyr, \citealp{Lam_2017}) and low amount of ionizing radiation.

Similarly, \citet{Kirk_2020} identified WASP-177\,b as a good candidate for the detection of \ion{He}{i} due to its low surface gravity, large scale height, and K-type host star. However, \citet{Kirk_2022}'s observations with Keck/NIRSPEC resulted in an upper limit of 1.25\,\%. This limit could be lower due to the planet's grazing transit, and the authors noted that this tentative detection may indicate a weaker relationship between stellar high-energy activity and the formation of metastable helium. Our upper limit with HET/HPF of 1.10\,\% agrees with that of \citet{Kirk_2022}.

Meanwhile, we have confirmed several previous robust and tentative detections of \ion{He}{i} for HD189733\,b, HD209458\,b, WASP-52\,b, WASP-69\,b, and WASP-76\,b. We compiled those previous spectroscopic results from the ExoAtmospheres database\footnote{\url{https://research.iac.es/proyecto/exoatmospheres/index.php}}, and we compare them in Figure \ref{Fig: he_obs}.

A \ion{He}{i} triplet absorption ranging 0.35--0.70\,\% has been detected with several instruments for HD189733\,b (\citealp[CARMENES]{He_HD189733bSalz2018}; \citealp[GIANO-B]{He_HD189733_Gloria2020}; \citealp[Keck/NIRSPEC]{He_HD189733_Zhang2022}; \citealp[SPIRou]{Allart_Helium_Survey, SPIROU_He_survey_Masson2024}). However, we were only able to detect the signal (0.45 $\pm$ 0.08\,\%) when using out-of-transit baseline contemporaneous to the in-transit observations. Otherwise, stellar variability over month timescales is greater than the planetary signal, which prevented us from initially detecting the \ion{He}{i} triplet absorption. Therefore, our analysis for HD189733\,b illustrates the limitations of the HET/HPF observing strategy for planetary atmospheric studies where stellar variability is a comparable signal or noise source to the signal.

HD209458\,b also displays variability in its \ion{He}{i} signal strength. \citet[CARMENES]{Nortmann2018_WASP-69} and \citet[SPIRou]{SPIROU_He_survey_Masson2024} set their upper limits at 0.84\,\% and 0.50\,\%, respectively. However, \citet{AF_2019} using CARMENES found an excess absorption of 0.91\,\%. Using HET/HPF, we also find an excess absorption of 0.67\,\%, but we note that our signal's amplitude could be attributed to correlated noise (see Sect.\,\ref{sect: Results} for further details). A variable planetary signal may have contributed to this discrepancy in detections and non-detections. Therefore, we need more spectroscopic transit observations to study the behavior of the \ion{He}{i} triplet in HD209458\,b's atmosphere.

WASP-52\,b has also been studied with a variety of different instruments. \citet{Kirk_2022} and \citet{FT_2025} reported detections of 3.44\,\% and 4.10\,\% using Keck/NIRSPEC and JWST/NIRISS, respectively. On the other hand, \citet{Allart_Helium_Survey} reported an upper limit of 1.69\,\%. We find evidence of \ion{He}{i} absorption, but at a level of 1.93\,$\pm$0.20\%, which is compatible with the upper limit reported by \citet{Allart_Helium_Survey} in the same year, but in tension with the detections. The wide spread in the reported absorptions may indicate that WASP-52\,b \ion{He}{i} signal is variable, as has been found for other exoplanets \citep{GJ3470_Ninan2020, He_HD189733_Zhang2022}. Interestingly, the reported excess absorption levels tentatively increase with later observing dates, as seen in Figure \ref{Fig: he_obs}. This long-term variability might be related to stellar activity \citep{WASP69b_he_variability_stellar}. We encourage the exoplanet community to continue observing this target in order to track its \ion{He}{i} triplet line behavior over the next few years.

As the first exoplanet with a ground-based helium detection (\citealp{Nortmann2018_WASP-69}), WASP-69\,b is well-studied in the current literature. Past observations with a variety of instruments detect excess absorption levels ranging from 2.35--4.31\,\% (3.59 $\pm$ 0.19\,\%, CARMENES, \citealp{Nortmann2018_WASP-69}; 2.21 $\pm$ 0.46\,\%, SPIRou, \citealp{Allart_Helium_Survey}; 2.70 $\pm$ 0.40\,\%, Keck/NIRSPEC, \citealp{Tyler_2024}; 3.91 $\pm$ 0.22\,\%, GIANO-B, \citealp{Guilluy_WASP_69b}; 3.00 $\pm$ 0.80\,\%, SPIRou, \citealp{SPIROU_He_survey_Masson2024}; 3.17 $\pm$ 0.05\,\%, SPIRou, \citealp{Allart_WASP_69b_2025}). We studied WASP-69b with HET/HPF for the first time. Although we retrieved a noisy TS for visit 1, it showed a tentative absorption signal. Visit 2 showed a clear excess absorption of 3.50\,\%, consistent with previous detections in the literature.

Transmission spectroscopic observations of WASP-76\,b have reported ambiguous results. With CARMENES, \citet{WASP-76_He_Casasayas2021} found a tentative absorption signal of 0.55\,\% but reported a conservative upper limit of 0.88\,\%. \citet{SPIROU_He_survey_Masson2024} using SPIRou derived an upper limit of 0.40\,\% from the amplitude of the residuals near the \ion{He}{i} triplet line. Our combined TS showed a clear absorption feature of 0.48\,\%, which is consistent with the previous observations. The discrepancy between results may be due to stellar variability and atmospheric changes that may have caused a decrease in the strength of the helium planetary signal, as \citet{SPIROU_He_survey_Masson2024} suggested.

\subsection{Helium circumstellar structures and stellar variability}
\label{sect: He variability}

Following \citet{HAT-P-32b_Zhang2023} and \citet{HAT-P-67b_Gully2024}, we computed the \ion{He}{i} triplet line stellar light curve to inspect the presence of circumstellar structures (e.g., \citealp{HAT-P-32b_Zhang2023, HAT-P-67b_Gully2024, WASP-107_He_Vignesh, WASP-121_He_Allart}). Figures\,\ref{Fig: He LC plot 1} and \ref{Fig: He LC plot 2} do not show evidence of extremely extended outflows of planetary origin, like in HAT-P-32\,b and HAT-P-67\,b. If such outflows existed, the tails would extend for only a small fraction of the total planetary orbit and/or would not be well sampled. Alternatively, the signal could be diluted within the stellar variability.

We find that stellar helium variability is typically below or of the order of 1\,\%, ranging from 0.5\,\% to 2--4\,\% in some extreme cases. These results are consistent with previous work by \citet{Krolikowski2024_He_variability} on He variability with HPF. They monitored the stellar \ion{He}{i} triplet line for a sample of young host stars and found that stars older than 300\,Myr seem to behave similarly to mature and old stars. These young stars exhibited intrinsic stellar variability of 1\,\%, ranging 0--2\,\%.
The HET/HPF observational methodology can be used to study atmospheres of exoplanets, even those around moderately young stars.

\subsection{Constraints on mass loss rates and helium absorption signals}
\label{sect: mass loss rates}

This work presents the first analyses of the \ion{He}{i} triplet for HAT-P-12\,b, HAT-P-17\,b, HD118203\,b, TrES-1\,b, and WASP-156\,b. Although their transmission spectra do not show evidence of planetary absorption, we can estimate the expected helium absorption signal based on the empirical relations from \citet[see Sect.\,6.3]{MOPYS_Orell-Miquel2024}. However, we need to compute or estimate the X-ray ($\sim$5--100\,\AA) and extreme ultraviolet (EUV, $\sim$100--504\,\AA) irradiation that the planets receive from their host star.  Since there are no observations of these new targets in the XUV wavelengths\footnote{In this section, XUV refers to wavelengths of X-rays and EUV defined as between $\sim$5--504\,\AA.}, we compared their stellar properties with those of the available stars in the MUSCLES survey \citep{MUSCLES_survey} and its extension for \emph{JWST} atmospheric transmission spectroscopy targets \citep{Behr2023_MUSCLES} to find a suitable comparison star. Then, we scaled the fluxes from the proxy star to our target, accounting for their stellar radii and distances. All MUSCLES data products are available on the Barbara A. Mikulski Archive for Space Telescopes (MAST) at \citet{dataset_MUSCLES}\footnote{\dataset[10.17909/T9DG6F]{https://doi.org/10.17909/T9DG6F}}. We want to emphasize that since we scale the target's luminosity from a proxy star, these results should be considered an order-of-magnitude approximation.

HAT-P-12 ($R_\star$ = 0.701\,R$_\odot$, \citealp{Hartmann2009}) is part of the MUSCLES extension survey and its analysis is presented in \citet{Behr2023_MUSCLES}. However, the star was not observed at wavelengths below 2500\,\AA\ due to its distance ($d$ = 143\,pc, \citealp{GAIA_DR2}) and low activity. \citet[Table\,2]{Behr2023_MUSCLES} reports a 3$\sigma$ upper limit of 8.79$\times$10$^{27}$\,erg\,s$^{-1}$ for the X-ray luminosity $L_{\rm X}$. We derive $L_{\rm XUV}$ $\simeq$ 2.40$\times$10$^{28}$\,erg\,s$^{-1}$ using the luminosity relation between X-rays and EUV from \citet{Jorge_Helium2025}. Those upper limits are consistent with the luminosities from the HAT-P-12 MUSCLE model, where the X-ray and far ultraviolet fluxes are based on MUSCLES observations of HD85512. We computed $L_{\rm X}$ $\simeq$ 7.12$\times$10$^{26}$\,erg\,s$^{-1}$, and $L_{\rm XUV}$ $\simeq$ 7.81$\times$10$^{27}$\,erg\,s$^{-1}$.
Using the $L_{\rm XUV}$ from MUSCLES and the XUV and \ion{He}{i} triplet signal relation from \citet{MOPYS_Orell-Miquel2024}, we obtain an EW = 22$\pm$3\,m\AA, consistent with our derived upper limit ($<$21.2\,m\AA). The noise level reached by our observations is below this limit. Therefore, an uncontaminated transit of HAT-P-12\,b could accurately measure its mass loss. If we consider the upper limit XUV values, we estimate a much larger signal (EW $\sim$ 36\,m\AA) which is not supported by our observations.

HAT-P-17 ($R_\star$ = 0.838\,R$_\odot$, \citealp{Howard2012}; $d$ = 92.4\,pc, \citealp{GAIA_DR2}) is not part of the MUSCLES survey, so we estimated its XUV flux scaling the luminosities of HD97658 ($R_\star$ = 0.728\,R$_\odot$, \citealp{radius_hd97658}; $d$ = 21.1\,pc, \citealp{DISTANCE_hd97658}) and HAT-P-26 ($R_\star$ = 0.79\,R$_\odot$, $d$ = 141.8\,pc, \citealp{Hartmann2011}). We computed a $L_{\rm XUV}$ $\simeq$ 2.5$\times$10$^{28}$\,erg\,s$^{-1}$ from the scaled HD97658 MUSCLES spectrum (\citealp{Loyd2016_MUSCLES_hd97658}), while $L_{\rm XUV}$ $\simeq$ 2.5$\times$10$^{27}$\,erg\,s$^{-1}$ is obtained when using the HAT-P-26 spectrum (\citealp{Behr2023_MUSCLES}). According to these $L_{\rm XUV}$, we computed an expected \ion{He}{i} triplet signal of $\sim$3--10\,m\AA, which is consistent with the observational upper limit of $<$11.9\,m\AA.

For HD118203 ($R_\star$ = 2.03\,R$_\odot$, \citealp{Maciejewski2024}; $d$ = 92.04\,pc, \citealp{GAIA_DR2}), we used as a proxy HD149026 ($R_\star$ = 1.34\,R$_\odot$, $d$ = 75.0\,pc, \citealp{HD149026_VALUES}). We computed a $L_{\rm XUV}$ $\simeq$ 1.26$\times$10$^{28}$\,erg\,s$^{-1}$ from the scaled HD149026 spectrum (\citealp{Behr2023_MUSCLES}). Although our observations imposed a restrictive upper limit on the presence of \ion{He}{i} (EW $<$2.1\,m\AA), the expected absorption is estimated to be of the same order ($\sim$2\,m\AA).

For TrES-1 ($R_\star$ = 0.807\,R$_\odot$, \citealp{Bonomo2017}; $d$ = 159.622\,pc, \citealp{GAIA_DR2}), we estimated its XUV flux using HD97658 as a proxy. We computed a $L_{\rm XUV}$ $\simeq$ 2.35$\times$10$^{28}$\,erg\,s$^{-1}$ from the scaled HD97658 MUSCLES spectrum. From the XUV flux and the empirical relation from \citet{MOPYS_Orell-Miquel2024}, we derive an EW $\simeq$ 9$\pm$1.5\,m\AA. The expected EW is consistent and a bit below the upper limit imposed from the observations ($<$13.8\,m\AA).

For WASP-156\,b ($R_\star$ = 0.826\,R$_\odot$, \citealp{Polanski2024}; $d$ = 121.843\,pc, \citealp{GAIA_DR2}), we scaled the HAT-P-26 spectrum as a proxy. We obtained a $L_{\rm XUV}$ $\simeq$ 2.46$\times$10$^{27}$\,erg\,s$^{-1}$. We estimated an EW absorption of $\sim$ 3$\pm$1\,m\AA, well bellow the upper limit from our observations ($<$9.5\,\AA). Therefore, more transit observations are required to determine more restrictive limits to WASP-156\,b's mass loss.

\subsection{Helium detections and trends in the gas giant population}
\label{sect: gas giants trends}

\begin{figure*}[t!]
   \centering
   \includegraphics[width=\linewidth]{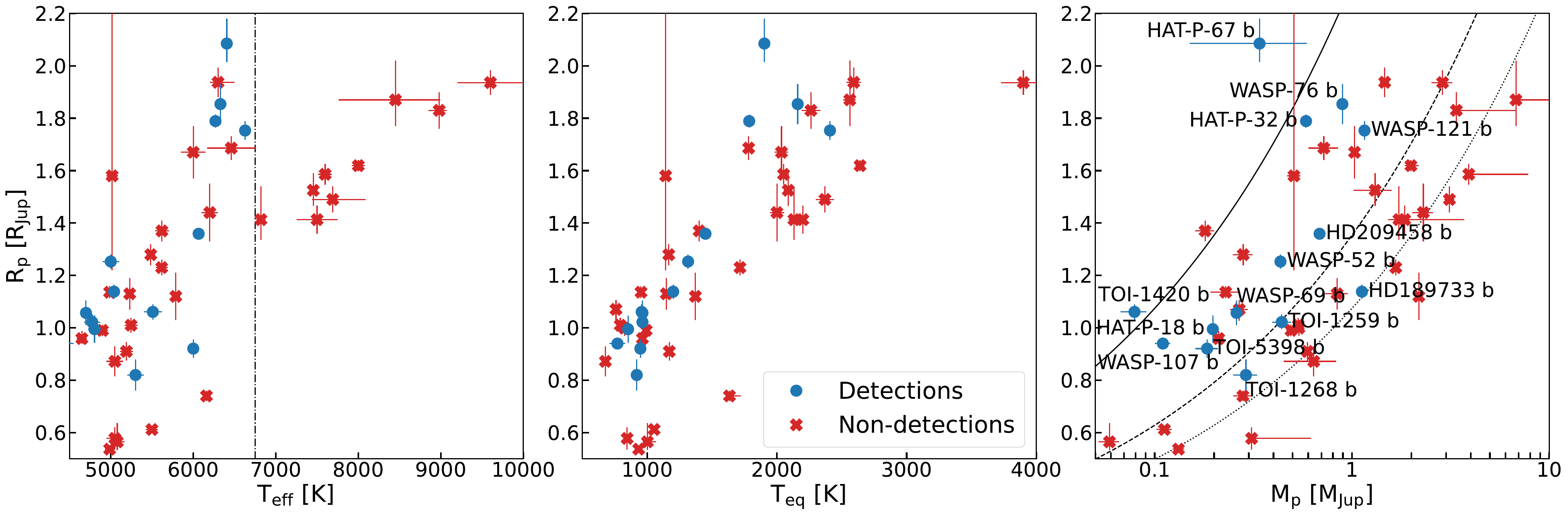}
   \caption{ \label{Fig: Gas giants trends} Distribution of \ion{He}{i} triplet observations for gas giants ($R_{\rm p}$ $>$ 0.5 R$_{\rm Jup}$) as a function of host star effective temperature ($T_{\rm eff}$, left panel), planet equilibrium temperature ($T_{\rm eq}$, central panel), and planetary mass (right panel). Planets with detection of helium are marked with blue dots, while those with non-detections/upper limits are marked with red crosses.
   In the left panel, the vertical dot-dashed line marks $T_{\rm eff}$ = 6750\,K.
   In the right panel, we plotted the constant density curves at 0.1, 0.5, and 1\,g\,cm$^{-3}$ with black solid, dashed, and dotted lines, respectively. We labeled all the planets with He detections, in the right panel.
    }
\end{figure*}

We combined the results on gas giants from this work, from \citet{MOPYS_Orell-Miquel2024}, and from the ExoAtmospheres database\footnote{\url{https://research.iac.es/proyecto/exoatmospheres/index.php}; Accessed on 1 August 2025.}. Our goal is to explore whether the \ion{He}{i} triplet detections in the gas giant population are related to planetary and stellar properties. We obtained a list of 46 planets with $R_{\rm p}$ $>$ 0.5 R$_{\rm Jup}$ and \ion{He}{I} studies. Of these, less than one-third (14) showed evidence of He absorption.

Figure\,\ref{Fig: Gas giants trends} presents the planet radius of gas giants with helium studies as a function of their host star's effective temperature ($T_{\rm eff}$), $T_{\rm eq}$, and mass.
There are no \ion{He}{i} detections for gas giants orbiting stars with $T_{\rm eff}$ higher than $\sim$6750\,K. Half of the detections are in systems with $T_{\rm eff}$ $\sim$5000\,K, whose planets have $T_{\rm eq}$ $\simeq$1000\,K.
The four planets (HAT-P-32\,b, \citealp{HAT-P-32b_Zhang2023}; HAT-P-67\,b, \citealp{HAT-P-67b_Gully2024}; WASP-76\,b, and WASP-121\,b, \citealp{WASP-121_He}) showing \ion{He}{i} outflow orbiting $\sim$6750\,K stars are also among the largest sampled planets ($R_{\rm p}$ $>$ 1.6 R$_{\rm Jup}$). HAT-P-32\,b and HAT-P-67\,b have $T_{\rm eq}$ slightly below the UHJ limit of $\sim$2000\,K, while WASP-76\,b, and WASP-121\,b can be considered within the UHJ category. No helium detections have been made for exoplanets more massive than $\sim$1.1\,M$_{\rm Jup}$.

Figure\,\ref{Fig: Gas giants trends} depicts the three different scenarios for warm, hot, and ultra-hot Jupiter-sized planets. The $T_{\rm eff}$ of the host star, used as a proxy for the spectral type and the spectral energy distribution in the X-ray and UV range, plays a major role in the detectability of helium.
The detection trend for WJs may be similar to that of the general exoplanet population \citep{MOPYS_Orell-Miquel2024}. WJs mainly orbit $\sim$5000\,K host stars, e.g. K-type stars. These stars have a spectral energy distribution that favors the population of the \ion{He}{i} metastable state and its detection \citep{Oklopcic2019_He_Kstar}. 
UHJs and some HJs typically orbit hotter stars than WJs, making them less suitable for populating metastable helium. Therefore, their non-detections are likely related to the stellar type of their host rather than a lack of helium atoms in their atmosphere. Inspection of evaporation through other spectral lines, such as H$\alpha$, would be more appropriate for UHJs orbiting A-type stars \citep{MOPYS_Orell-Miquel2024}.
However, the balance between XUV and near-UV irradiation for F-type stars appears to be enough to make the \ion{He}{i} outflow detectable for low-density HJ and UHJ planets ($<$0.5\,g\,cm$^{-3}$), such as HAT-P-32\,b, HAT-P-67\,b, WASP-76\,b, and WASP-121\,b. These low-density, irradiated gas giants are expected to evaporate at a low rate until they reach a bulk density of $\sim$0.1\,g\,cm$^{-3}$. At that point, they undergo runaway mass loss, which strips away their outer atmospheric layers \citep{HotSaturns_evaporation}. This large outflow would be detectable even if the metastable population mechanism is not efficient, such as in the F-type stars environment \citep{Oklopcic2019_He_Kstar}. This evaporation process is more intense during the first few hundred Myr (e.g., \citealp{JorgeSanz2011, Owen2013, Gupta2020_corepower}), but it can act at Gyr time scales as well (e.g., see animation of \citealp[Fig 3]{HotSaturns_evaporation}). All four HJs and UHJs with helium detections orbit stars older than 1\,Gyr.

We also explored the stellar age parameter, but we have not found a clear relation with the \ion{He}{i} triplet detections.
There are only three gas giants with helium detection around stars younger than 1\,Gyr (TOI-5398\,b, \citealp{TOI-5398bMattia}; TOI-1268\,b, \citealp{MOPYS_Orell-Miquel2024}; and WASP-52\,b), but there are also less studied planets, 12 in total. Thus, the ratio of detections to non-detections in the young planet population is similar to that of the old population.


\section{Conclusions}

In this work, we present the results of a helium survey that aims to investigate the atmospheric mass loss in individual irradiated gas giants, but also explore population trends. We analyzed HPF high-resolution transmission spectroscopic time-series of sixteen gas giants searching for \ion{He}{i} triplet absorption. We are able to confirm previous robust or tentative detections for HD189733\,b, \hd209\,b, WASP-52\,b, WASP-69\,b, and WASP-76\,b, confirming the HET/HPF ability to precisely measure atmospheric absorption. However, HD189733\,b analysis highlights the limitations of the HET/HPF observing strategy and emphasizes the importance of contemporaneous baseline observations for this method.

For the first time, we provide constraints on the helium line of HAT-P-12\,b, HAT-P-17\,b, HD118203\,b, TrES-1\,b, and WASP-156\,b. For these targets, our observational upper limits are consistent with the empirical helium predictions from estimated XUV fluxes. Additionally, we do not detect significant absorption in HAT-P-3\,b, WASP-11\,b, WASP-80\,b, WASP-127\,b, and WASP-177\,b, which is consistent with previous studies. HAT-P-26\,b had a previous detection using narrow-band photometric observations \citep{Vissapragada2022_narrowband} that we are unable to reproduce with our high-resolution observations.

We compiled a list of 46 irradiated gas giants for which helium studies were conducted, combining the results presented here with those available in the literature. Using this dataset, we explore \ion{He}{i} triplet trends within the populations of warm, hot, and ultra-hot Jupiters. We note differences in the detectability of helium among these three populations. Warm Jupiter detections seem to be more closely linked to the host star type, with all of the detections found in planets orbiting K-type stars, which are considered the most favorable stellar type according to theory \citep{Oklopcic2019_He_Kstar}. Although there are only four hot Jupiter detections, they come from low-density planets orbiting F-type stars. To date, no helium absorption signal has been reported for ultra-hot Jupiters.
Unfortunately, our results are based in limited number of explored atmospheres and further observations of irradiated gas giants are needed to increase this sample and improve our understanding of atmospheric mass loss through \ion{He}{i} triplet observations.

\begin{acknowledgments}

We thank the anonymous referee for the discussion and comments that improved the clarity and readability of the manuscript.

This material is based upon work supported by the National Aeronautics and Space Administration under Grant Number  80NSSC20K0257 for the XRP program issued through the Science Mission Directorate.

C.V.M. acknowledges support from the Alfred P. Sloan Foundation under grant number FG-2021-16592.

Based on observations obtained with the Hobby–Eberly Telescope (HET), which is a joint project of the University of Texas at Austin, the Pennsylvania State University, Ludwig-Maximillians-Universit\"at M\"unchen, and Georg-August Universit\"at G\"ottingen. The HET is named in honor of its principal benefactors, William P. Hobby and Robert E. Eberly.
These results are based on observations obtained with the Habitable-zone Planet Finder Spectrograph on the HET with the Wide-Field Upgrade (\citealp{HPF_2021_acknowledgment}).
The authors acknowledges the support and resources from the Texas Advanced Computing Center. We are grateful to the HET Resident Astronomers and Telescope Operators for valuable assistance in gathering our HPF data.

The Center for Exoplanets and Habitable Worlds is supported by the Pennsylvania State University and the Eberly College of Science.

This research has made use of the NASA Exoplanet Archive, which is operated by the California Institute of Technology, under contract with the National Aeronautics and Space Administration under the Exoplanet Exploration Program.
This research has made use of NASA’s Astrophysics Data System.

We acknowledge the use of the ExoAtmospheres database during the preparation of this work.

J.O-M is grateful for the support, assistance, and encouragement he has always received from Padrina Conxa, Padrina Merc\`e, Jeroni, Merc\`e, and more relatives and friends. And especially from you-drigueta, Yess.

\end{acknowledgments}





%
\facilities{HET/HPF, NASA Exoplanet Archive, ExoAtmospheres database}


\software{ {\tt emccee} (\citealp{emcee}), {\tt jupyter }(\citealp{jupyter_notebook}) , {\tt matplotlib } (\citealp{matplotlib}), {\tt muler} (\citealp{muler_Gully2022}), {\tt numpy} (\citealp{numpy}), {\tt pandas} \citep{pandas}, {\tt scipy} (\citealp{scipy}). }


\appendix

\section{Extra tables}
\label{App: Tables}

\startlongtable 
\begin{deluxetable*}{ccccccc}
\tablecaption{\label{table - Log Table}
Log of the observations for our targets explaining for each track: the night of the observations (year-month-day), observing category (IN: in-transit, OUT: out-of-transit, or P4) where the number indicates the visit associated, number of spectra, median phase covering, mean exposure time (Exp. Time), and mean S/N.}
\tablehead{
Planet & Date & Class  & Num & Phase & Exp. Time &  S/N
}
\startdata
HAT-P-3\,b & 2021- 1- 2 &   P4      & 4 & $-$0.32 & 618 & 109 \\ 
           & 2021- 1-31 &  OUT1  & 4 & $-$0.35 & 618 & 94 \\
           & 2021- 2- 1 &  IN1      & 8 & $-$0.01 & 639 & 95 \\
           & 2021- 2- 2 &  OUT1  & 4 & 0.34 & 618 & 87 \\
\hline  \noalign{\smallskip}
HAT-P-12\,b & 2019-12-18 &   P4  & 2 & $-$0.35 & 916 & 96 \\
    & 2020- 3-26 &  OUT1   & 2 & 0.38 & 777 & 34 \\
    & 2020- 3-30 &  OUT1   & 1 & $-$0.32 & 958 & 31 \\
    & 2020- 3-31 &  OUT1   & 3 & $-$0.07 & 958 & 32 \\
    & 2020- 3-31 &   IN1  & 5 & 0.0 & 958 & 57 \\
    & 2020- 4- 1 &   OUT1  & 4 & 0.31 & 958 & 54 \\
     & 2025- 5-26 &   OUT2  & 4 & $-$0.32 & 618 & 60 \\
    & 2025- 5-27 &   IN2  & 8 & 0.0 & 618 & 47 \\
\hline  \noalign{\smallskip}
HAT-P-17\,b & 2020- 5- 7 &   P4  & 4 & $ -0.1 $ & 309 & 86 \\
    & 2020- 6- 7 &  OUT1  & 4 & $ -0.11 $ & 309 & 77 \\
    & 2020- 6- 8 &  IN1   & 11 & $ -0.01 $ & 309 & 73 \\
    & 2020- 6-10 &  OUT1   & 4 & $ 0.18 $ & 309 & 88\\
    & 2020- 6-11 &  OUT1   & 4 & $ 0.28 $ & 309 & 76 \\
    & 2020- 7-23 &  P4   & 2 & $ 0.33 $ & 511 & 102 \\
    & 2020- 7-23 &  P4   & 2 & $ 0.35 $ & 511 & 159 \\
    & 2020- 7-28 &  OUT2   & 4 & $ -0.19 $ & 309 & 122 \\
    & 2020- 7-29 &  OUT2   & 4 & $ -0.07 $ & 309 & 122 \\
    & 2020- 7-30 &  IN2   & 10 & $ 0.01 $ & 309 & 109 \\
    & 2020- 7-30 &  OUT2   & 4 & $ 0.03 $ & 309 & 127 \\
    & 2020- 7-31 &  OUT2   & 4 & $ 0.1 $ & 309 & 105 \\
    & 2020- 8- 1 &  OUT2   & 4 & $ 0.2 $ & 309 & 99 \\
    & 2020- 8- 1 &  OUT2   & 2 & $ 0.22 $ & 511 & 141 \\
\hline  \noalign{\smallskip}
HAT-P-26\,b & 2021- 1-31 &  OUT1   & 4 & $ -0.24 $ & 660 & 96 \\
    & 2021- 2- 1 & IN1    & 5 & $ -0.01 $ & 660 & 87 \\
    & 2021- 2- 2 & OUT1    & 4 & $ 0.23 $ & 660 & 83 \\
    & 2021- 2-24 &  P4   & 2 & $ 0.41 $ & 820 & 110 \\
\hline  \noalign{\smallskip}
HD118203\,b    & 2021- 1- 5 &   OUT1  & 2 & $ -0.31 $ & 309 & 304\\
    & 2021- 1- 6 &  OUT1   & 2 & $ -0.15 $ & 309 & 215 \\
    & 2021- 1- 7 &  IN1   & 18 & $ 0.01 $ & 309 & 344 \\
    & 2021- 1- 8 &  OUT1   & 2 & $ 0.18 $ & 309 & 331 \\
    & 2021- 1- 9 &  OUT1   & 2 & $ 0.34 $ & 309 & 290 \\
    & 2021- 1-26 &  P4   & 2 & $ 0.1 $ & 213 & 228 \\
    & 2021- 2-19 &  IN2   & 18 & $ 0.01 $ & 309 & 305 \\
    & 2021- 2-19 &  OUT2   & 9 & $ 0.03 $ & 309 & 322 \\
    & 2021- 2-20 &  OUT2   & 2 & $ 0.17 $ & 309 & 328 \\
\hline  \noalign{\smallskip}
HD189733\,b  & 2019- 4-17 &  OUT4   & 11 & $ 0.37 $ & 309 & 395 \\
    & 2019- 5- 4 &  IN1   & 9 & $ 0.01 $ & 309 & 403 \\
             & 2019- 5-12 &  P4 (OUT1)   & 2 & $ -0.42 $ & 213 & 320 \\
    & 2019- 6-13 &  IN2   & 9 & $ -0.01 $ & 309 & 292 \\
    & 2019-10-30 &  P4   & 2 & $ -0.43 $ & 618 & 737 \\
    & 2020- 7-24 &  P4   & 2 & $ 0.49 $ & 213 & 402 \\
    & 2020- 8- 1 &  IN3   & 2 & $ -0.02 $ & 106 & 231 \\
    & 2020- 8-19 &  P4   & 2 & $ 0.18 $ & 213 & 479 \\
    & 2020-11-24 &  P4   & 2 & $ -0.22 $ & 213 & 377 \\
\hline  \noalign{\smallskip}
HD209458\,b    & 2019-10- 1 &  OUT1   & 1 & $ -0.34 $ & 447 & 315  \\
    & 2019-10- 2 &  IN1   & 5 & $ 0.0 $ & 522 & 478  \\
    & 2019-10- 4 &  OUT2   & 2 & $ -0.43 $ & 522 & 159  \\
    & 2019-10- 5 &  OUT2   & 4 & $ -0.2 $ & 309 & 233  \\
\hline  \noalign{\smallskip}
TrES-1\,b    & 2019- 8-12 &  P4   & 2 & $ 0.07 $ & 618 & 110  \\
    & 2020- 7- 5 &  OUT1   & 1 & $ 0.35 $ & 799 & 96  \\
    & 2020- 7- 7 &  IN1   & 5 & $ 0.01 $ & 799 & 104  \\
    & 2020- 7- 8 & OUT1    & 4 & $ 0.34 $ & 799 & 63  \\
    & 2020- 7-10 & OUT1\&2    & 1 & $ -0.08 $ & 820 & 93  \\
    & 2020- 7-11 & OUT2    & 4 & $ 0.33 $ & 820 & 81  \\
    & 2020- 7-12 & OUT2    & 4 & $ -0.34 $ & 820 & 97  \\
    & 2020- 7-13 &  OUT2   & 5 & $ -0.09 $ & 820 & 99  \\
    & 2020- 7-13 &  IN2   & 5 & $ -0.01 $ & 820 & 93  \\
    & 2020- 7-14 & OUT2    & 4 & $ 0.24 $ & 820 & 88  \\
    & 2020- 7-23 & P4    & 2 & $ 0.28 $ & 820 & 116  \\
    & 2020- 8- 1 & P4    & 1 & $ 0.24 $ & 820 & 85  \\
    & 2020-10-13 &  P4   & 2 & $ 0.26 $ & 820 & 91  \\
    & 2020-11- 9 &  P4   & 2 & $ 0.15 $ & 820 & 103  \\
\hline  \noalign{\smallskip}
WASP-11\,b  & 2019-12- 6 &   OUT1  & 4 & $ -0.21 $ & 820 & 103  \\
    & 2019-12- 7 &  IN1   & 5 & $ -0.01 $ & 820 & 101  \\
    & 2019-12- 7 &  OUT1   & 5 & $ 0.05 $ & 820 & 106  \\
    & 2019-12- 8 &  OUT1   & 4 & $ 0.26 $ & 820 & 102  \\
    & 2019-12-18 &  IN2   & 5 & $ 0.0 $ & 756 & 109  \\
    & 2019-12-20 &  OUT2   & 3 & $ -0.46 $ & 820 & 101  \\
    & 2019-12-21 &  OUT2   & 1 & $ -0.26 $ & 820 & 110  \\
    & 2019-12-21 &  OUT2   & 3 & $ -0.2 $ & 820 & 114  \\
    & 2019-12-24 &  P4   & 2 & $ -0.39 $ & 916 & 96  \\
    & 2020- 7-25 &  P4   & 2 & $ 0.14 $ & 820 & 129  \\
    & 2020-11- 8 &  P4   & 2 & $ -0.4 $ & 820 & 86  \\
    & 2020-11-12 &  P4   & 2 & $ -0.32 $ & 820 & 150  \\
\hline  \noalign{\smallskip}
WASP-52\,b   & 2019- 8-13 & IN1    & 4 & $ -0.02 $ & 820 & 93   \\
    & 2019- 8-14 & OUT1    & 1 & $ -0.35 $ & 916 & 96  \\
    & 2019- 8-15 & OUT1    & 4 & $ 0.22 $ & 916 & 69  \\
    & 2019-10- 7 & OUT2    & 1 & $ 0.34 $ & 767 & 84  \\
    & 2019-10- 8 & OUT2    & 1 & $ -0.1 $ & 767 & 102  \\
    & 2019-10- 8 & IN2    & 5 & $ 0.0 $ & 639 & 76  \\
    & 2019-10-10 &  OUT2   & 2 & $ 0.04 $ & 767 & 86  \\
    & 2019-10-10 &  OUT2   & 3 & $ 0.13 $ & 767 & 95  \\
\hline  \noalign{\smallskip}
WASP-69\,b    & 2019- 9-20 &  IN1   & 1 & $ 0.01 $ & 628 & 40   \\
    & 2019-10-19 &  OUT2   & 8 & $ 0.49 $ & 268 & 140   \\
    & 2019-10-20 &  OUT2   & 4 & $ -0.26 $ & 532 & 273   \\
    & 2019-10-21 &  IN2   & 7 & $ 0.0 $ & 437 & 215   \\
    & 2019-10-22 &  OUT2   & 4 & $ 0.26 $ & 532 & 245   \\
\hline  \noalign{\smallskip}
WASP-76\,b   & 2020-10- 1 &  OUT1   & 4 & $ 0.36 $ & 309 & 149  \\
    & 2020-10- 2 &  OUT1   & 8 & $ -0.08 $ & 320 & 124  \\
    & 2020-10- 2 &  IN1   & 10 & $ -0.02 $ & 320 & 142  \\
    & 2020-10- 4 &  IN2   & 10 & $ 0.02 $ & 320 & 123  \\
    & 2020-10- 5 &  OUT2   & 4 & $ -0.43 $ & 309 & 166  \\
    & 2020-10- 6 &  OUT2   & 4 & $ 0.13 $ & 309 & 131  \\
\hline  \noalign{\smallskip}
WASP-80\,b      & 2019- 5-17 &  P4   & 10 & $ 0.44 $ & 309 & 109  \\
    & 2019- 7-12 & OUT1    & 1 & $ -0.33 $ & 309 & 90 \\
    & 2019- 7-13 & OUT1    & 2 & $ -0.03 $ & 309 & 84 \\
    & 2019- 7-13 &  IN1  & 10 & $ 0.0 $ & 309 & 118 \\
    & 2020- 7-23 &   P4  & 4 & $ -0.47 $ & 511 & 130 \\
    & 2020- 8- 1 & P4    & 2 & $ 0.45 $ & 511 & 140 \\
\hline  \noalign{\smallskip}
WASP-127\,b   & 2020- 2-13 &   OUT1  & 10 & $ -0.49 $ & 415 & 124  \\
    & 2020- 2-14 & OUT1    & 5 & $ -0.25 $ & 415 & 141  \\
    & 2020- 2-15 &  IN1   & 10 & $ -0.02 $ & 415 & 139  \\
    & 2020- 2-16 &  OUT1   & 5 & $ 0.23 $ & 415 & 94  \\
    & 2020- 2-17 &  OUT1   & 5 & $ 0.47 $ & 415 & 85  \\
    & 2020- 3- 6 & P4    & 2 & $ -0.24 $ & 415 & 47  \\
    & 2020-12- 8 &   P4  & 2 & $ 0.11 $ & 511 & 195  \\
\hline  \noalign{\smallskip}
WASP-156\,b & 2020-10-23 &  OUT1   & 4 & $ -0.26 $ & 618 & 75  \\
    & 2020-10-24 & IN1    & 6 & $ 0.0 $ & 554 & 63  \\
    & 2020-10-25 & OUT1    & 4 & $ 0.23 $ & 618 & 93  \\
    & 2020-11-20 & IN2    & 6 & $ -0.01 $ & 554 & 98  \\
    & 2020-11-20 &  OUT2   & 4 & $ 0.02 $ & 554 & 91  \\
    & 2020-11-21 &  OUT2   & 4 & $ 0.25 $ & 618 & 117  \\
    & 2020-11-22 &  OUT2   & 4 & $ -0.46 $ & 618 & 122  \\
\hline  \noalign{\smallskip}
WASP-177\,b     & 2020- 8-18 &  OUT1   & 3 & $ 0.36 $ & 831 & 84  \\
    & 2020- 8-19 & OUT1    & 4 & $ -0.32 $ & 831 & 98  \\
    & 2020- 8-20 &  IN1   & 4 & $ 0.0 $ & 831 & 87  \\
    & 2020- 8-20 &  OUT1   & 4 & $ 0.03 $ & 831 & 86  \\
    & 2020- 8-21 &  OUT1\&2   & 4 & $ 0.33 $ & 831 & 86  \\
    & 2020- 8-22 &  OUT2   & 1 & $ -0.34 $ & 831 & 40  \\
    & 2020- 8-23 &   OUT2  & 4 & $ -0.02 $ & 831 & 65  \\
    & 2020- 8-23 &   IN2  & 4 & $ 0.0 $ & 831 & 75  \\
    & 2020- 8-24 &   OUT2  & 3 & $ 0.3 $ & 831 & 84  \\
    & 2020- 8-25 &  OUT2   & 4 & $ -0.37 $ & 831 & 87  \\
\hline  \noalign{\smallskip}
\enddata
\end{deluxetable*}

\begin{table*}[h!]
\caption{Transit and system parameters used to compute the transmission spectra for each planet analyzed in this work. \label{table - TRANSIT PARAMETERS}}
\resizebox{\textwidth}{!}{\begin{tabular}{l c c c c c c c c c}
\hline \hline 
\noalign{\smallskip} 

       & $P$ & $T_0$\,$^{(a)}$ & $T_{14}$\,$^{(b)}$ & $T_{12}$\,$^{(b)}$ & $a_{\rm p}$ & $i_{\rm p}$ & $\gamma_{\rm RV}$ & $K_{\star}$ & $K_{\rm p}$\,$^{(c)}$ \\
Planet & [d] & [d] & [h] & [min] & [au] & [deg] & [km\,s$^{-1}$] & [m\,s$^{-1}$] & [km\,s$^{-1}$] \\ \noalign{\smallskip}

\hline
\noalign{\smallskip}


HAT-P-3\,b &  2.89973815\,(13)\,$^{(1)}$ &  6843.022438\,(81)\,$^{(1)}$ & 2.08$\pm$0.01\,$^{(2)}$ & 14.2 & 0.03878\,(65)\,$^{(3)}$ & 86.31$\pm$0.19\,$^{(3)}$  & $-$23.4\,(4)\,$^{(4)}$ & 90.63$\pm$0.58\,$^{(2)}$ & 145 \\  \noalign{\smallskip}

HAT-P-12\,b &  3.2130598\,(21)\,$^{(1)}$ &  4419.19556\,(20)\,$^{(1)}$ & 2.338$\pm$0.014\,$^{(1)}$ & 18.6 &  0.0384\,(3)\,$^{(1)}$ & 89.0$\pm$0.4\,$^{(1)}$  & $-$40.0\,(9)\,$^{(2)}$ & 35.8$\pm$1.9\,$^{(1)}$ & 130 \\  \noalign{\smallskip}

HAT-P-17\,b &  10.3385346\,(9)\,$^{(1)}$ &  6569.05972\,(5)\,$^{(1)}$ & 4.05$\pm$0.02\,$^{(2)}$ & 32.4 &  0.0882\,(14)\,$^{(2)}$ & 89.2$\pm$0.1\,$^{(2)}$  & 20.13\,(21)\,$^{(2)}$ & 58.8$\pm$0.9\,$^{(2)}$ & 93 \\  \noalign{\smallskip}

HAT-P-26\,b &  4.23450020\,(64)\,$^{(1)}$ &  6901.059458\,(94)\,$^{(1)}$ & 2.45$\pm$0.02\,$^{(2)}$ & 11.1 &  0.0479\,(6)\,$^{(2)}$ & 88.6$\pm$0.5\,$^{(2)}$  & 14.57\,(46)\,$^{(3)}$ & 8.5$\pm$1.0\,$^{(2)}$ & 123 \\  \noalign{\smallskip}

HD118203\,b &  6.1349890\,(13)\,$^{(1)}$ &  8712.66178\,(19)\,$^{(1)}$ & 5.638$\pm$0.013\,$^{(1)}$ & 21.50 &  0.0701\,(4)\,$^{(1)}$ & 88.9$\pm$1.0\,$^{(1)}$  & $-$29.177\,(8)\,$^{(1)}$ & 222.2$\pm$2.1\,$^{(1)}$ & 124 \\  \noalign{\smallskip}

HD189733\,b &  2.218574944\,(30)\,$^{(1)}$ &  6194.067619\,(34)\,$^{(1)}$ & 1.8036$\pm$0.0023\,$^{(2)}$ & 10.0 &  0.03099\,(60)\,$^{(1)}$ & 85.58$\pm$0.06\,$^{(1)}$  & $-$2.55\,(16)\,$^{(1)}$ & 205$\pm$6\,$^{(1)}$ & 151 \\  \noalign{\smallskip}

HD209458\,b &  2.218574944\,(30)\,$^{(1)}$ &  4560.80588\,(8)\,$^{(2)}$ & 2.978$\pm$0.0051\,$^{(3)}$ & 8.7 &  0.04707\,(45)\,$^{(1)}$ & 86.78$\pm$0.07\,$^{(1)}$  & $-$14.741\,(2)\,$^{(4)}$ & 84.27$\pm$0.70\,$^{(1)}$ & 145\,$^{(3)}$ \\  \noalign{\smallskip}

TrES-1\,b &  3.030069476\,(72)\,$^{(1)}$ &  6822.891157\,(63)\,$^{(1)}$ & 2.5\,$^{(2)}$ & 18.8 &  0.03926\,(58)\,$^{(3)}$ & 90.0\,$^{(3)}$  & $-$20.40\,(44)\,$^{(4)}$ & 115.2$\pm$6.2\,$^{(5)}$ & 141 \\  \noalign{\smallskip}

WASP-11\,b &  3.72247919\,(18)\,$^{(1)}$ &  6646.984352\,(76)\,$^{(1)}$ & 2.556$\pm$0.030\,$^{(1)}$ & 19.283 &  0.04375\,(74)\,$^{(2)}$ & 89.03$\pm$0.34\,$^{(2)}$  & 4.74\,(54)\,$^{(3)}$ & 80.6$\pm$1.8\,$^{(4)}$ & 128 \\  \noalign{\smallskip}

WASP-52\,b & 1.74978117\,(16)\,$^{(1)}$ &  6784.057988\,(68)\,$^{(1)}$ & 1.85\,$^{(2)}$ & 23.5\,$^{(2)}$ &  0.0272\,(3)\,$^{(3)}$ & 85.35$\pm$0.20\,$^{(1)}$  & 0.48\,(33)\,$^{(4)}$ & 84.3$\pm$3.0\,$^{(3)}$ & 167 \\  \noalign{\smallskip}

WASP-69\,b & 3.8681382\,(6)\,$^{(1)}$ &  5748.83344\,(18)\,$^{(1)}$ & 1.85\,$^{(1)}$ & 23.5\,$^{(1)}$ &  0.04525\,(75)\,$^{(1)}$ & 86.71$\pm$0.20\,$^{(1)}$  & $-$9.37215\,(23)\,$^{(1)}$ & 38.1$\pm$2.4\,$^{(1)}$ & 127 \\  \noalign{\smallskip}

WASP-76\,b & 1.80988198\,(56)\,$^{(1)}$ &  8080.6261\,(4)\,$^{(1)}$ & 3.83\,$^{(1)}$ & 23.6\,$^{(1)}$ &  0.0330\,(2)\,$^{(1)}$ & 89.62$\pm$0.03\,$^{(1)}$  & $-$1.0733\,(2)\,$^{(2)}$ & 116.0$\pm$1.3\,$^{(1)}$ & 197\,$^{(3)}$ \\  \noalign{\smallskip}

WASP-80\,b & 3.06785234\,(79)\,$^{(1)}$ &  6487.425006\,(25)\,$^{(1)}$ & 2.131$\pm$0.003\,$^{(1)}$ & 20\,$^{(1)}$ &  0.0344\,(10)\,$^{(1)}$ & 89.02$\pm$0.10\,$^{(1)}$  & 9.82\,(77)\,$^{(2)}$ & 109.0$\pm$3.1\,$^{(1)}$ & 122 \\  \noalign{\smallskip}

WASP-127\,b & 4.17806203\,(53)\,$^{(1)}$ & 6776.62124\,(23)\,$^{(1)}$ & 4.353$\pm$0.014\,$^{(1)}$ & 24.9\,$^{(1)}$ &  0.0484\,(9)\,$^{(1)}$ & 87.84$\pm$0.33\,$^{(1)}$  & $-$8.25\,(89)\,$^{(2)}$ & 22$\pm$3\,$^{(1)}$ & 126 \\  \noalign{\smallskip}

WASP-156\,b & 3.8361623\,(15)\,$^{(1)}$ & 8414.13606\,(31)\,$^{(1)}$ & 2.404$\pm$0.019\,$^{(1)}$ & 9.8\,$^{(1)}$ &  0.0458\,(9)\,$^{(1)}$ & 89.1$\pm$0.9\,$^{(1)}$  & 9.30\,(61)\,$^{(2)}$ & 18.7$\pm$1.1\,$^{(1)}$ & 130 \\  \noalign{\smallskip}

WASP-177\,b & 3.071722\,(1)\,$^{(1)}$ & 7994.37140\,(28)\,$^{(1)}$ & 1.61$\pm$0.03\,$^{(1)}$ & $\sim$30\,$^{(1)}$ &  0.03957\,(58)\,$^{(1)}$ & 84.14$\pm$0.8\,$^{(1)}$  & $-$7.1434\,(41)\,$^{(1)}$ & 77.3$\pm$5.2\,$^{(1)}$ & 139 \\  \noalign{\smallskip}

\hline \hline 
\noalign{\smallskip}

\end{tabular}
}
\tablecomments{
$^{(a)}$ $T_0$ is given in BJD$-$2\,450\,000.
$^{(b)}$ $T_{14}$ is the total transit duration between the first ($T_1$) and fourth ($T_4$) contacts, and $T_{12}$ is the duration of the ingress or egress.
$^{(c)}$ Calculated from $K_{\rm p} = 2 \pi ~ a_{\rm p} ~ P^{-1} \sin{i_{\rm p}}$ using the parameters in this table, when there is no reference. \\
\textbf{References.}
HAT-P-3\,b: $^{(1)}$ \citet{Kokori2023}, $^{(2)}$ \citet{Bourrier2023}, $^{(3)}$ \citet{Mancini2018}, $^{(4)}$ \textit{Gaia}\,DR2 (\citealp{GAIA_DR2}).
HAT-P-12\,b: $^{(1)}$ \citet{Hartmann2009}, $^{(2)}$ \textit{Gaia}\,DR2 (\citealp{GAIA_DR2}).
HAT-P-17\,b: $^{(1)}$ \citet{Kokori2022}, $^{(2)}$ \citet[$e$ = 0.342$\pm$0.006; $\omega$ = 201$\pm$1]{Howard2012}.
HAT-P-26\,b: $^{(1)}$ \citet{Kokori2023}, $^{(2)}$ \citet[$e$ = 0.124$\pm$0.060; $\omega$ = 54$\pm$165]{Hartmann2011}, $^{(3)}$ \textit{Gaia}\,DR2 (\citealp{GAIA_DR2}).
HD118203\,b: $^{(1)}$ \citet[$e$ = 0.301$\pm$0.006; $\omega$ = 157.4$\pm$1.9]{Maciejewski2024}.
HD189733\,b:  $^{(1)}$ \citet{Kokori2023}, $^{(2)}$ \citet{Baluev2015}, $^{(3)}$ \citet{Torres2008}, $^{(4)}$ \textit{Gaia}\,DR2 (\citealp{GAIA_DR2}).
HD209458\,b: $^{(1)}$ \citet{Bonomo2017}, $^{(2)}$ \citet{Evans2015}, $^{(3)}$ \citet{Nuria2021_HD209}, $^{(4)}$ \citet{Naef2004}.
TrES-1\,b: $^{(1)}$ \citet{Kokori2023}, $^{(2)}$ \citet{Baluev2015}, $^{(3)}$ \citet{Bonomo2017}, $^{(4)}$ \textit{Gaia}\,DR2 (\citealp{GAIA_DR2}), $^{(5)}$ \citet{Stassun2017}.
WASP-11\,b: $^{(1)}$ \citet{Kokori2023}, $^{(2)}$ \citet{Mancini2015}, $^{(3)}$ \textit{Gaia}\,DR2 (\citealp{GAIA_DR2}), $^{(4)}$ \citet{Bonomo2017}.
WASP-52\,b: $^{(1)}$ \citet{Kokori2023}, $^{(2)}$ \citet{Chen2017}, $^{(3)}$ \citet{Hebrad2013}, $^{(4)}$ \textit{Gaia}\,DR2 (\citealp{GAIA_DR2}).
WASP-69\,b: $^{(1)}$ \citet{Anderson2014}.
WASP-76\,b: $^{(1)}$ \citet{Ehrenreich2020}, $^{(2)}$ \citet{West2016}, $^{(3)}$ \citet{WASP-76_He_Casasayas2021}.
WASP-80\,b: $^{(1)}$ \citet{Triaud2015}, $^{(2)}$ \textit{Gaia}\,DR2 (\citealp{GAIA_DR2}).
WASP-127\,b: $^{(1)}$ \citet{Seidel2020}, $^{(2)}$ \textit{Gaia}\,DR2 (\citealp{GAIA_DR2}).
WASP-156\,b: $^{(1)}$ \citet{Polanski2024}, $^{(2)}$ \textit{Gaia}\,DR2 (\citealp{GAIA_DR2}).
WASP-177\,b: $^{(1)}$ \citet{Turner2019}.
}
\end{table*}

\section{A note on \ion{Ca}{II} IRT inspection with HET/HPF }
\label{App: Ca IRT}

\begin{figure*}[ht!]
   \centering
   \includegraphics[width=\linewidth]{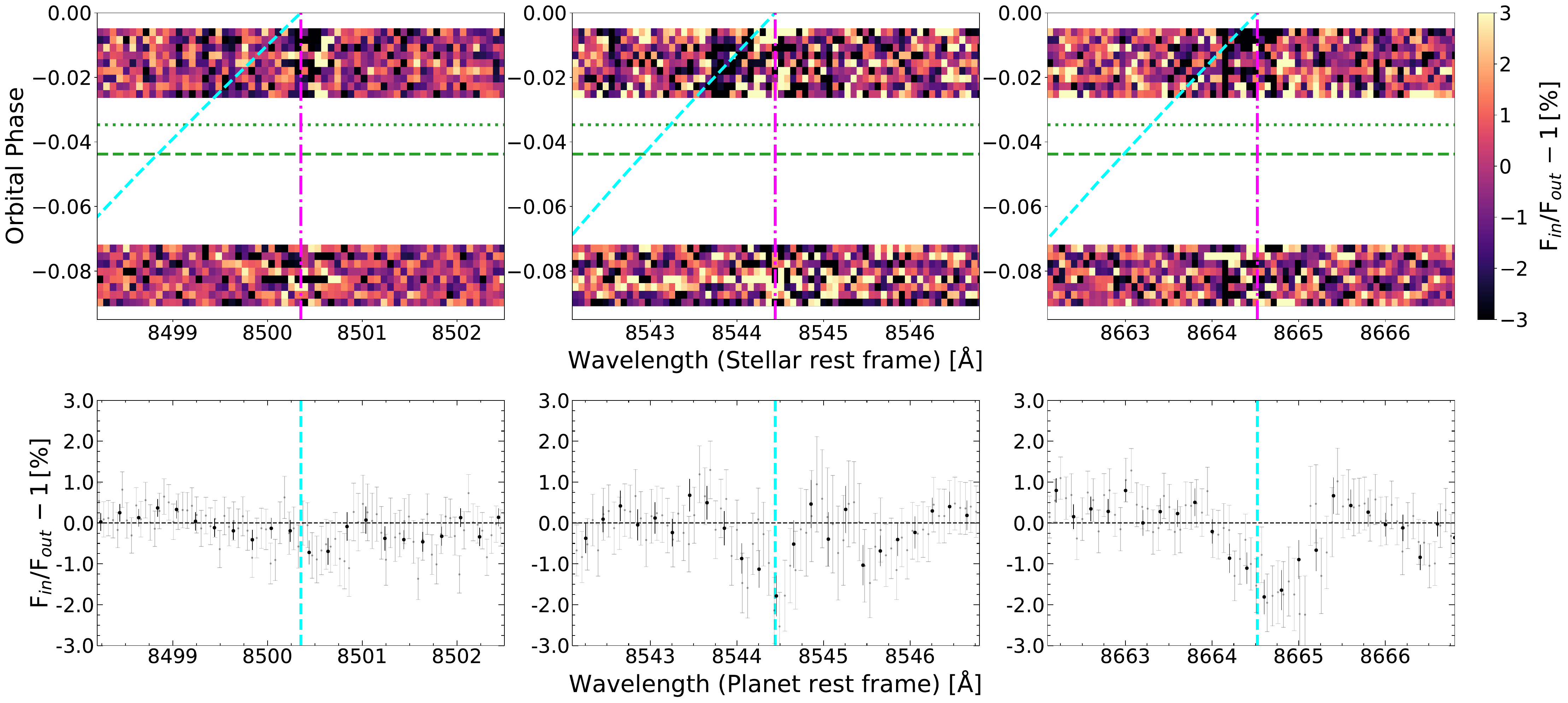}
   \caption{ \label{Fig: WASP-76 Ca IRT} Same as Figure\,\ref{Fig: TS plot 1} but for Ca\,{\sc ii}\,IRT lines of WASP-76\,b: $\lambda$8500\,\angstrom\ (left panel), $\lambda$8544\,\angstrom\ (central panel), $\lambda$8664\,\angstrom\ (right panel). In top panels, we also present the out-of-transit spectra from same night for reference, and marked the position of the stellar line (vertical magenta line).
    }
\end{figure*}

This work focuses on the \ion{He}{i} triplet line, but the \ion{Ca}{II} infrared triplet (IRT; $\lambda$8500\,\AA, $\lambda$8544\,\AA, and $\lambda$8664\,\AA) lines are also accessible with the HPF spectrograph.  In this section, we explore HPF's ability to detect and measure \ion{Ca}{II}\,IRT absorption signals in planetary atmospheres.
In particular, we focus on WASP-76\,b because \cite{WASP-76_He_Casasayas2021} already reported the detection of \ion{Ca}{II}\,IRT lines combining two transits observed by the CARMENES spectrograph.

We observed two consecutive transits of WASP-76\,b with HPF: visit 1 has out-of-transit baseline the night before and during the same night, and visit 2 has out-of-transit baseline one and two nights after. We find that the transmission spectra obtained using out-of-transit data from farther in time were noisier than those obtained using data from closer to the transiting night. The most reliable results are obtained from visit 1 using only the spectra taken on the same night, which is shown in Figure\,\ref{Fig: WASP-76 Ca IRT}. The TS from other out-of-transit selections are strongly affected by stellar variability. Our findings are expected since the \ion{Ca}{II}\,IRT lines are used as stellar activity tracers (\citealp{SERVAL}).

The $\lambda$8544\,\AA, and $\lambda$8664\,\AA\ TS exhibit $\sim$2\% features, much deeper than the 0.6\% and 0.35\% absorptions, respectively, reported by \cite{WASP-76_He_Casasayas2021}. The $\lambda$8500\,\AA\ TS does not have the precision to confirm the previous detection of 0.22\%, and we can only to rule out absorptions greater than 1\%.
If the signals at $\lambda$8544\,\AA, and $\lambda$8664\,\AA\ are real, e.g., due to planetary absorption variability, then the absorption at $\lambda$8500\,\AA\ should be deeper as well. However, there is no evidence for this scenario. Furthermore, the three residual maps show vertical structure at the line positions, likely originating from line core variability. The stellar \ion{Ca}{II}\,IRT lines are usually stronger and deeper than the He triplet line resulting in lower S/N at the core.
The residual maps from \citet[][Figure\,8]{WASP-76_He_Casasayas2021} show residuals from the line core as well, but the planetary trace is visible at phases far from the central transit time, due to the large velocity shift of the planet.
Our visit 1 with HPF only covered a fraction of the transit close to the central time, where the planetary trace intersects the core. Therefore, any planetary signal is buried in the core variability.
Similar results were obtained for HAT-P-67 b, where \cite{HAT-P-67b_Gully2024} with HPF did not detect the \ion{Ca}{II}\,IRT absorption reported by \cite{HAT-P-67_CaIRT_BelloArufe} with CARMENES.

We conclude that inspecting \ion{Ca}{II}\,IRT planetary absorption with HET/HPF is challenging due to line core variability. However, it could be feasible for planets with large velocity shifts, in-transit observations far from the central time, and out-of-transit observations on the same night.

\section{Stellar He triplet light curves}

\begin{figure*}[ht!]
   \centering
   \includegraphics[width=0.49\linewidth]{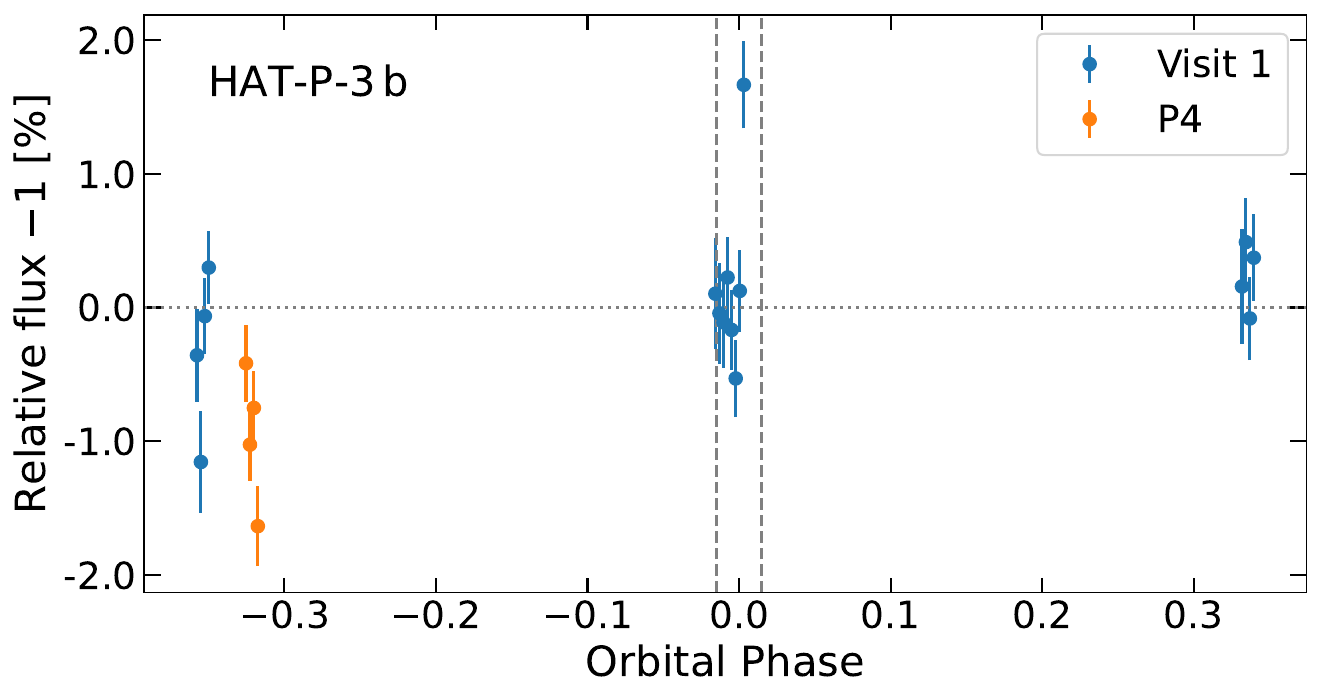}
   \includegraphics[width=0.49\linewidth]{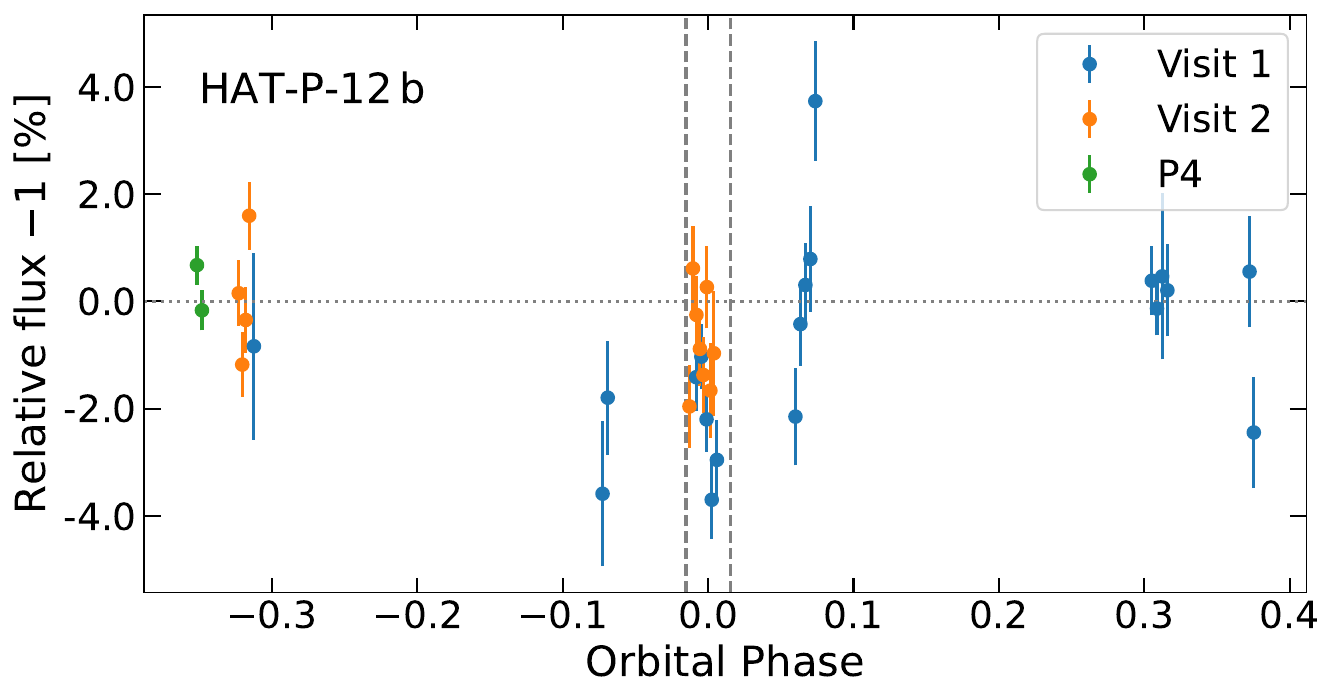}
   \includegraphics[width=0.49\linewidth]{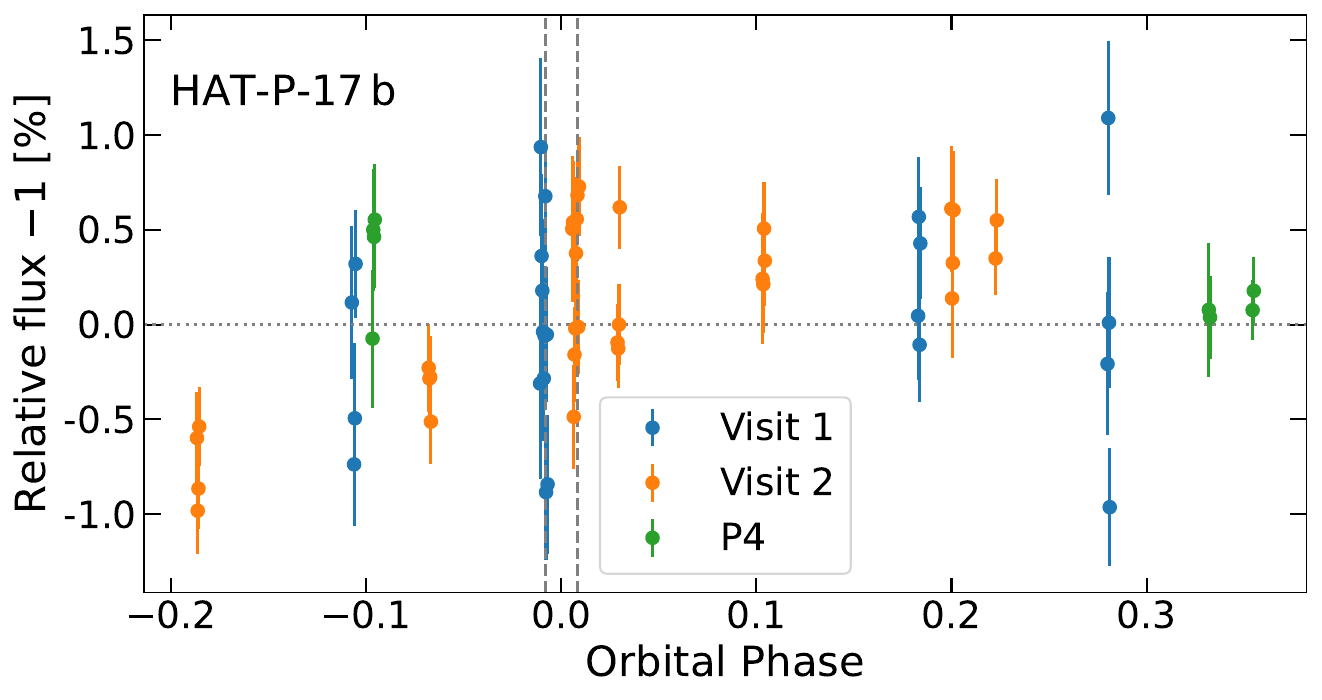}
   \includegraphics[width=0.49\linewidth]{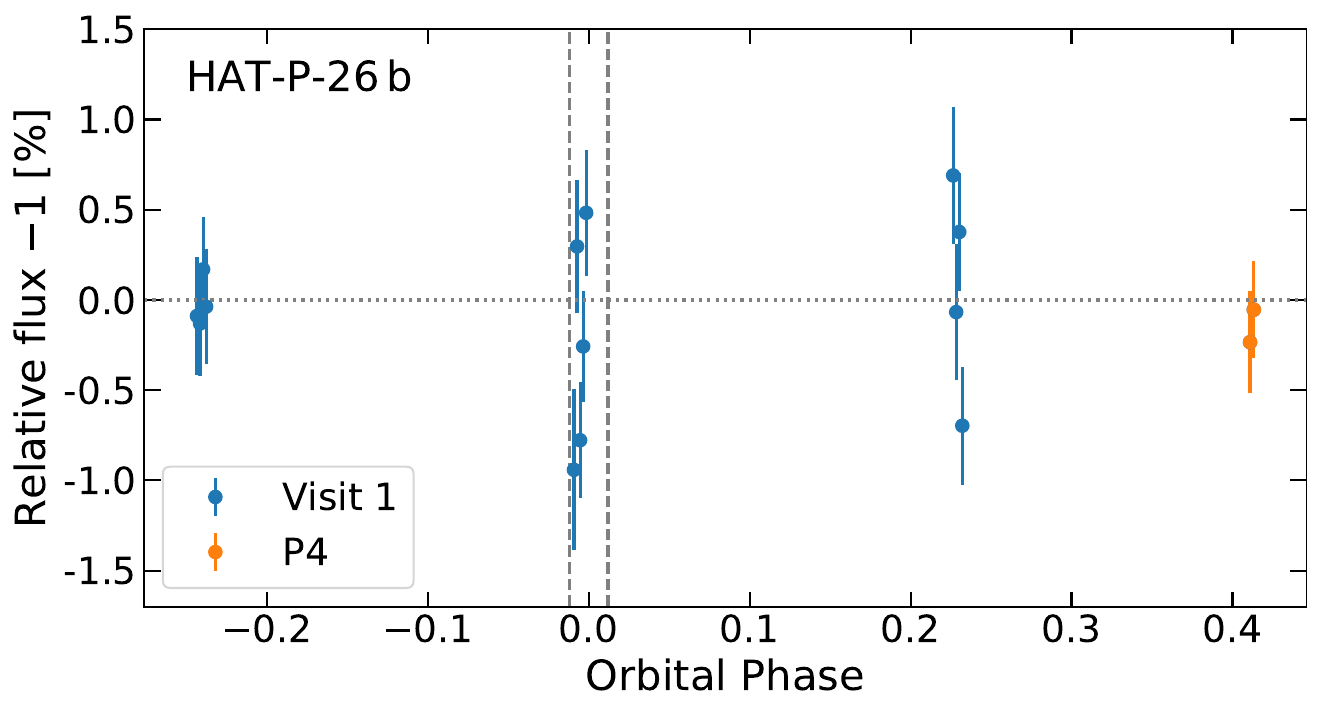}
   \includegraphics[width=0.49\linewidth]{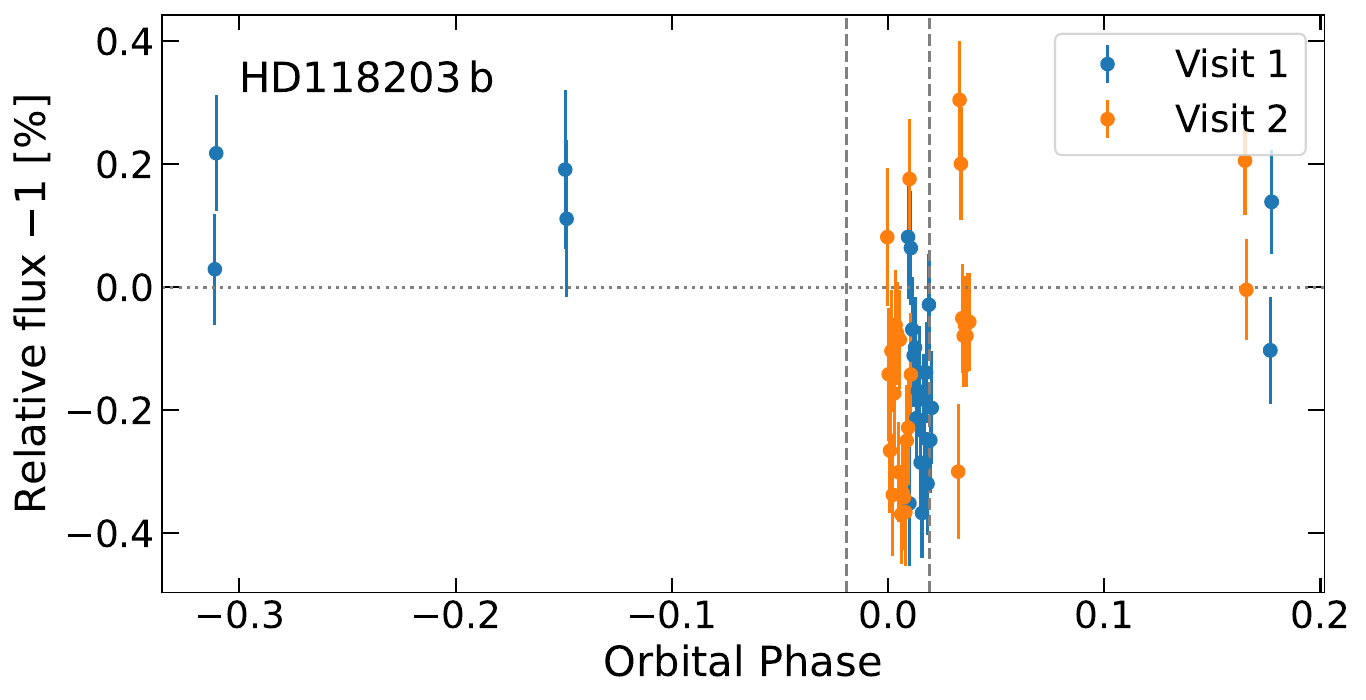}
   \includegraphics[width=0.49\linewidth]{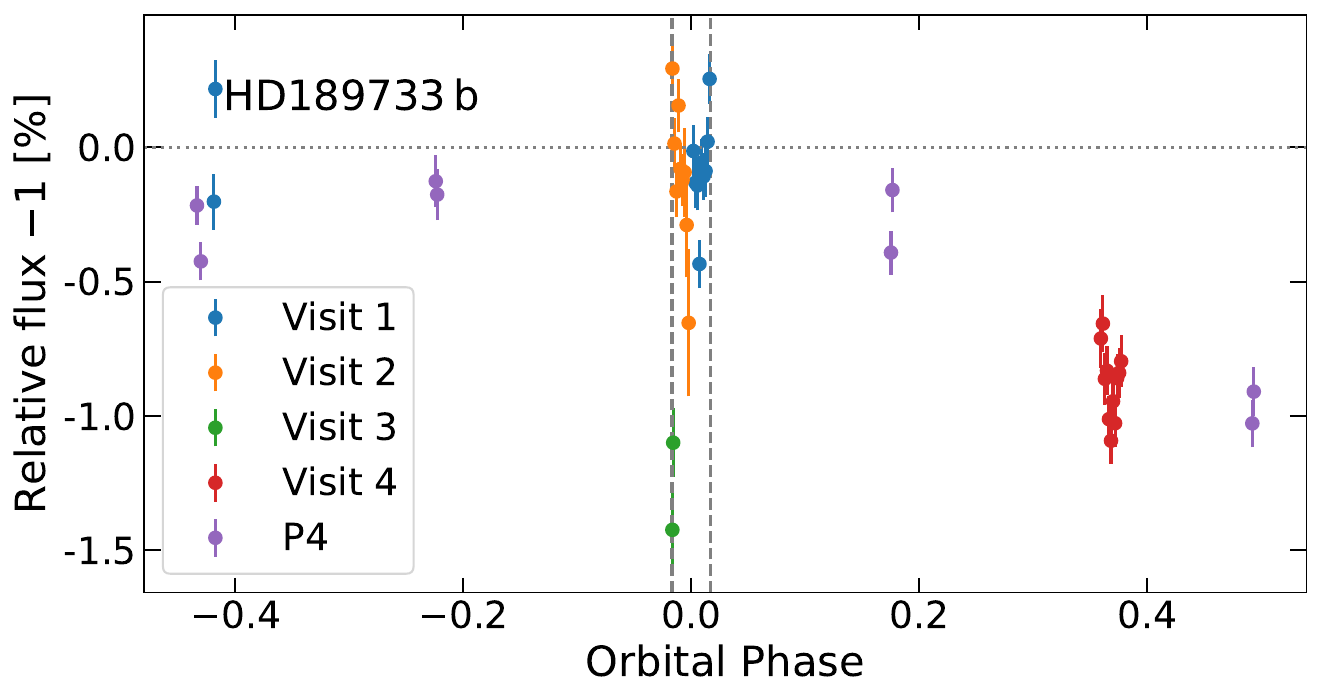}
   \includegraphics[width=0.49\linewidth]{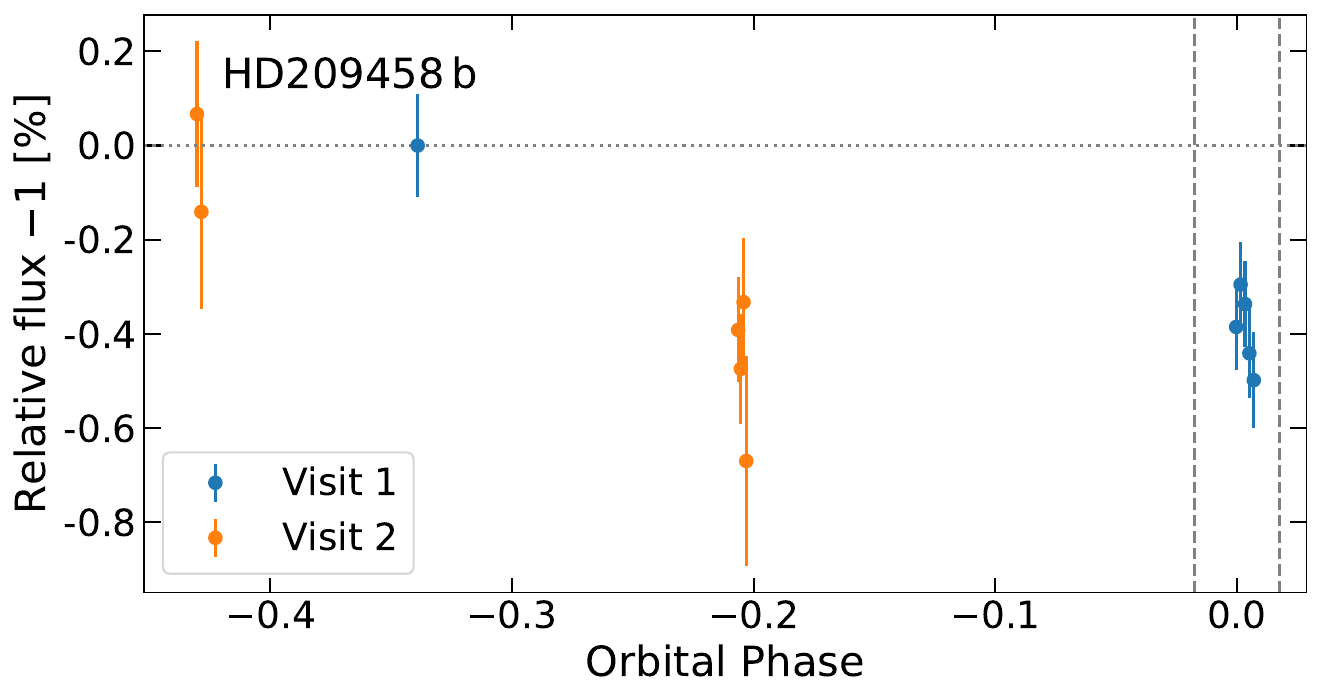}
   \includegraphics[width=0.49\linewidth]{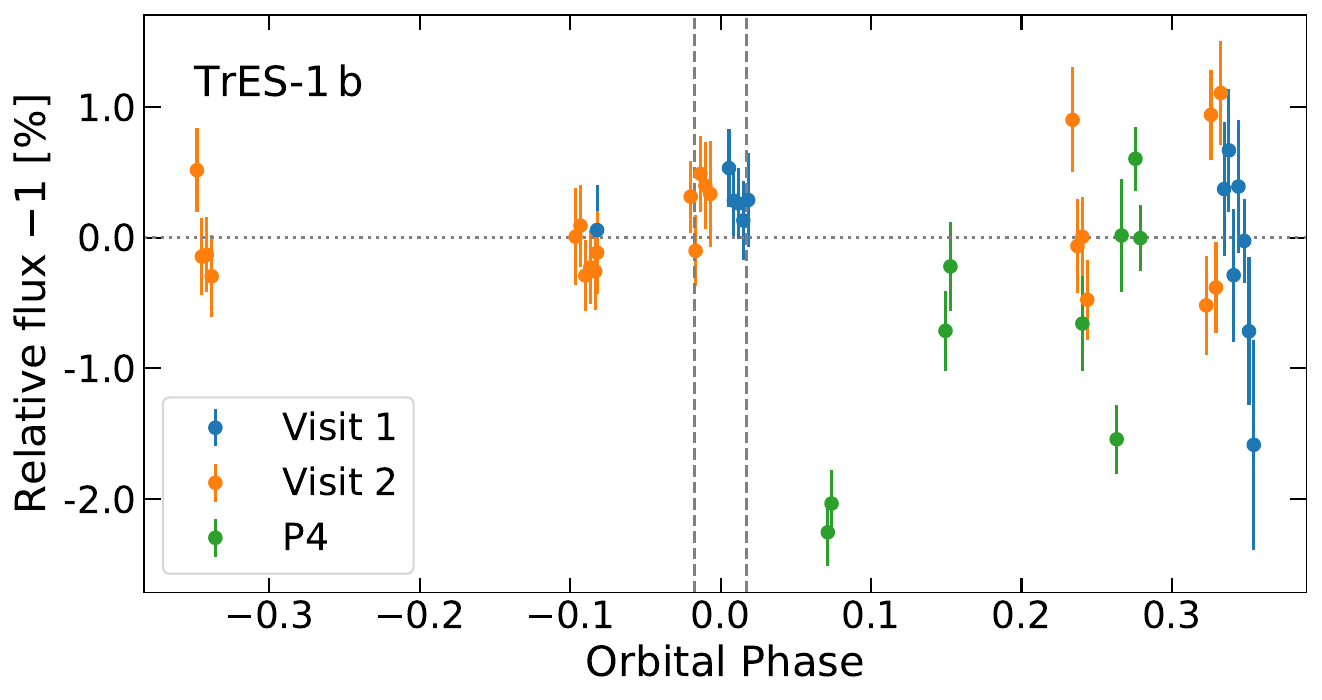}
   \caption{ \label{Fig: He LC plot 1}
   Stellar light curves of the \ion{He}{I} triplet line for HAT-P-3\,b, HAT-P-12\,b, HAT-P-17\,b, HAT-P-26\,b, HD118203\,b, HD189733\,b, \hd209\,b, and TrES-1\,b. The light curves have been constructed integrating the counts within a 1.5\,\angstrom\ band pass centered at the position of the doublet of the \he\ triplet line in the stellar rest frame. Data is color-coded by visit epoch. The dashed vertical lines indicate the transit duration.
   }
\end{figure*}

\begin{figure*}[ht!]
   \centering
   \includegraphics[width=0.49\linewidth]{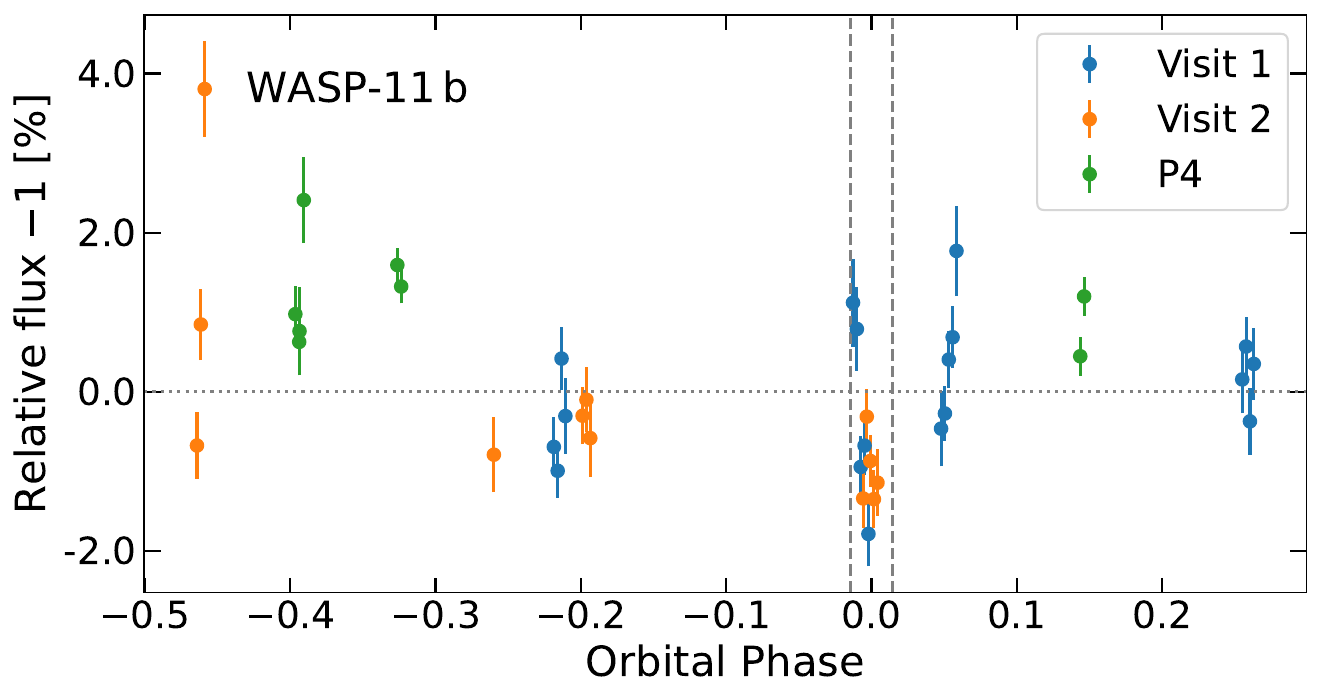}
   \includegraphics[width=0.49\linewidth]{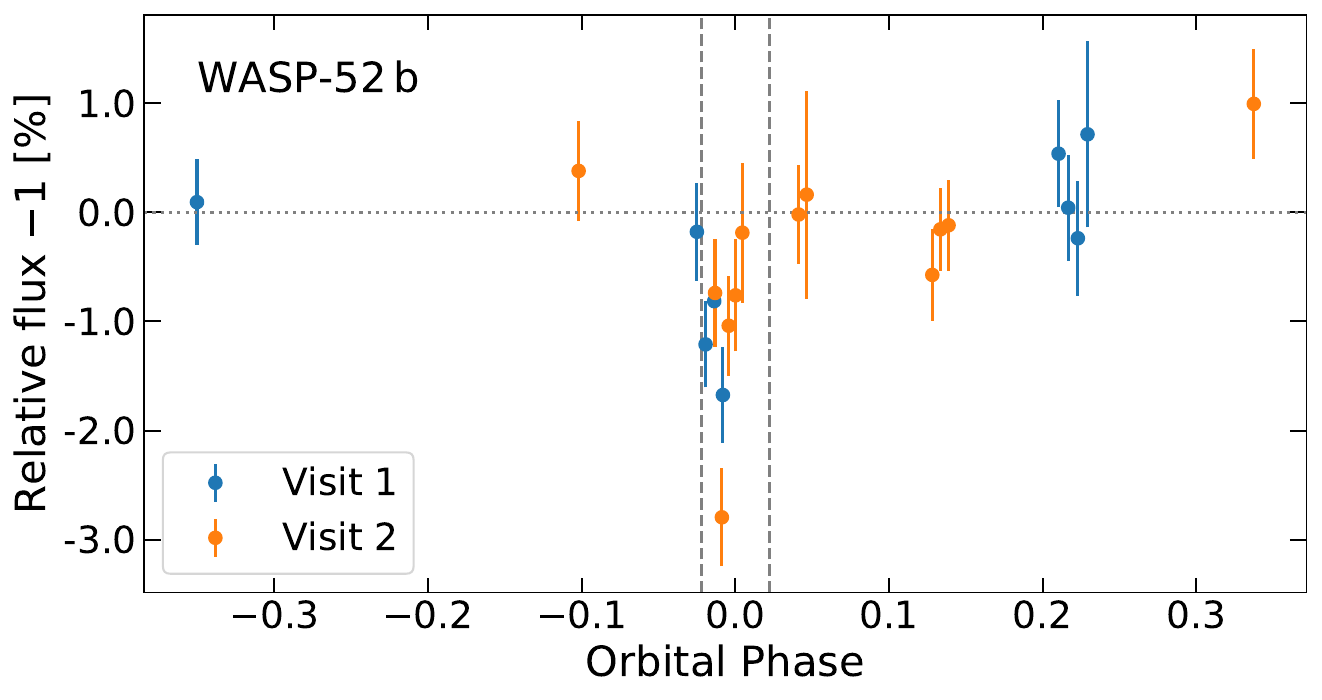}
   \includegraphics[width=0.49\linewidth]{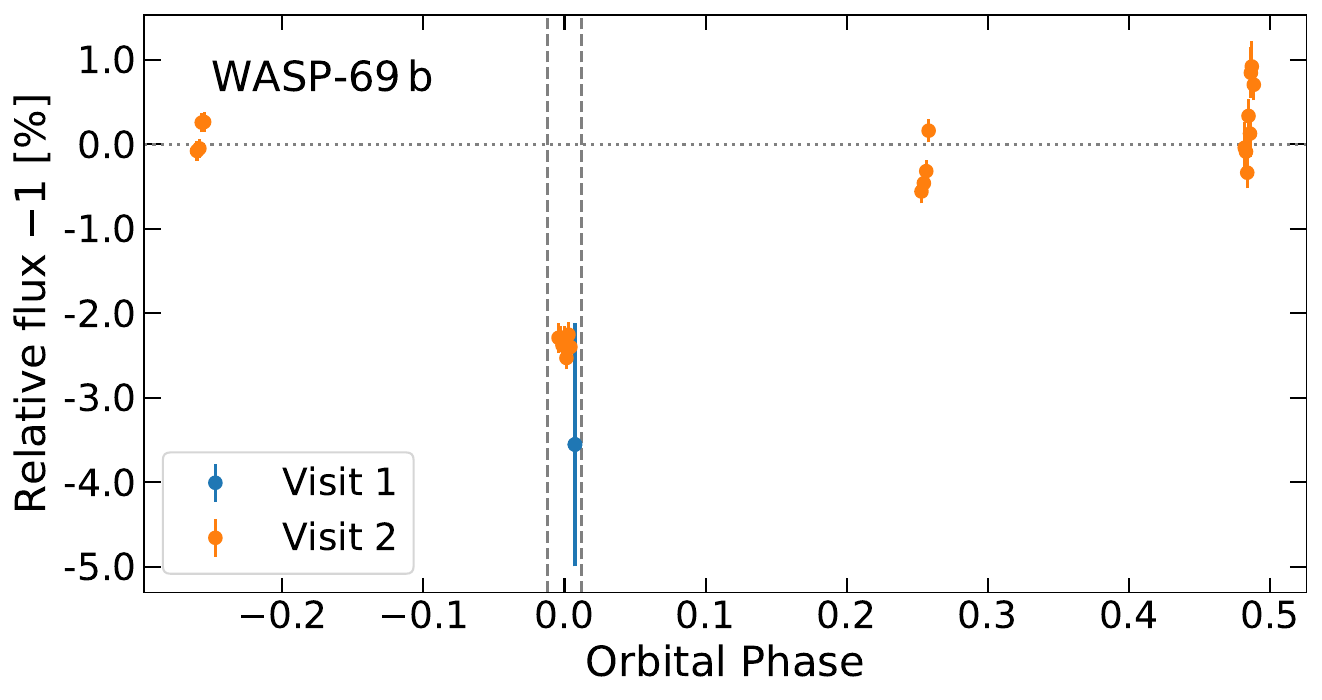}
   \includegraphics[width=0.49\linewidth]{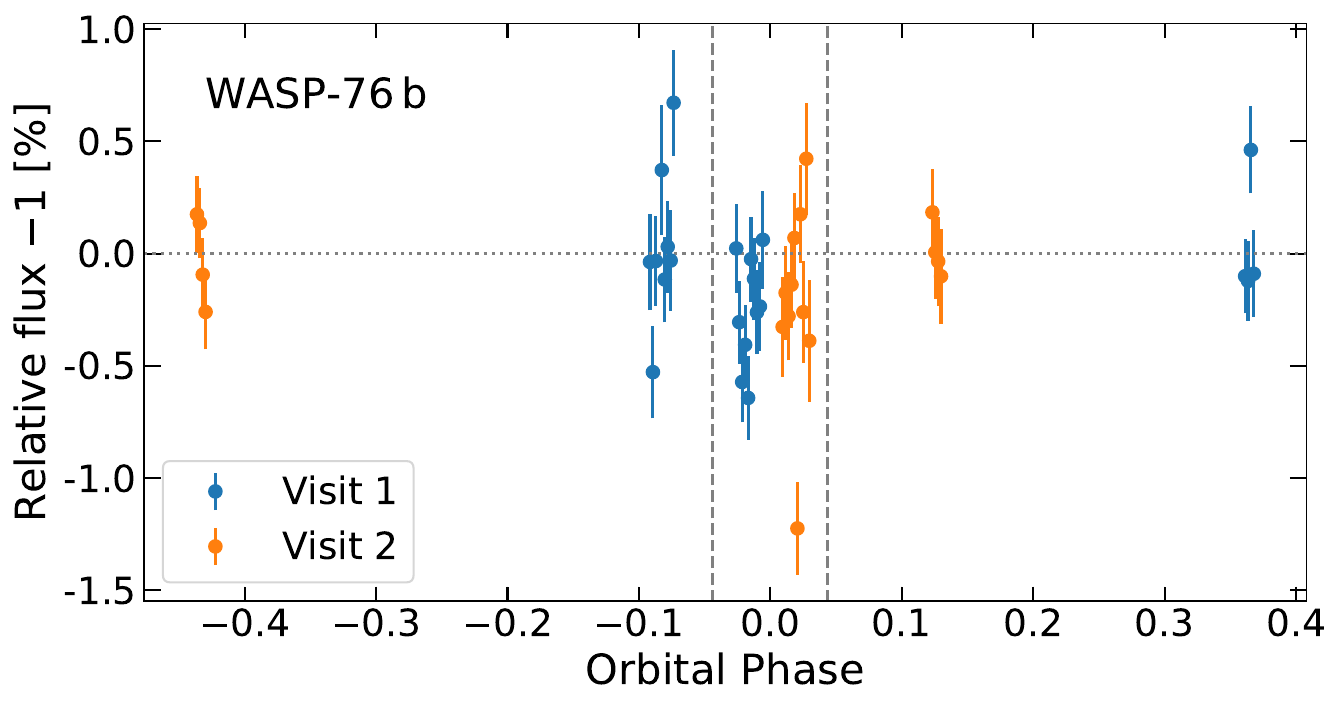}
   \includegraphics[width=0.49\linewidth]{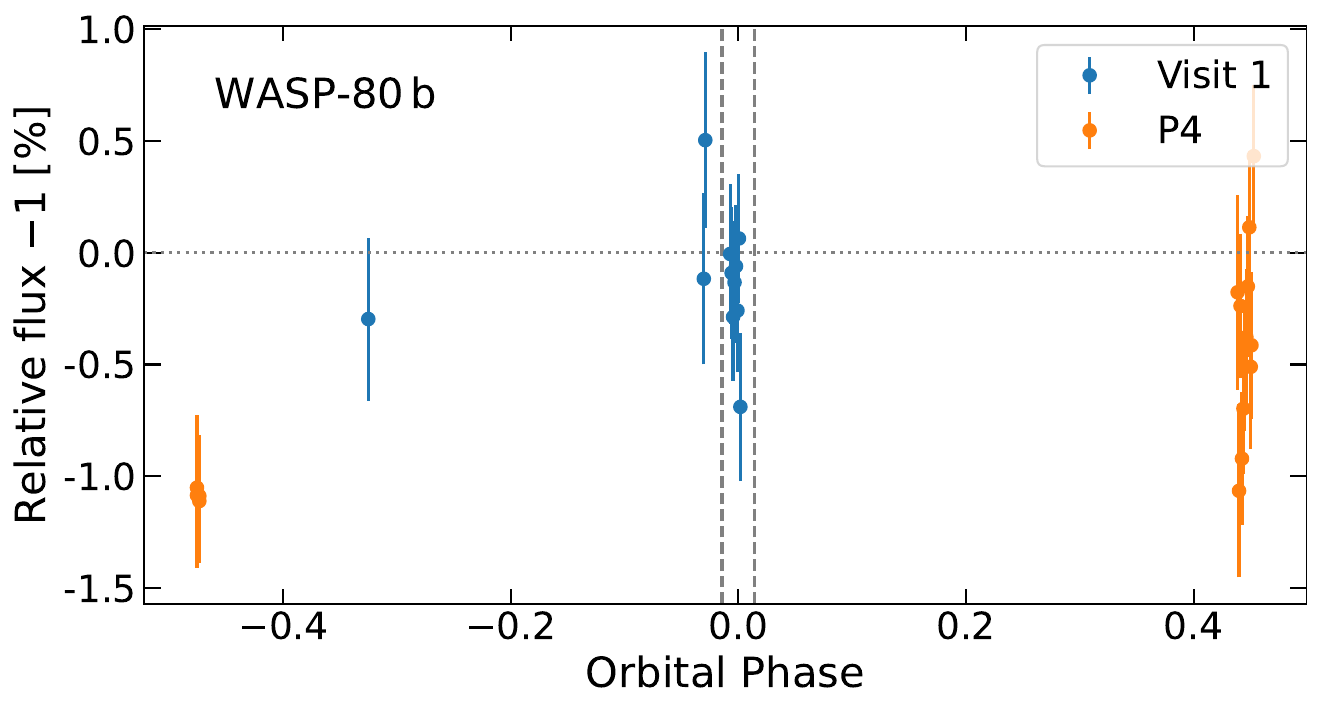}
   \includegraphics[width=0.49\linewidth]{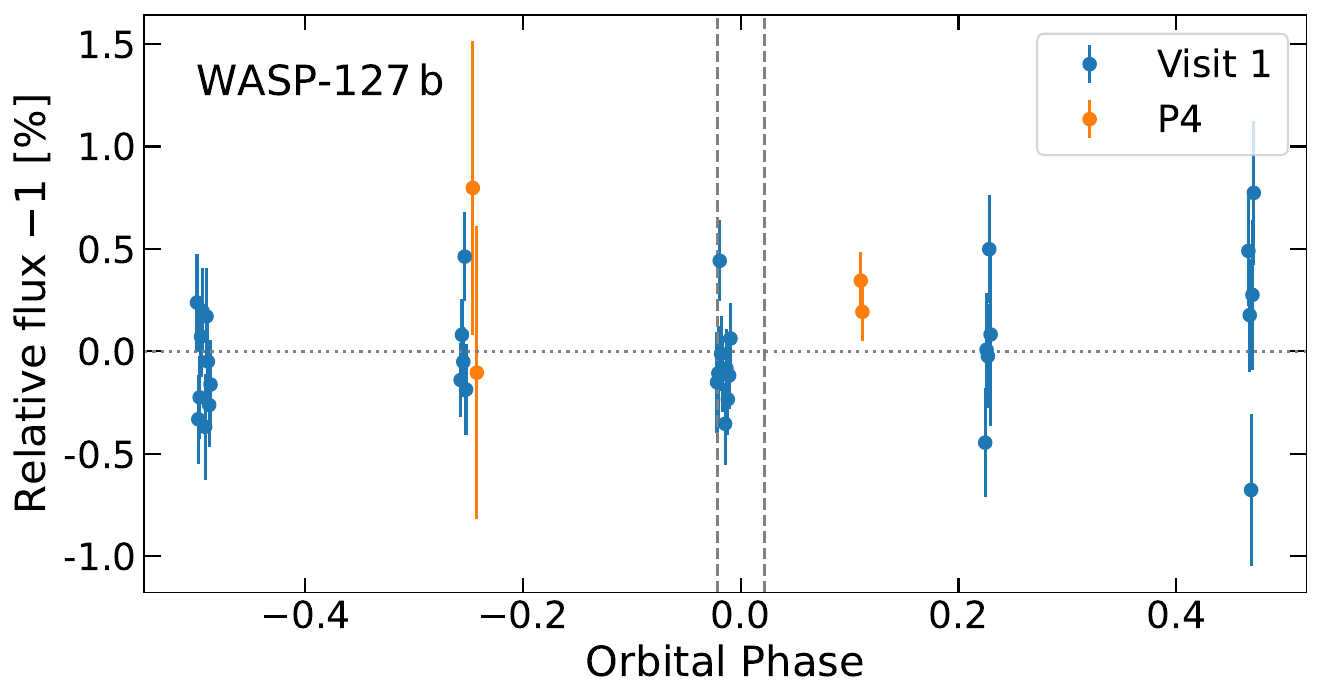}
   \includegraphics[width=0.49\linewidth]{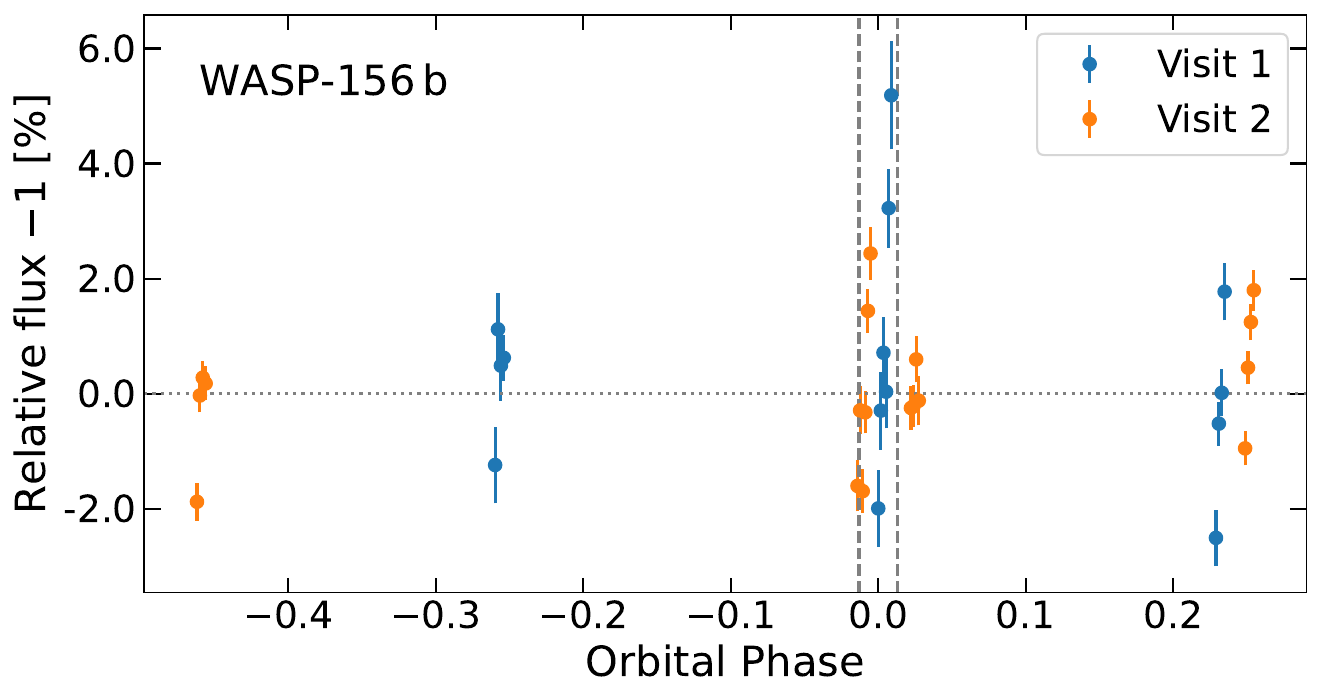}
   \includegraphics[width=0.49\linewidth]{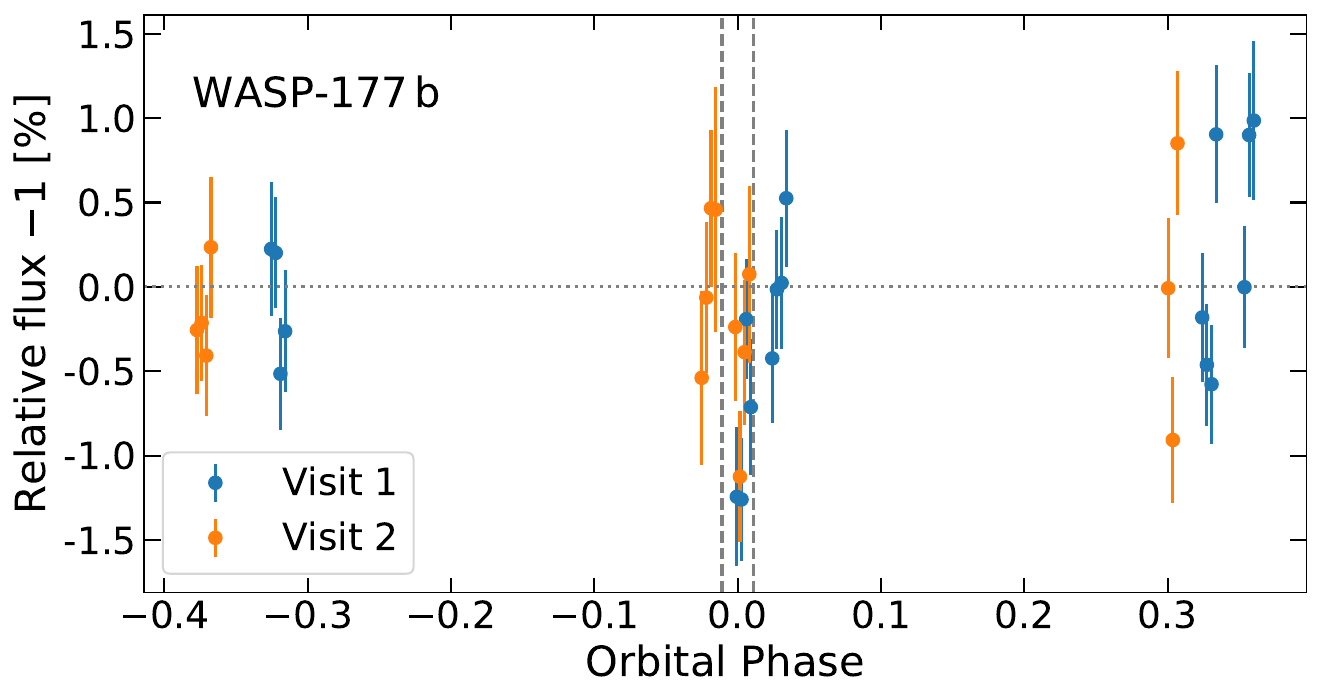}
   \caption{ \label{Fig: He LC plot 2}
   Same as Figure\,\ref{Fig: He LC plot 1} for WASP-11\,b, WASP-52\,b, WASP-69\,b, WASP-76\,b, WASP-80\,b, WASP-127\,b, WASP-156\,b, and WASP-177\,b.
   }
\end{figure*}

\section{Reference stellar spectra and individual transmission spectra}


\begin{figure*}[h!]
   \centering
   \includegraphics[width=0.32\linewidth]{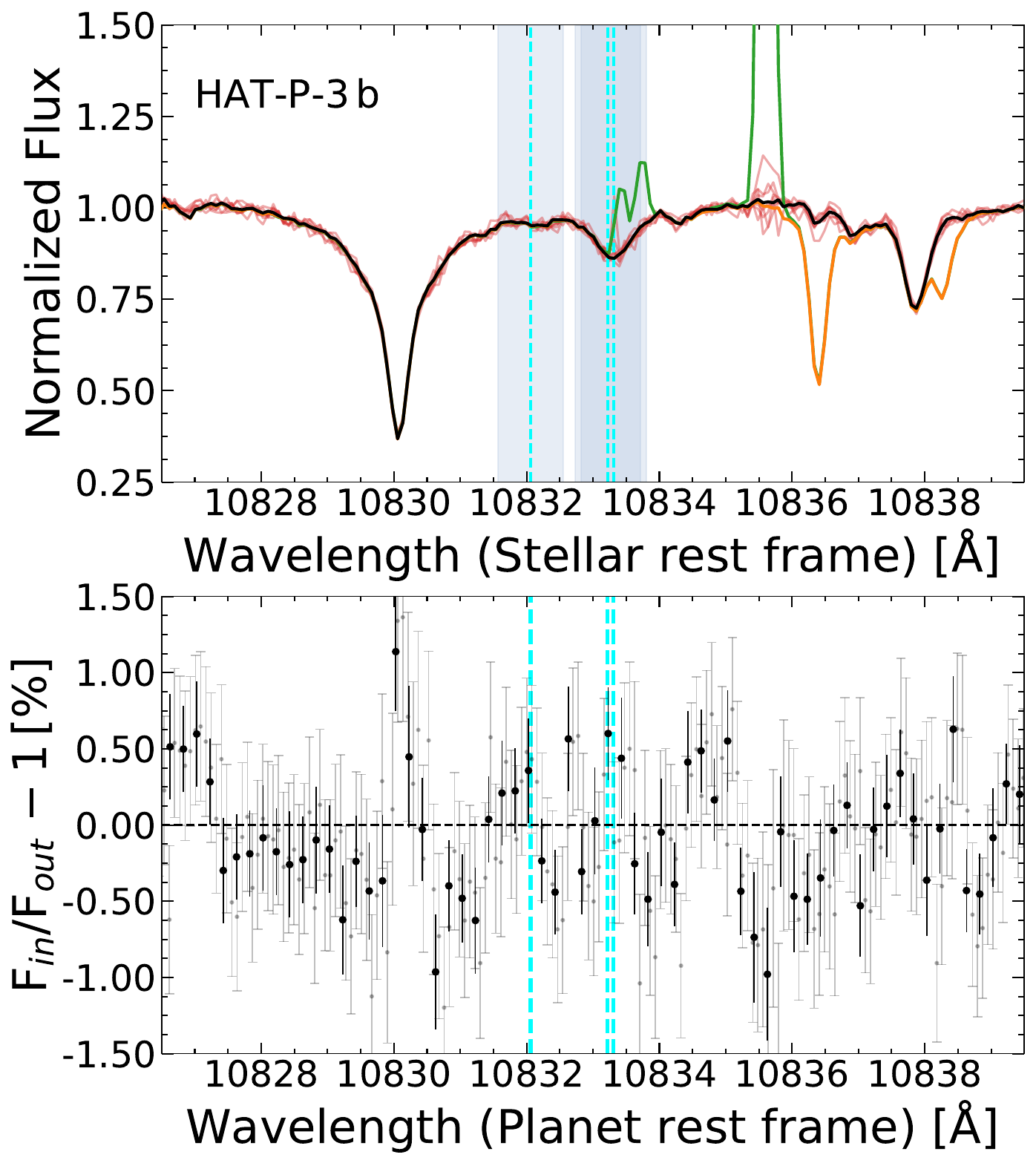}
   \includegraphics[width=0.32\linewidth]{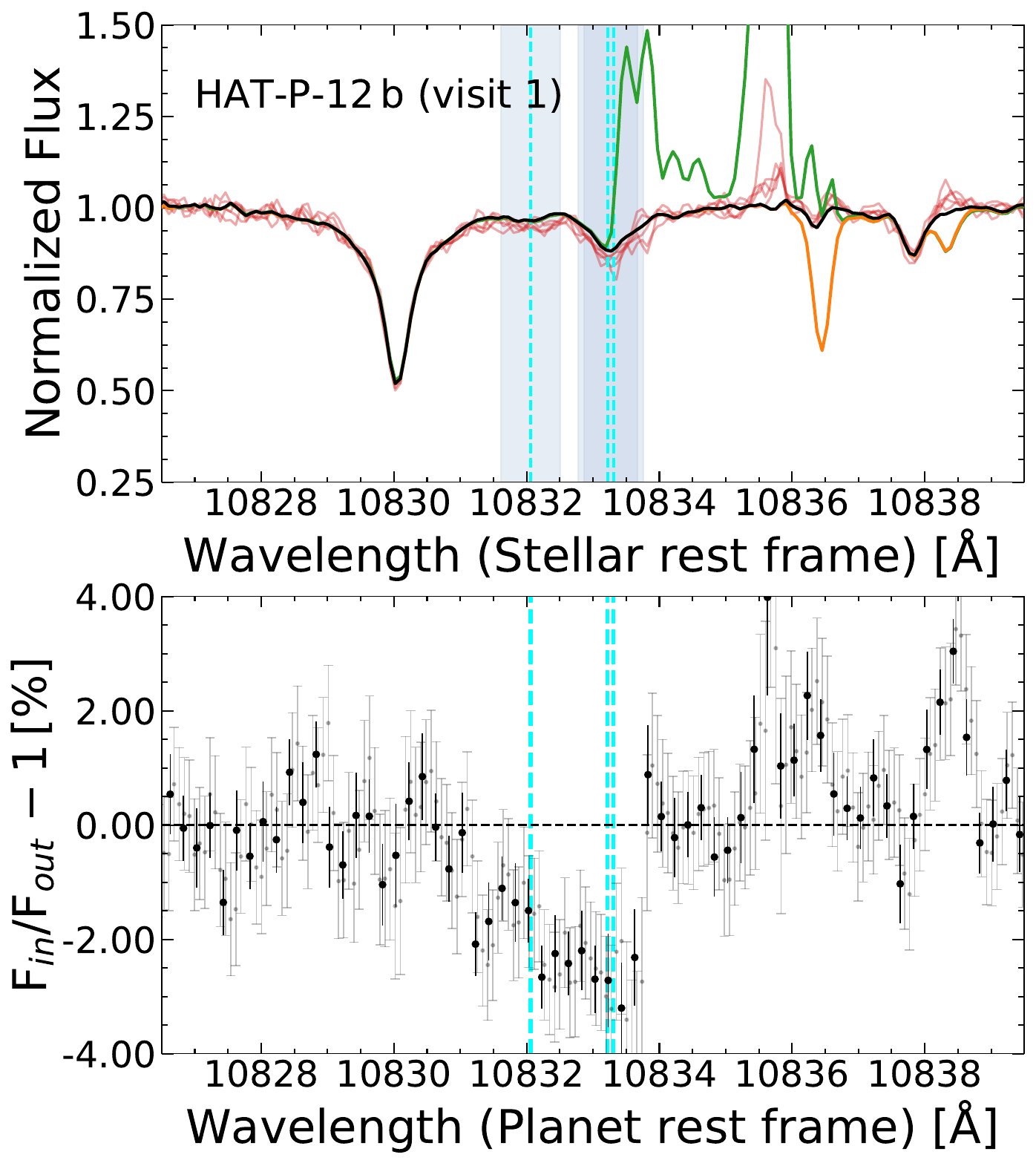}
   \includegraphics[width=0.32\linewidth]{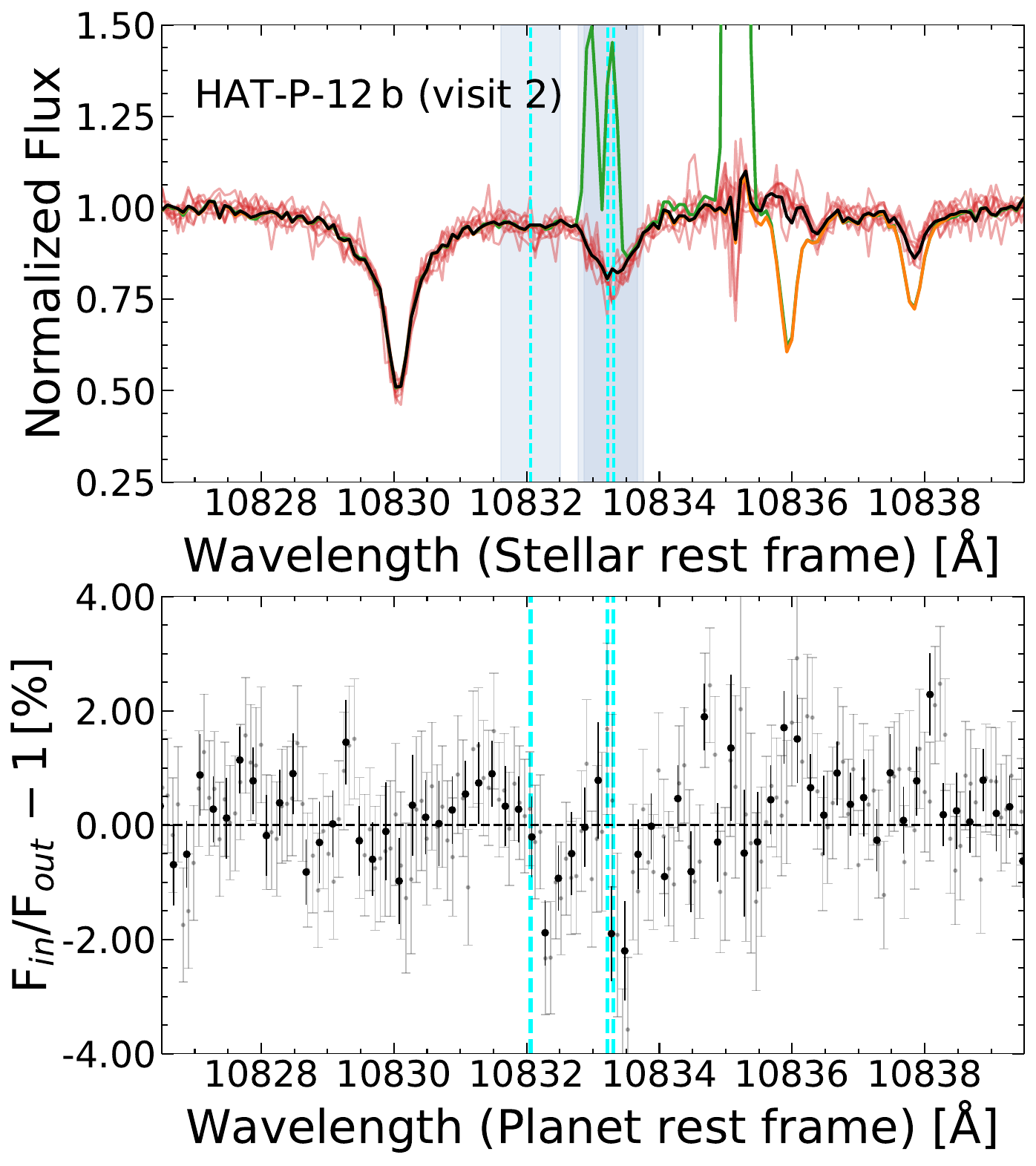}
   \includegraphics[width=0.32\linewidth]{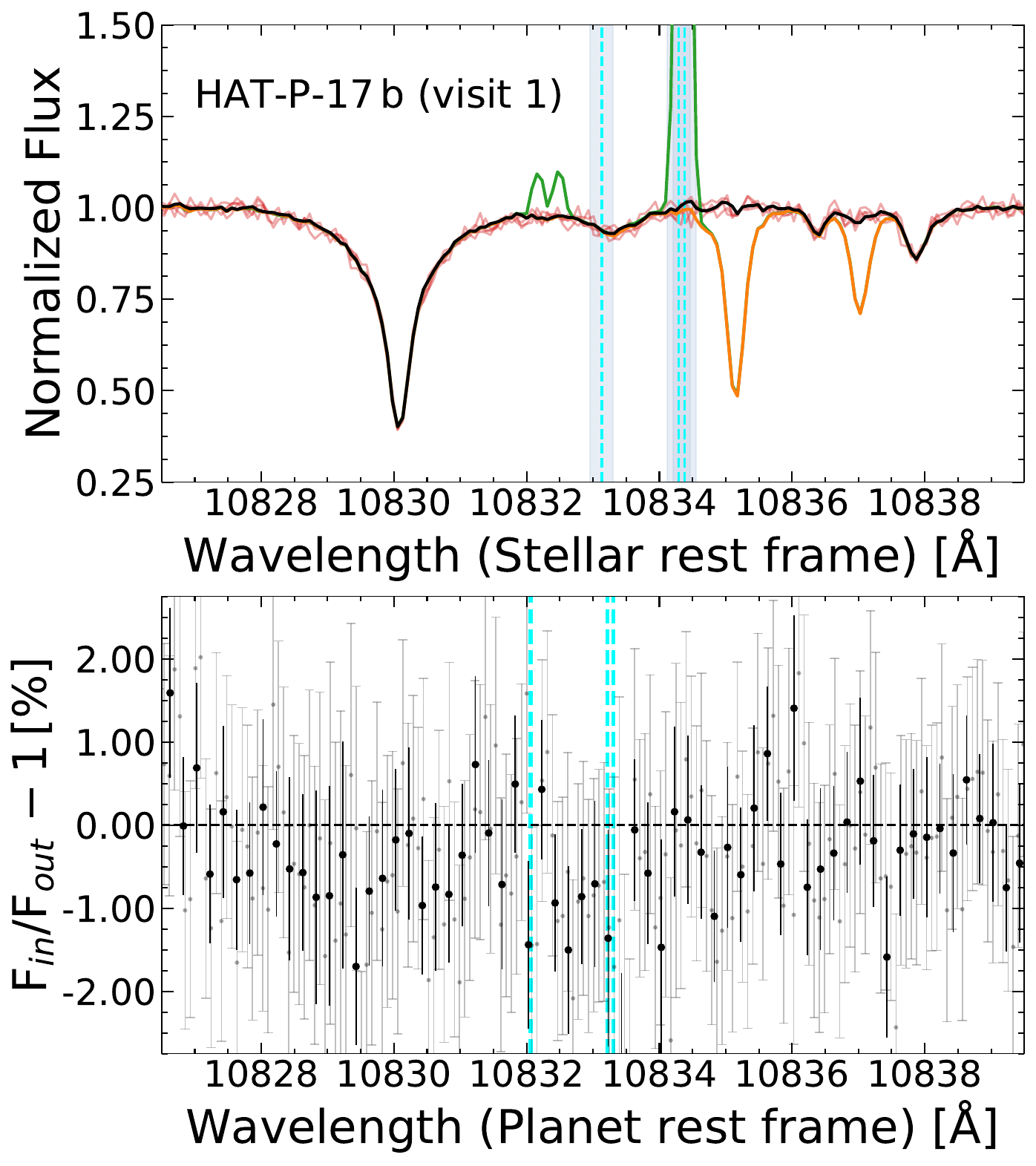}
   \includegraphics[width=0.32\linewidth]{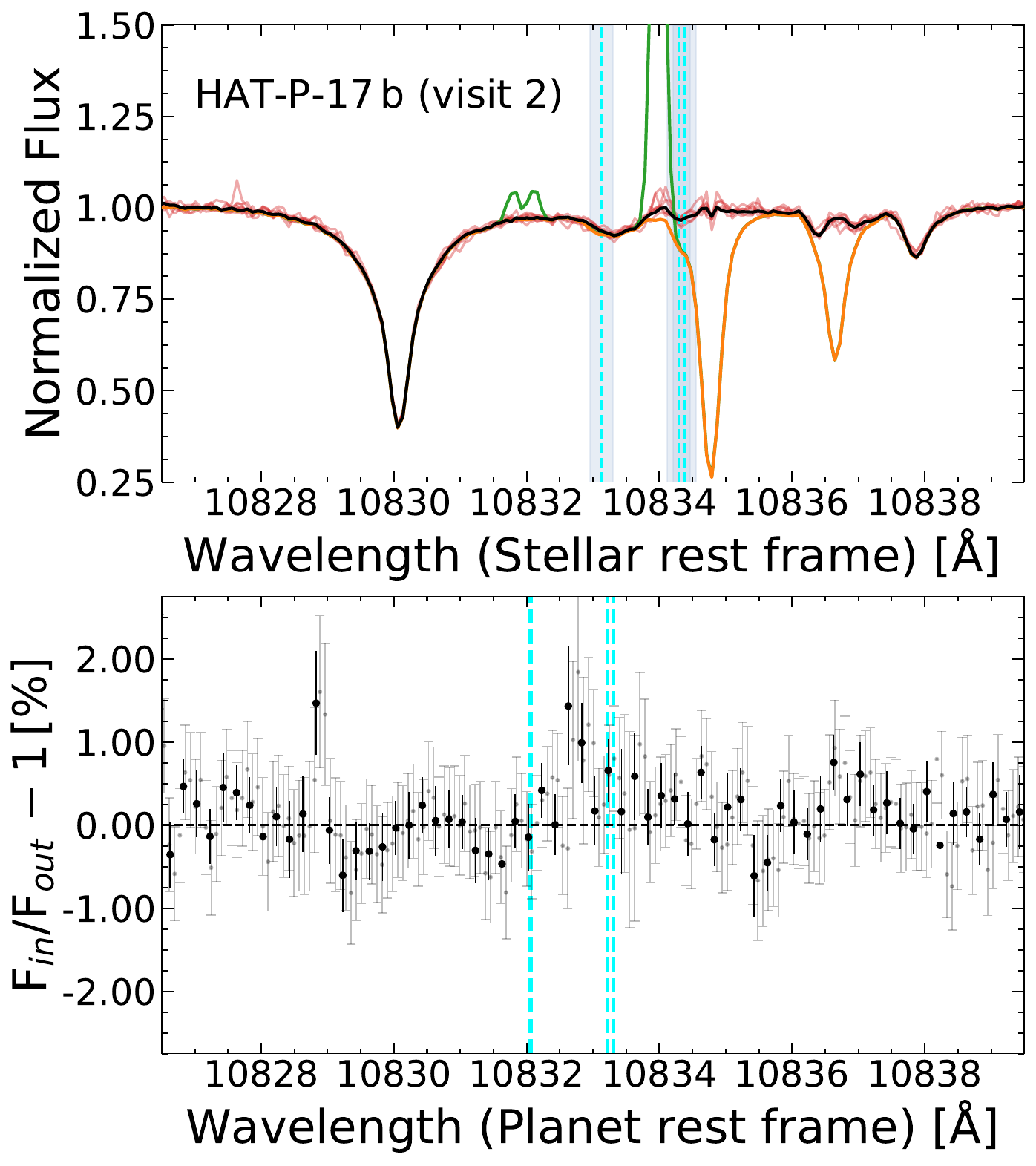}
   \includegraphics[width=0.32\linewidth]{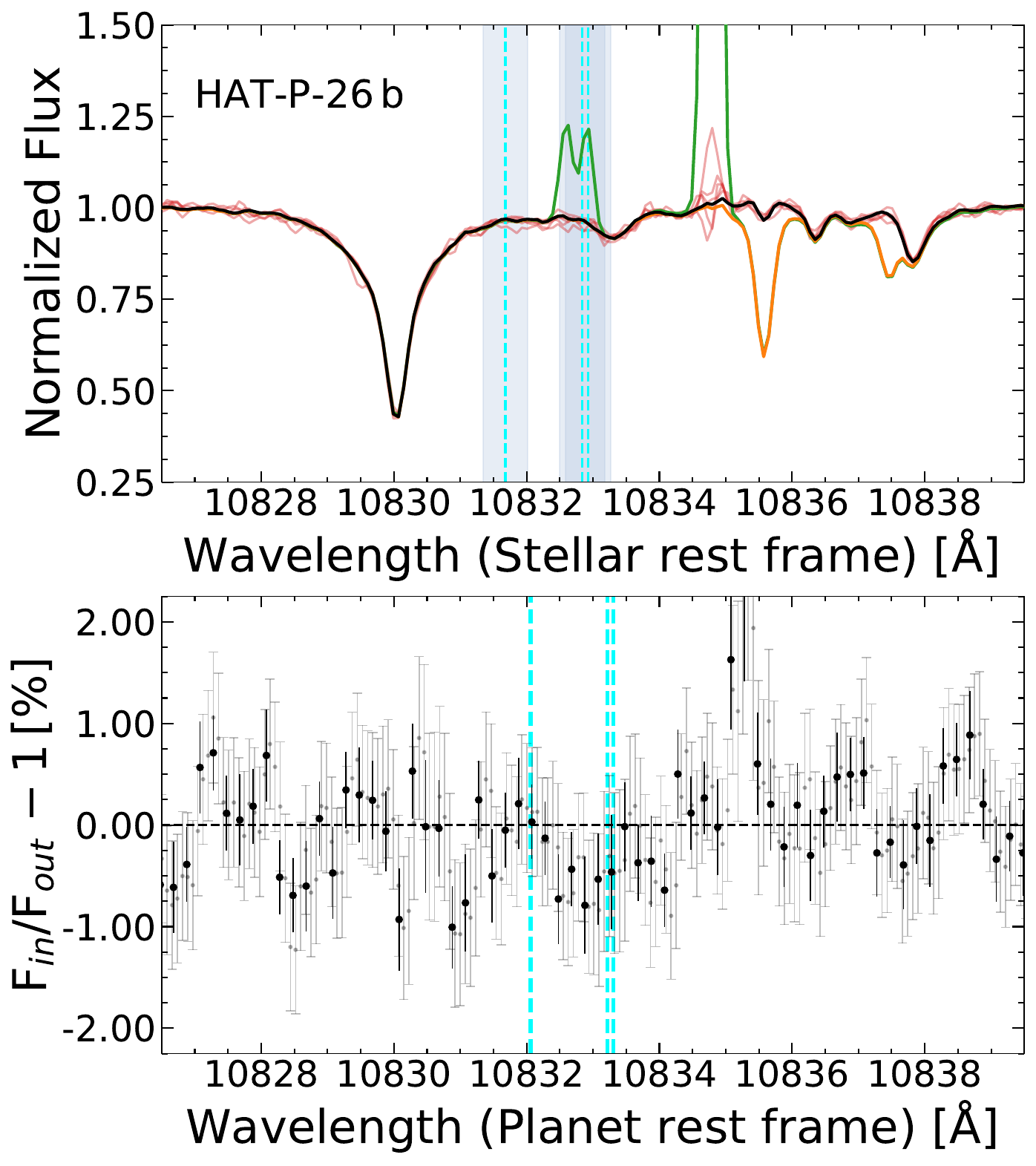}
   \includegraphics[width=0.32\linewidth]{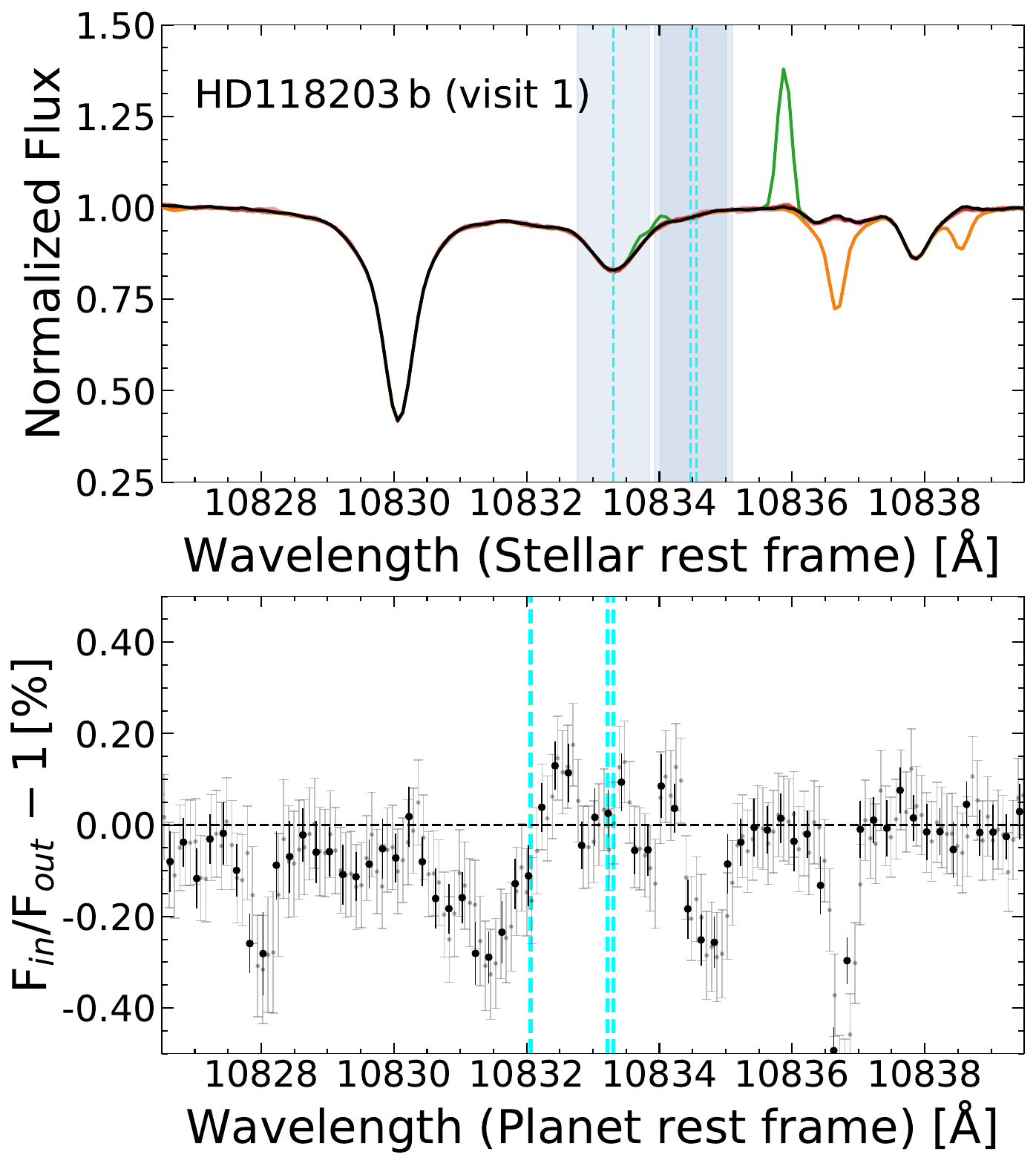}
   \includegraphics[width=0.32\linewidth]{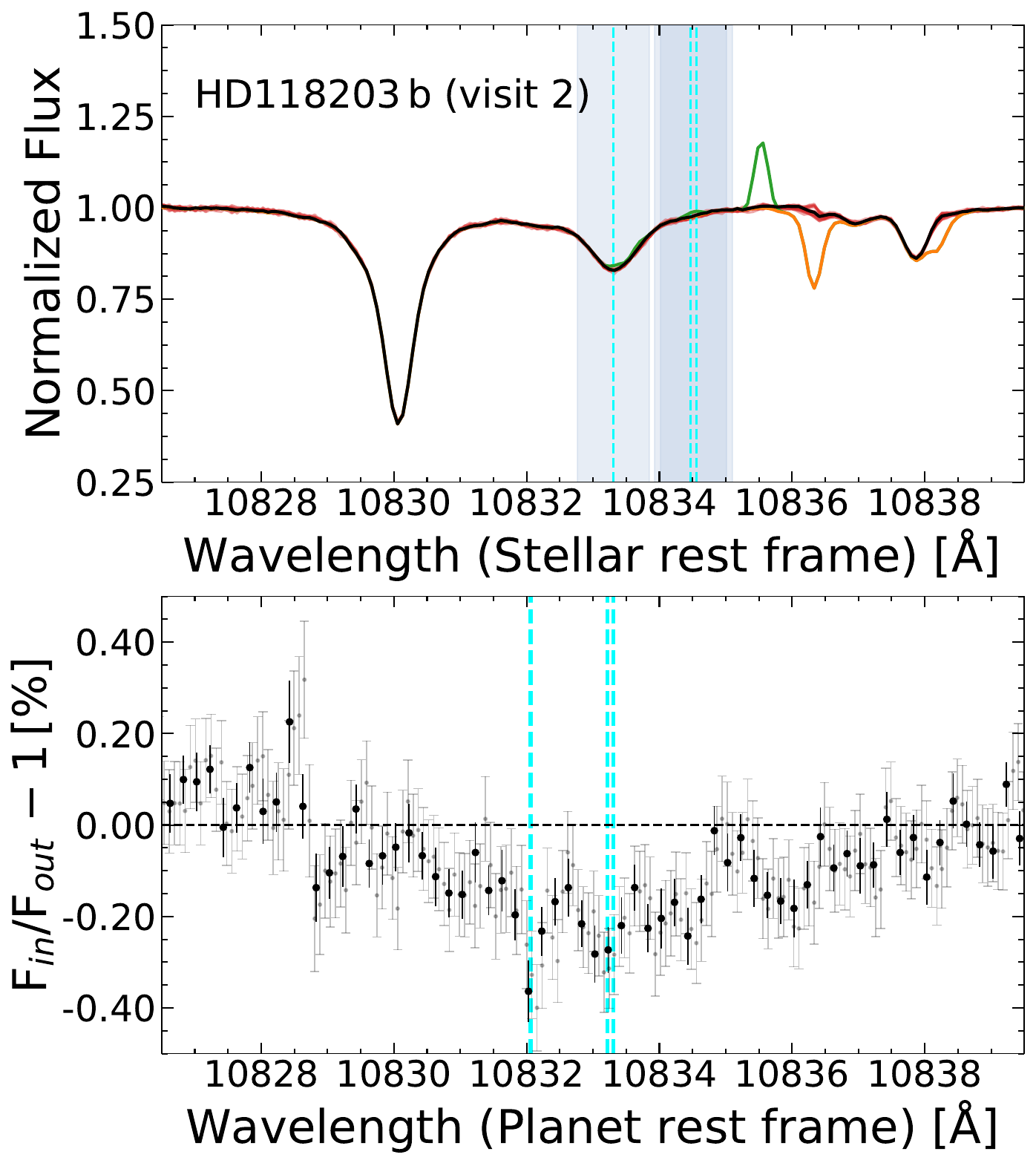}
   \caption{ \label{Fig: Visits plot 0}
   Reference stellar spectra for HAT-P-3\,b, HAT-P-12\,b, HAT-P-17\,b, HAT-P-26\,b, and HD118203\,b.
   \textit{Top panels}: Comparison of the out-of-transit combined spectrum from raw (green), telluric emission corrected (orange), and telluric absorption and emission corrected (black) spectra. In-transit individual spectra are superposed in red. The dashed cyan vertical lines indicate the planet rest frame position of the \ion{He}{I} triplet line, and the shaded area indicates the wavelength shift during a transit.
   \textit{Bottom panels}: Transmission spectra for each visit. Same as left panels in Figure\,\ref{Fig: TS plot 1}.
   }
\end{figure*}

\begin{figure*}[h!]
   \centering
   \includegraphics[width=0.32\linewidth]{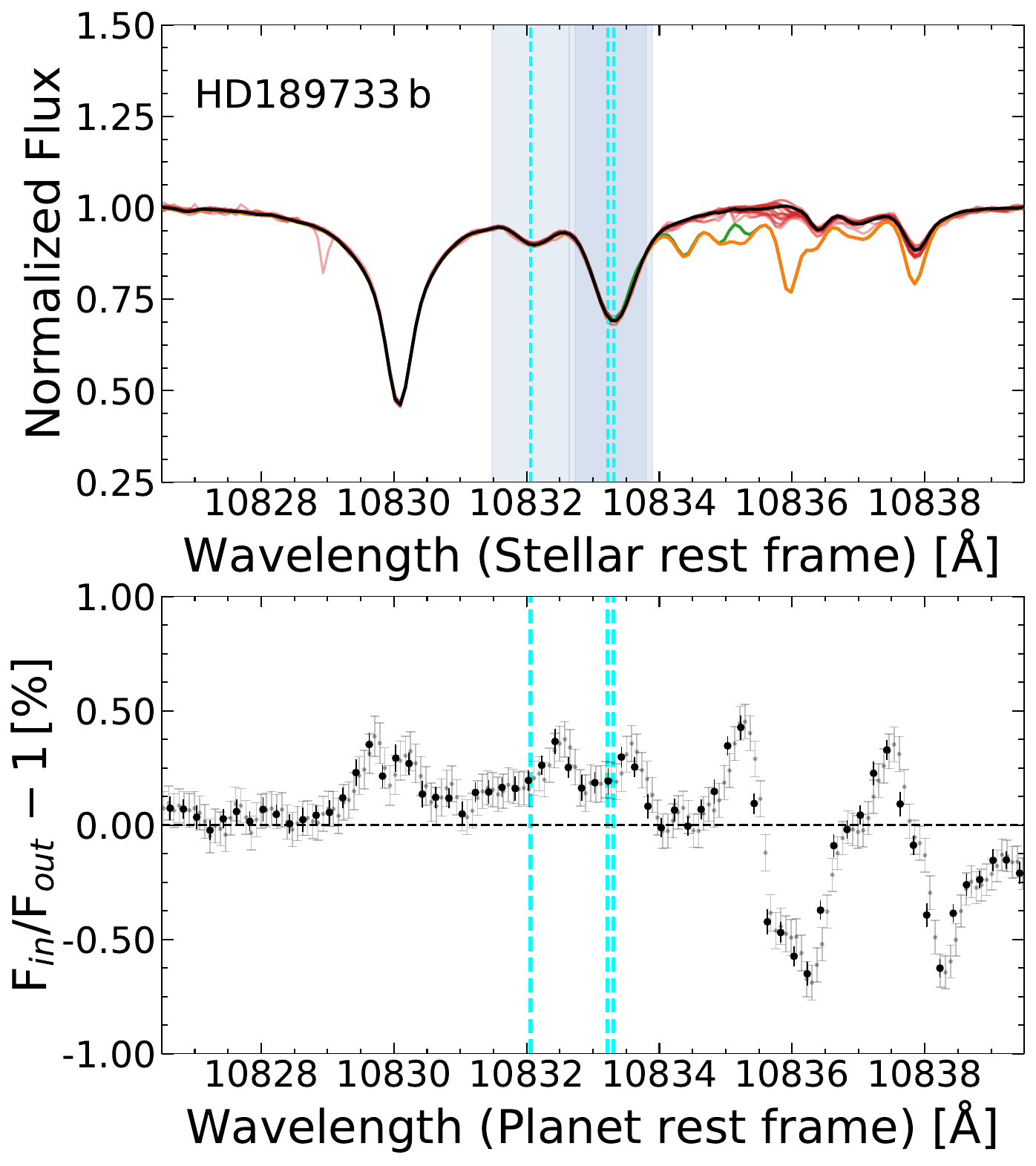}
   \includegraphics[width=0.32\linewidth]{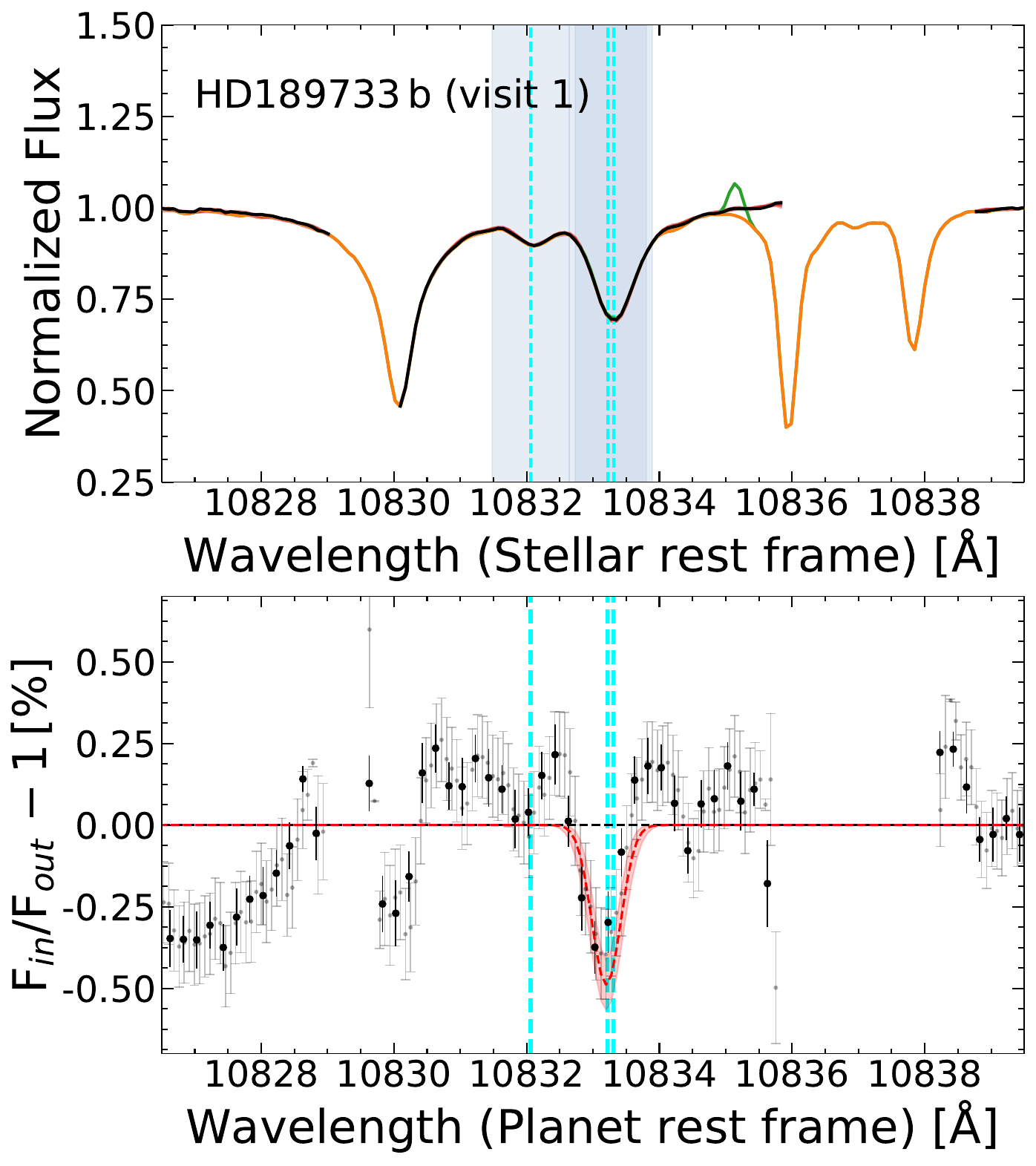}
   \includegraphics[width=0.32\linewidth]{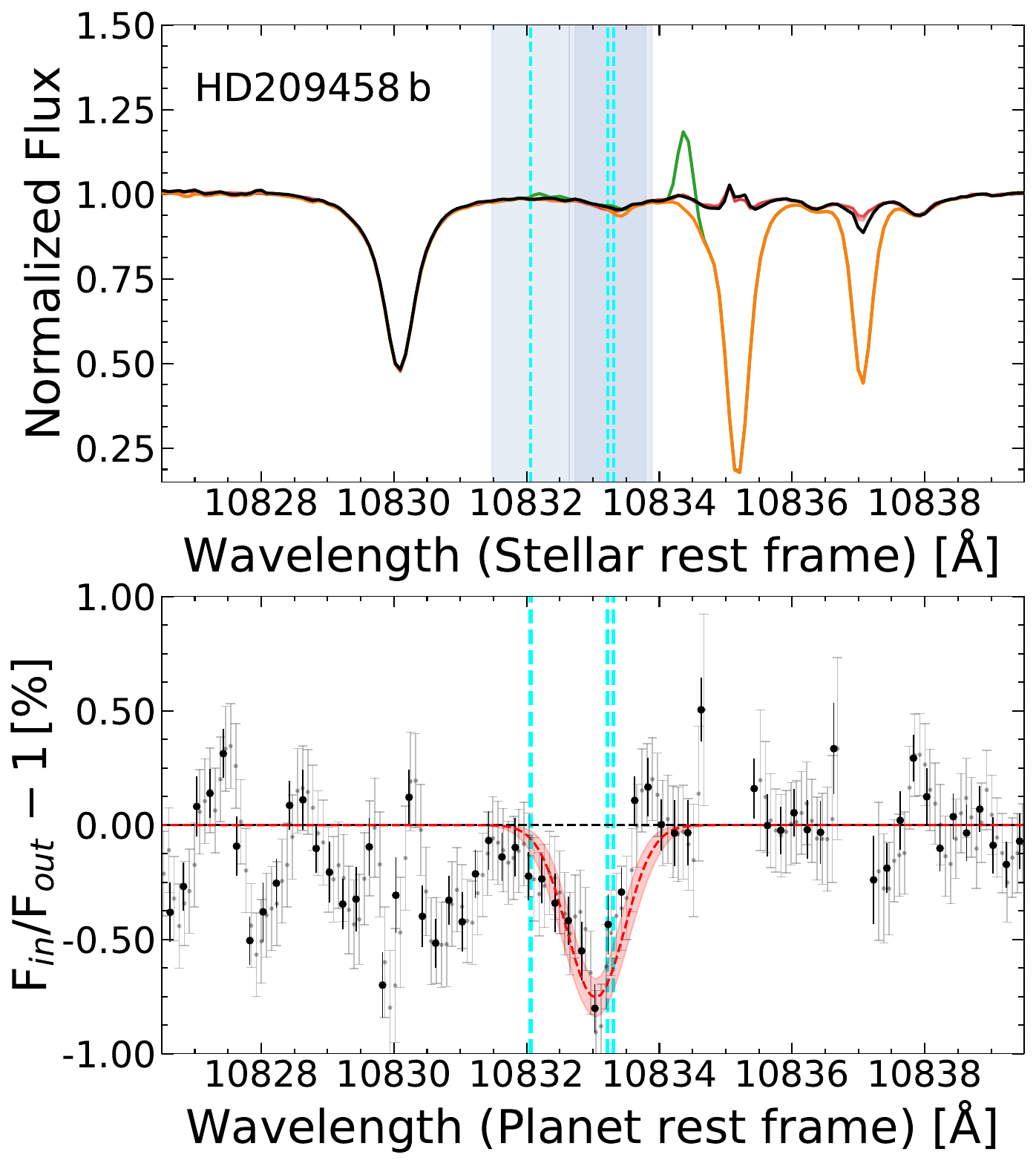}
   \includegraphics[width=0.32\linewidth]{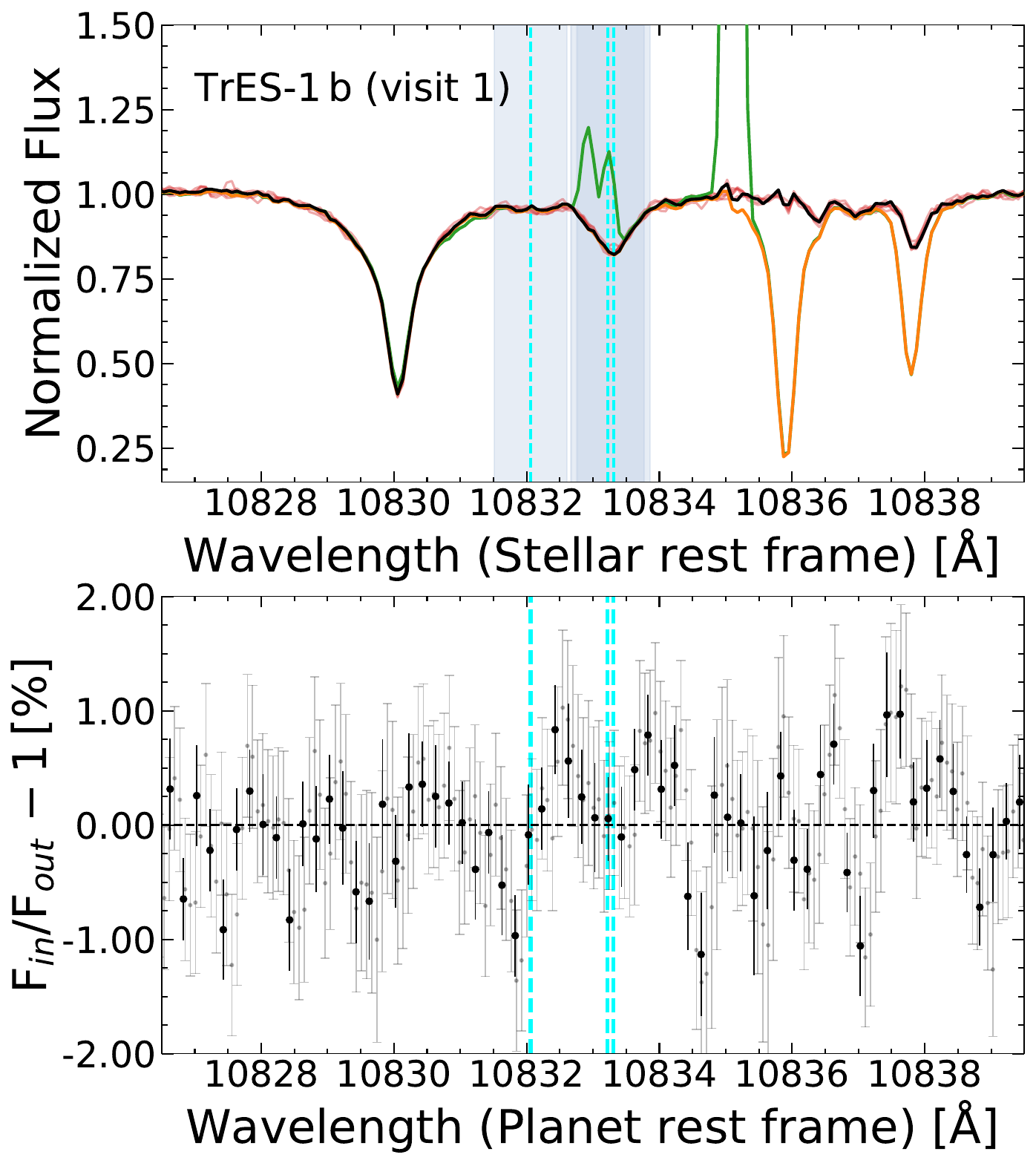}
   \includegraphics[width=0.32\linewidth]{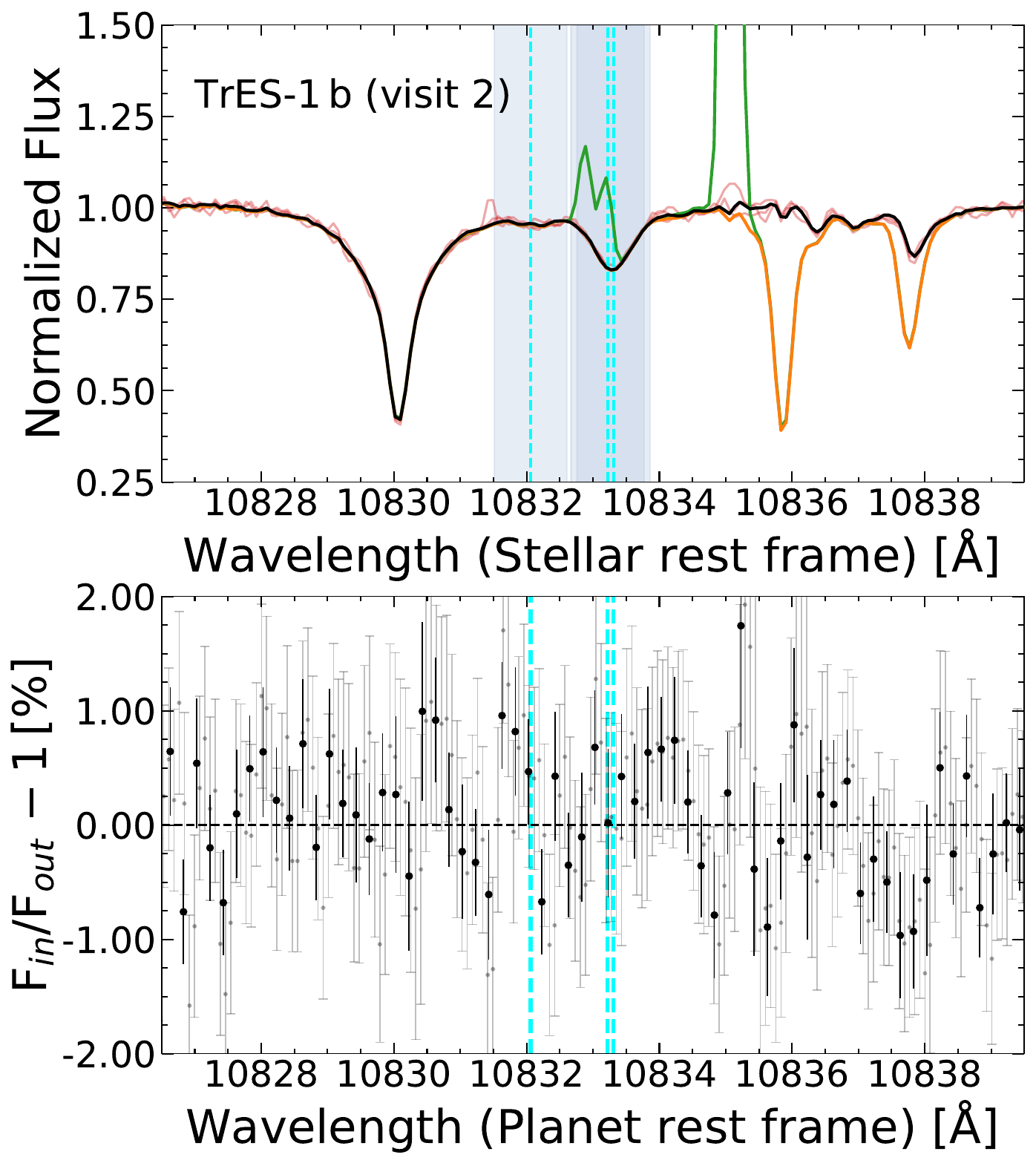}
   \includegraphics[width=0.32\linewidth]{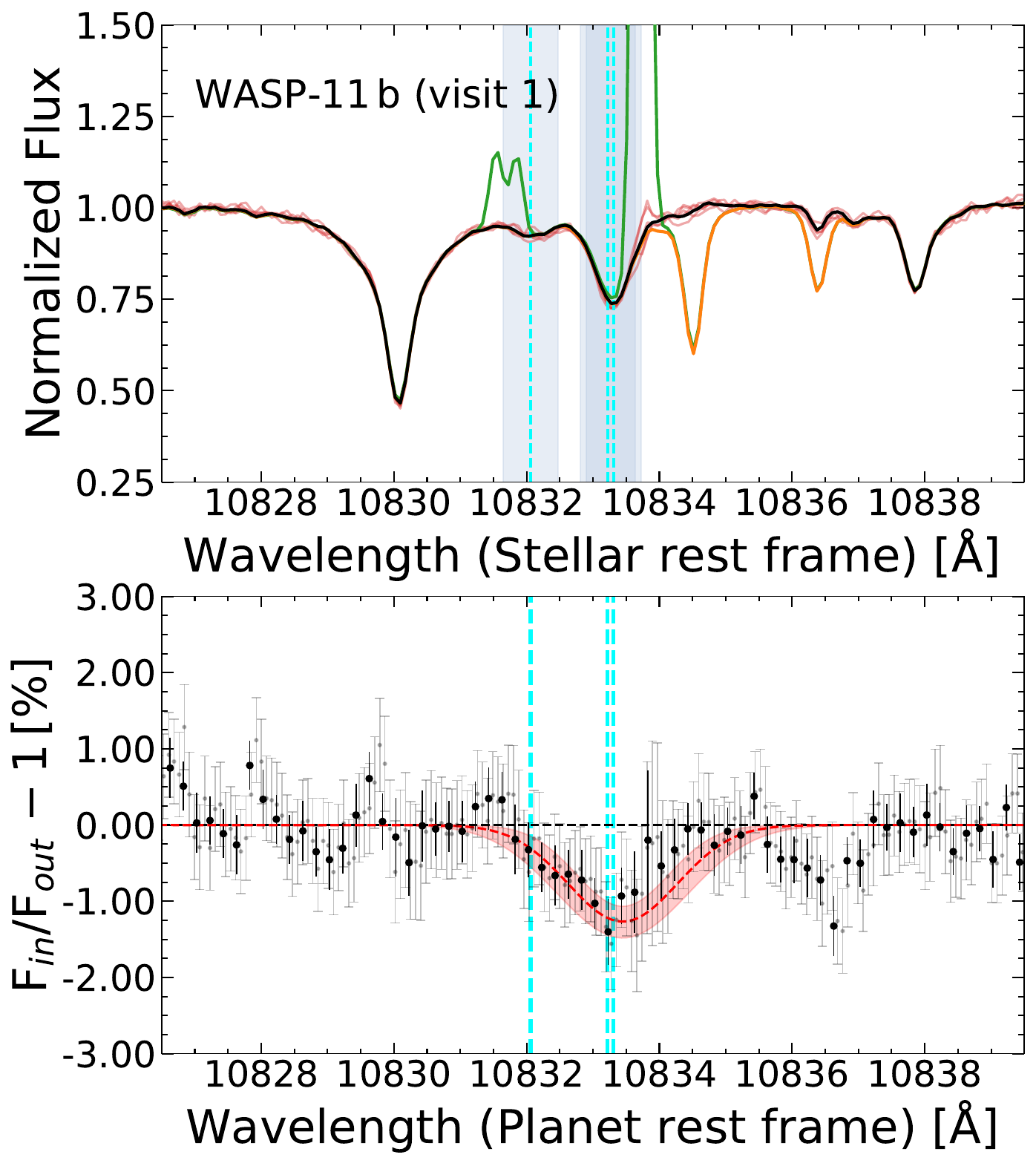}
   \includegraphics[width=0.32\linewidth]{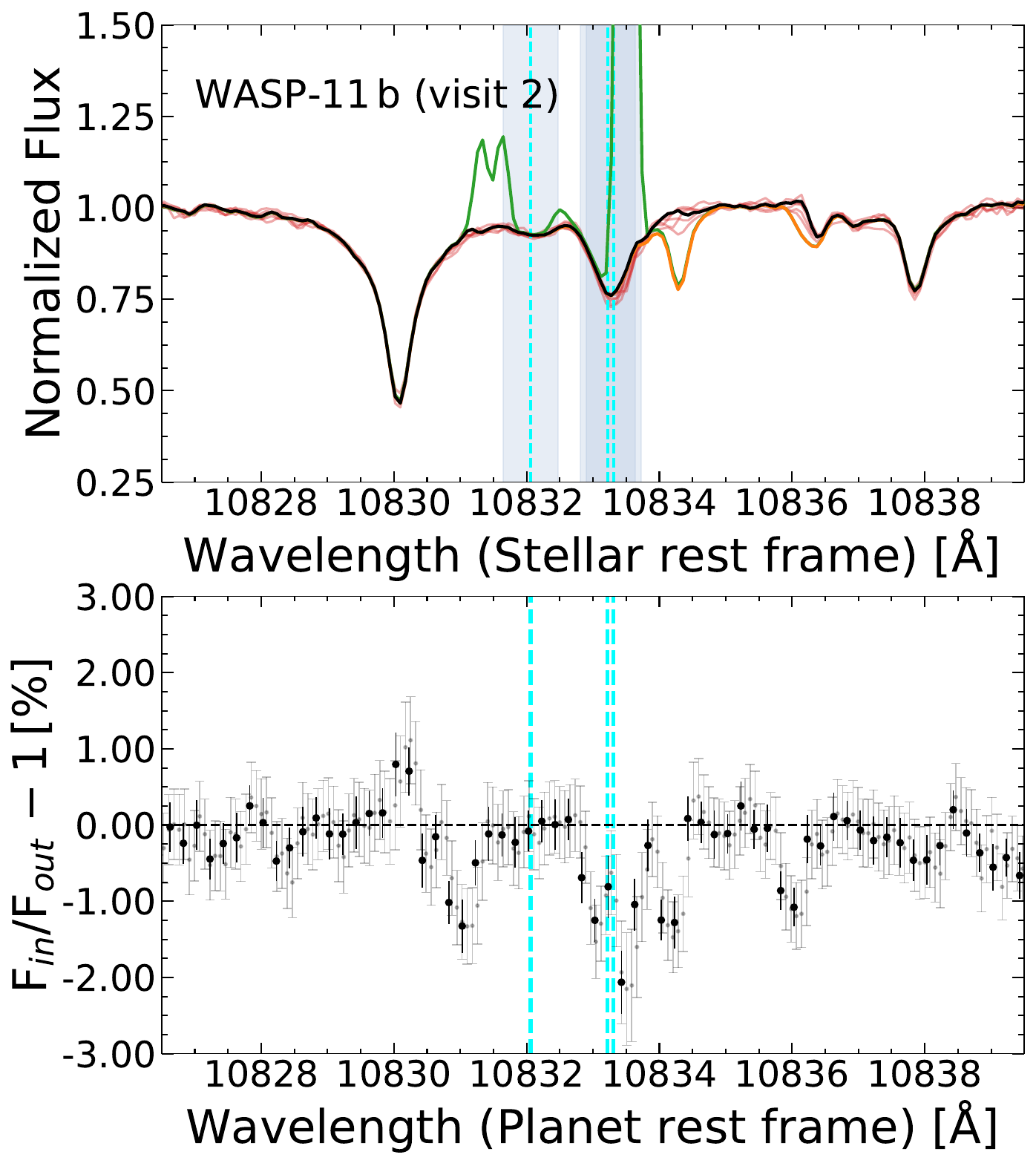}
   \includegraphics[width=0.32\linewidth]{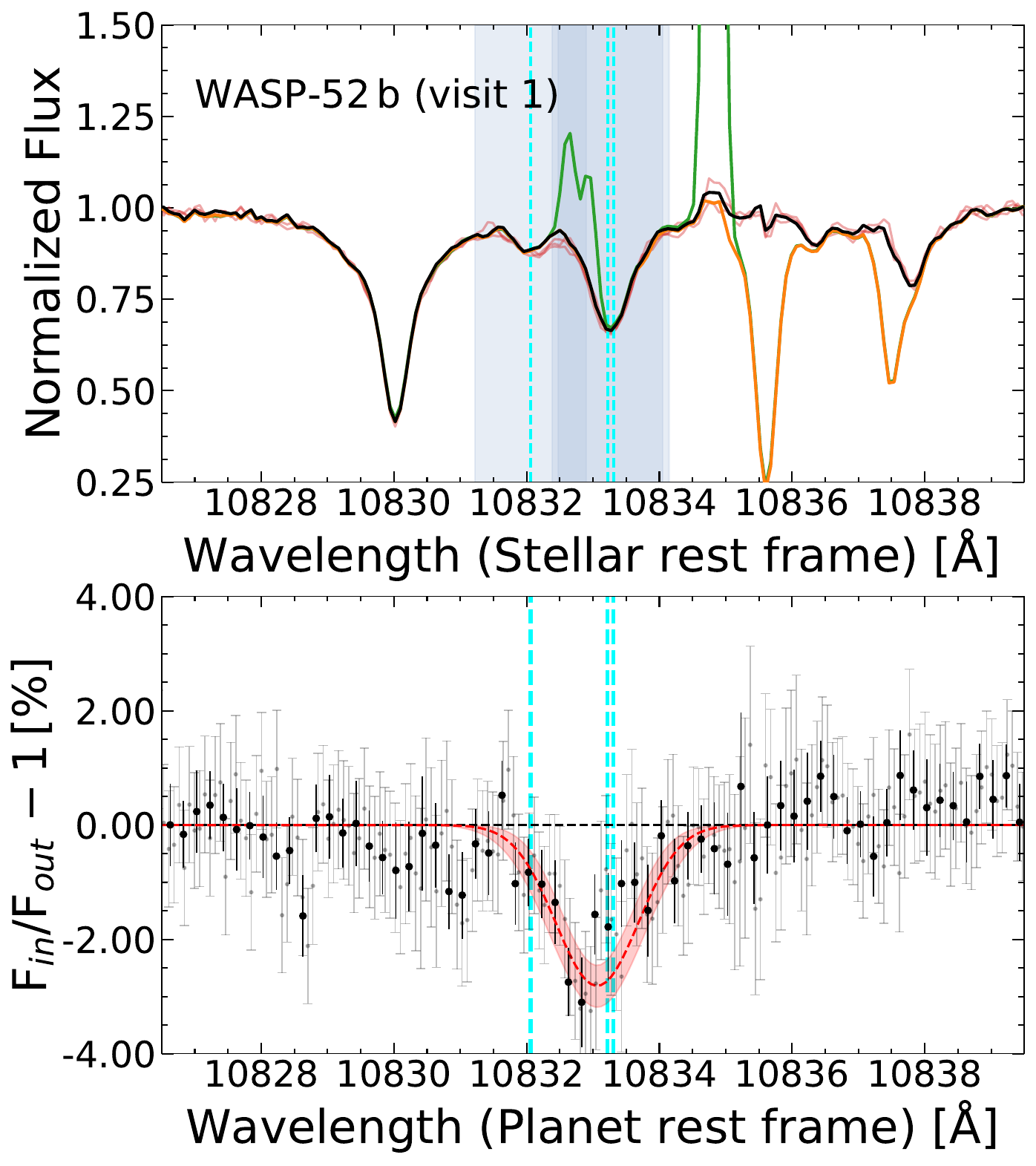}
   \includegraphics[width=0.32\linewidth]{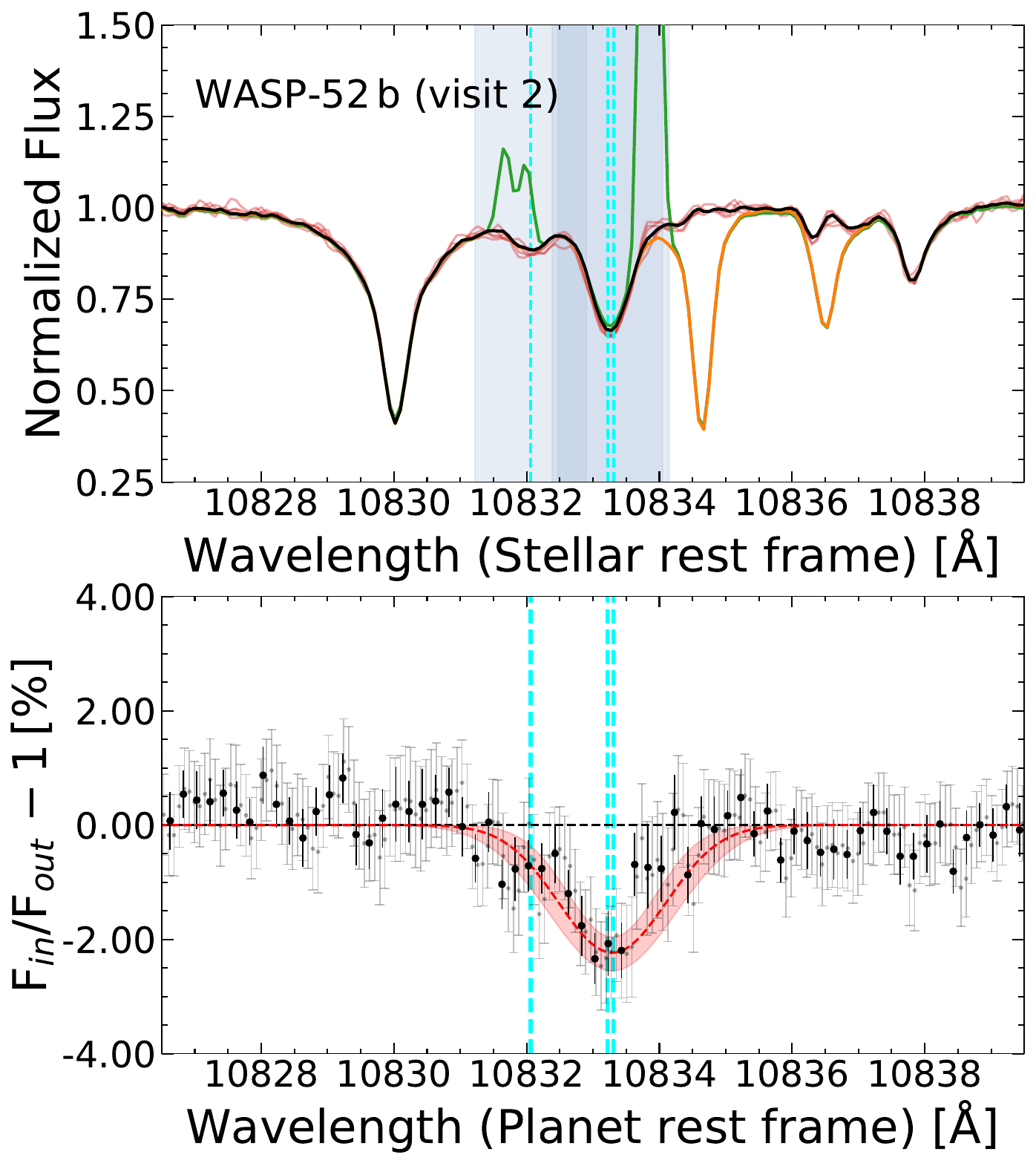}
   \caption{ \label{Fig: Visits plot 1}
   Same as Figure\,\ref{Fig: Visits plot 0} for the visits of HD189733\,b, \hd209\,b, TrES-1\,b, WASP-11\,b, and WASP-52\,b.
   }
\end{figure*}

\begin{figure*}[h!]
   \centering
   \includegraphics[width=0.32\linewidth]{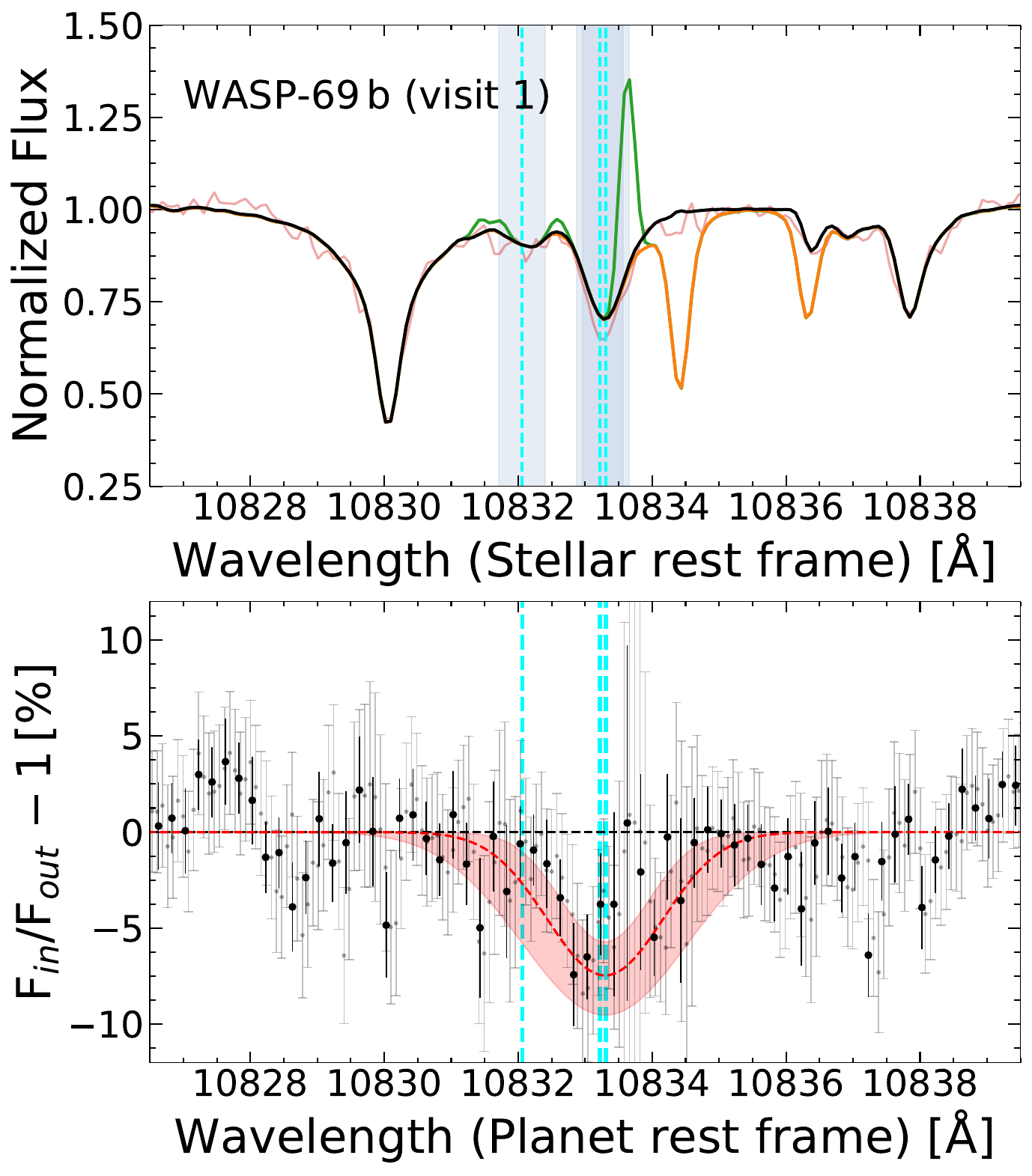}
   \includegraphics[width=0.32\linewidth]{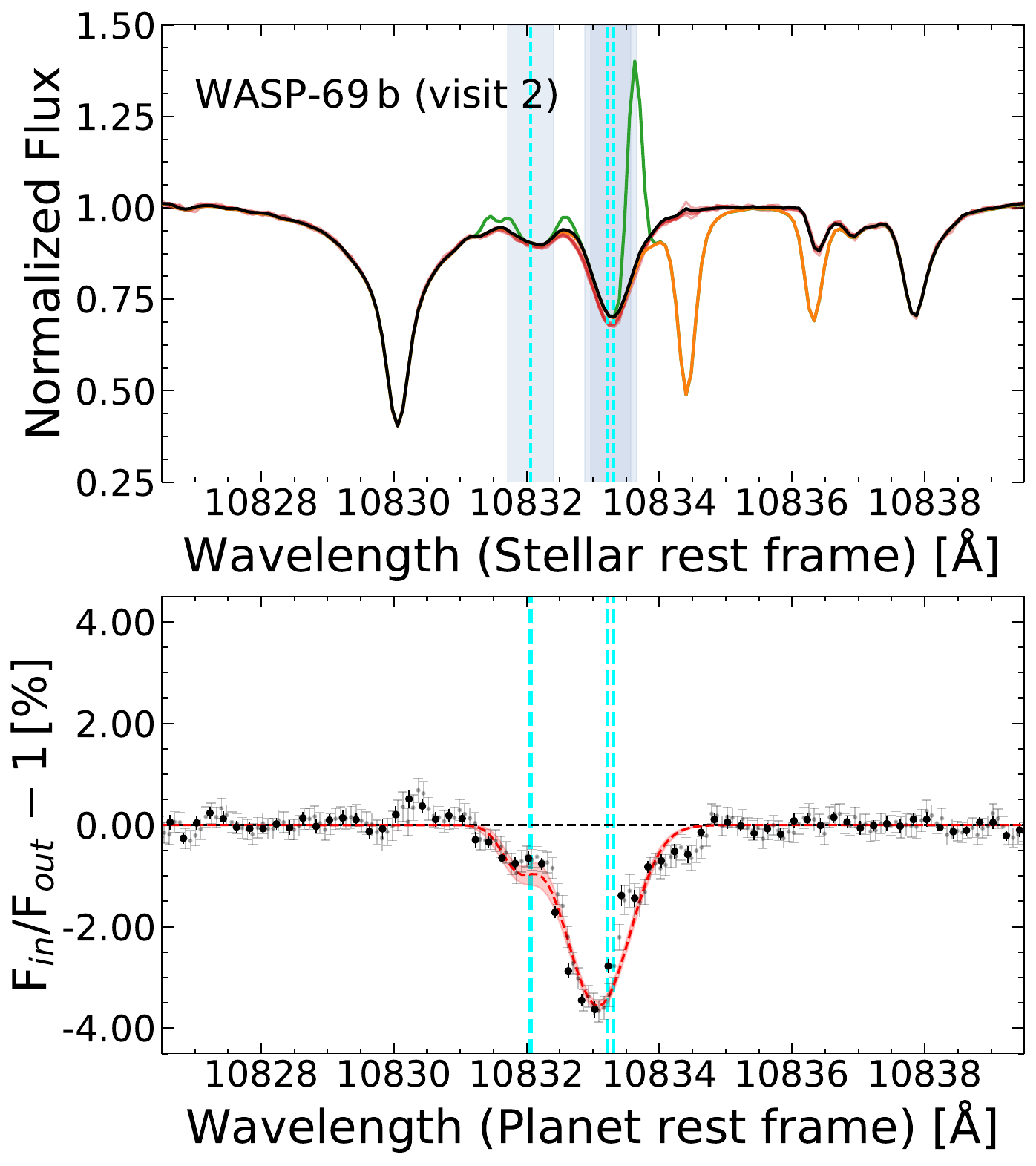}
   \includegraphics[width=0.32\linewidth]{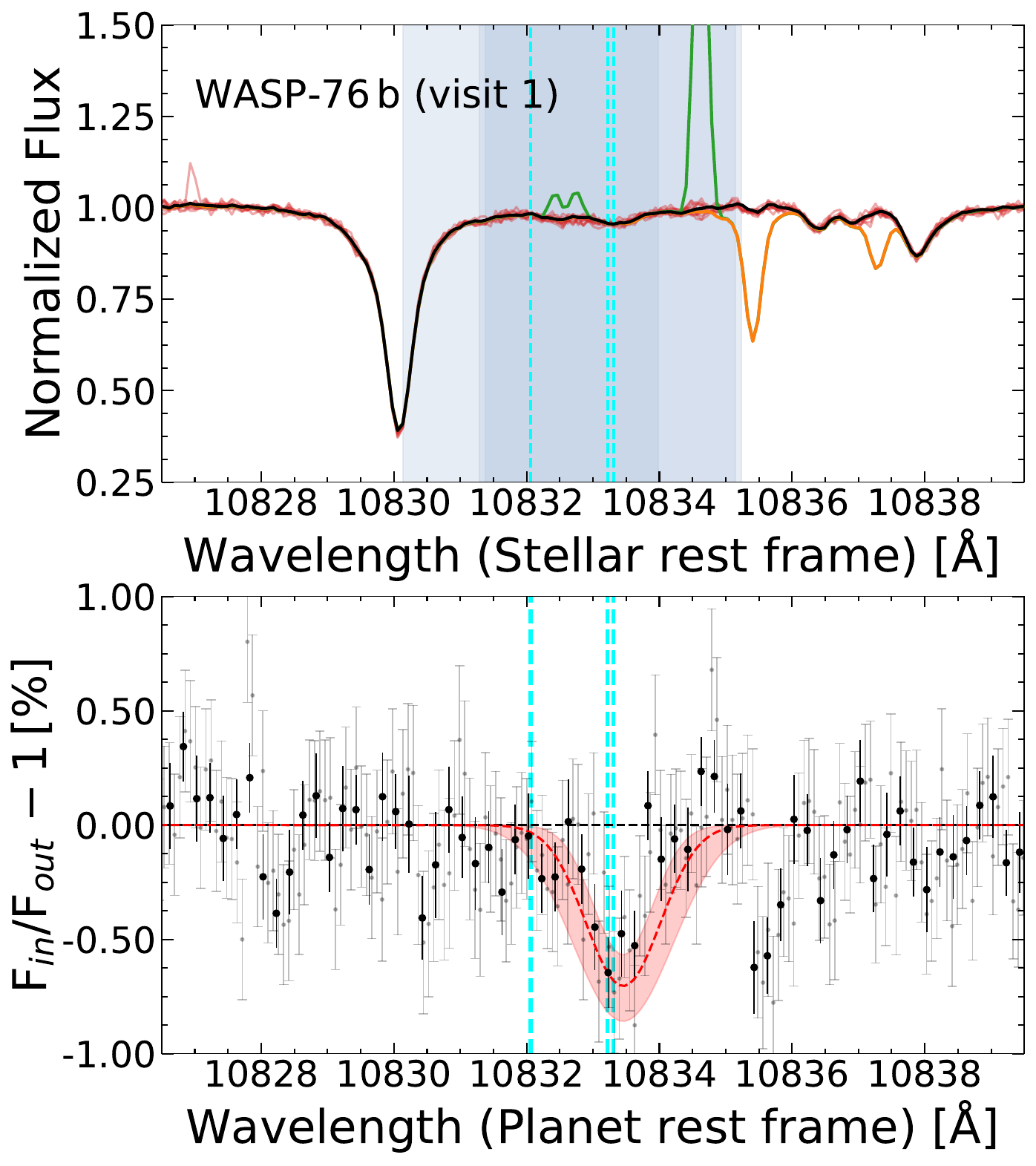}
   \includegraphics[width=0.32\linewidth]{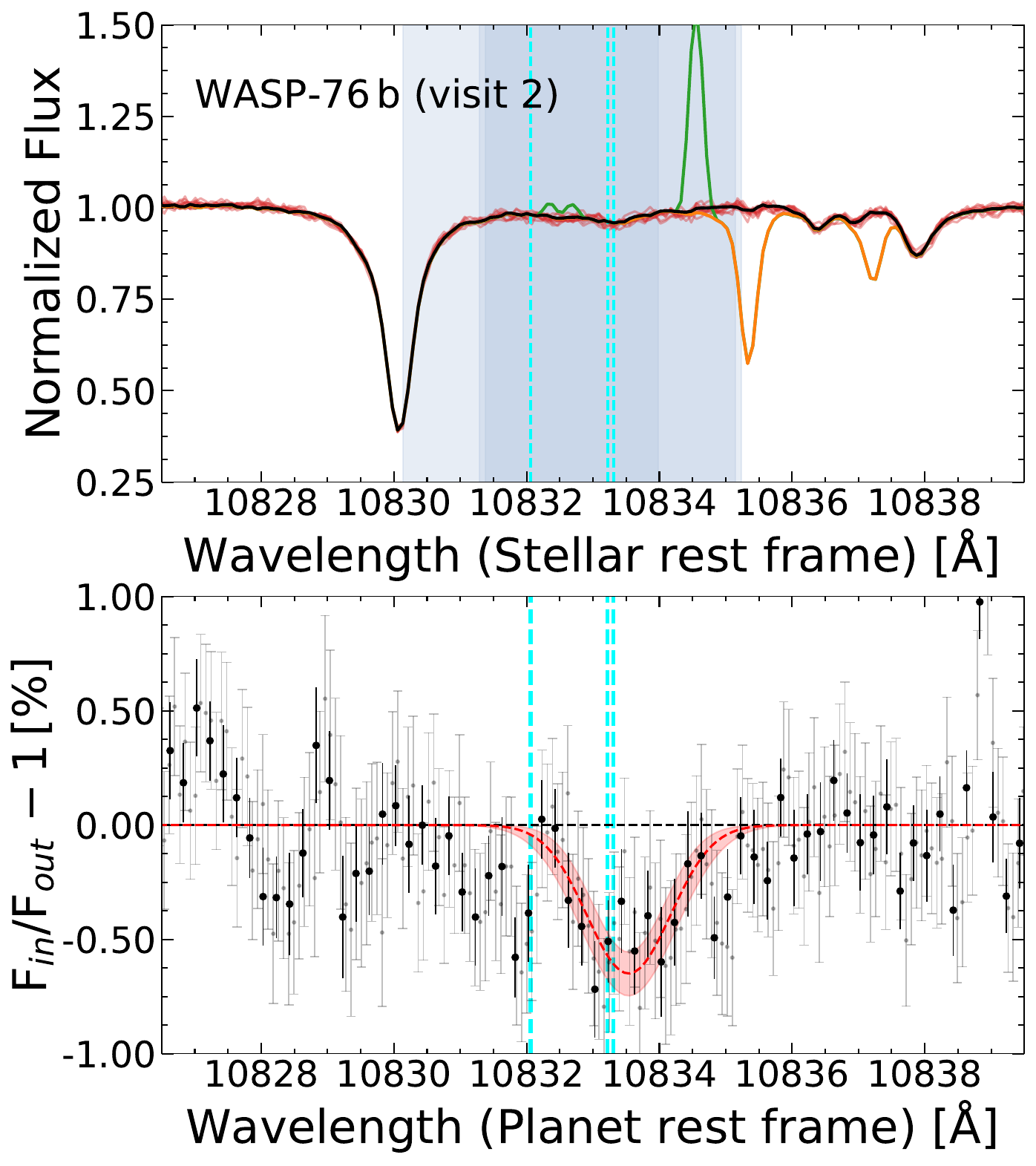}
   \includegraphics[width=0.32\linewidth]{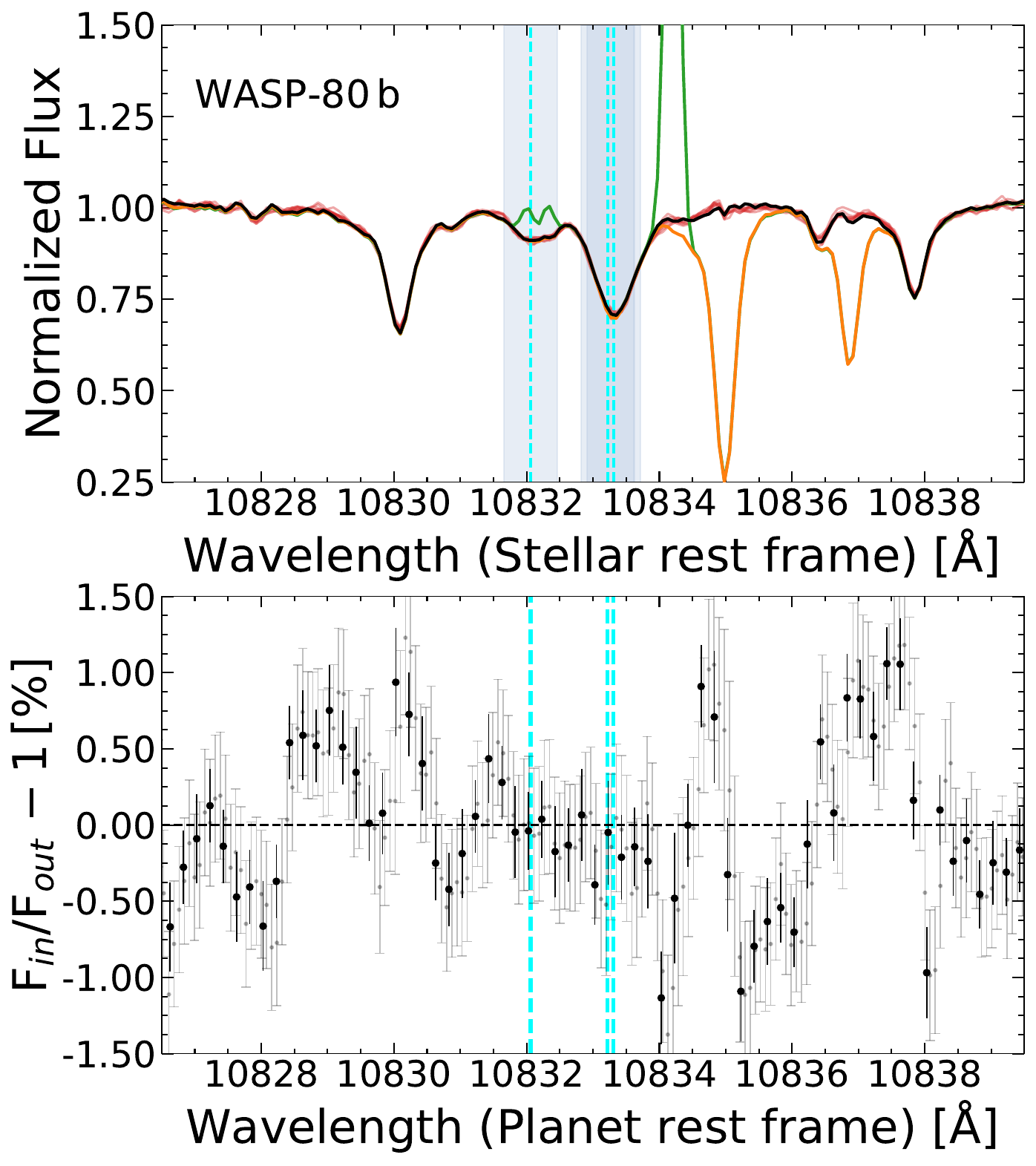}
   \includegraphics[width=0.32\linewidth]{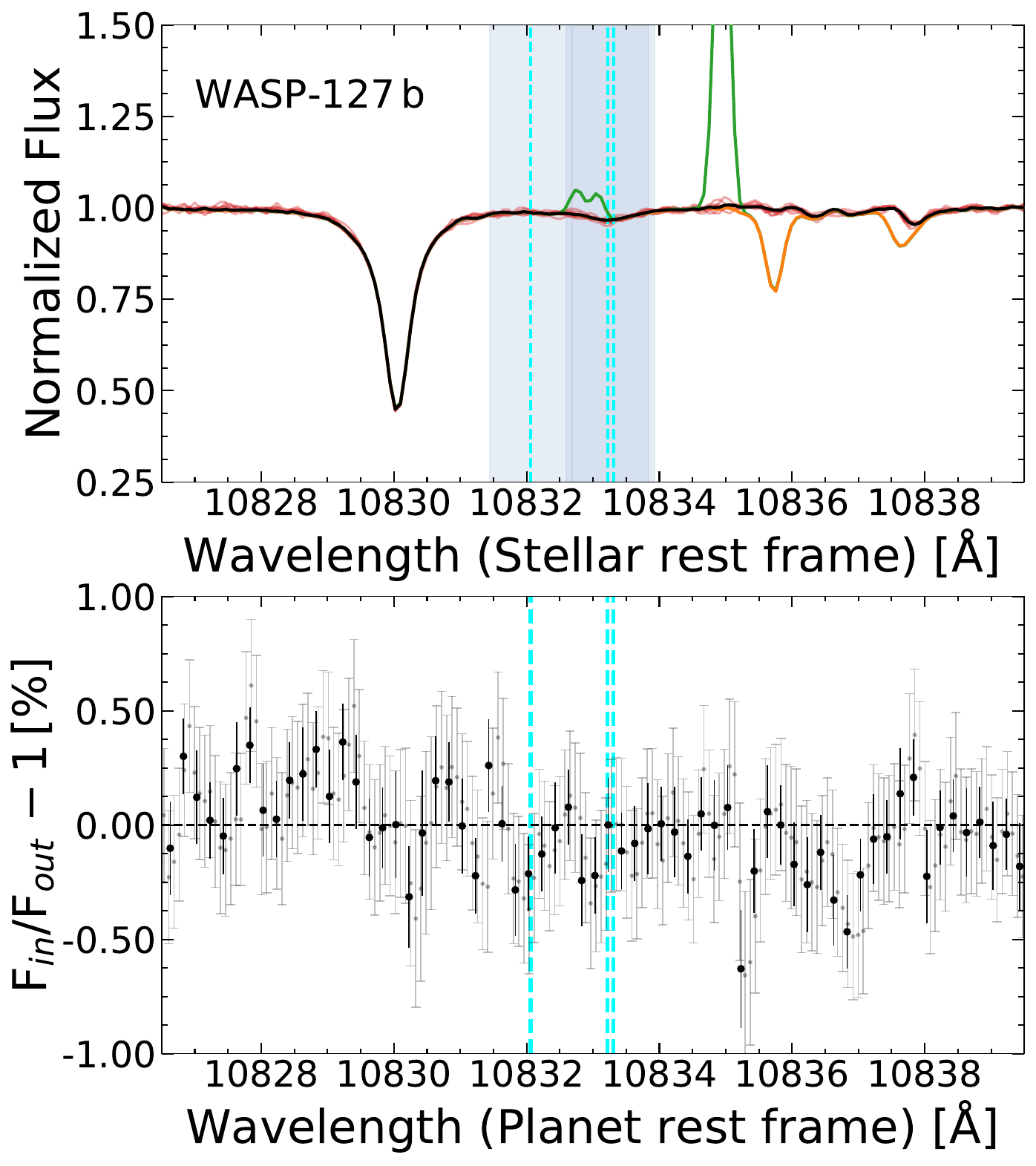}
   \includegraphics[width=0.32\linewidth]{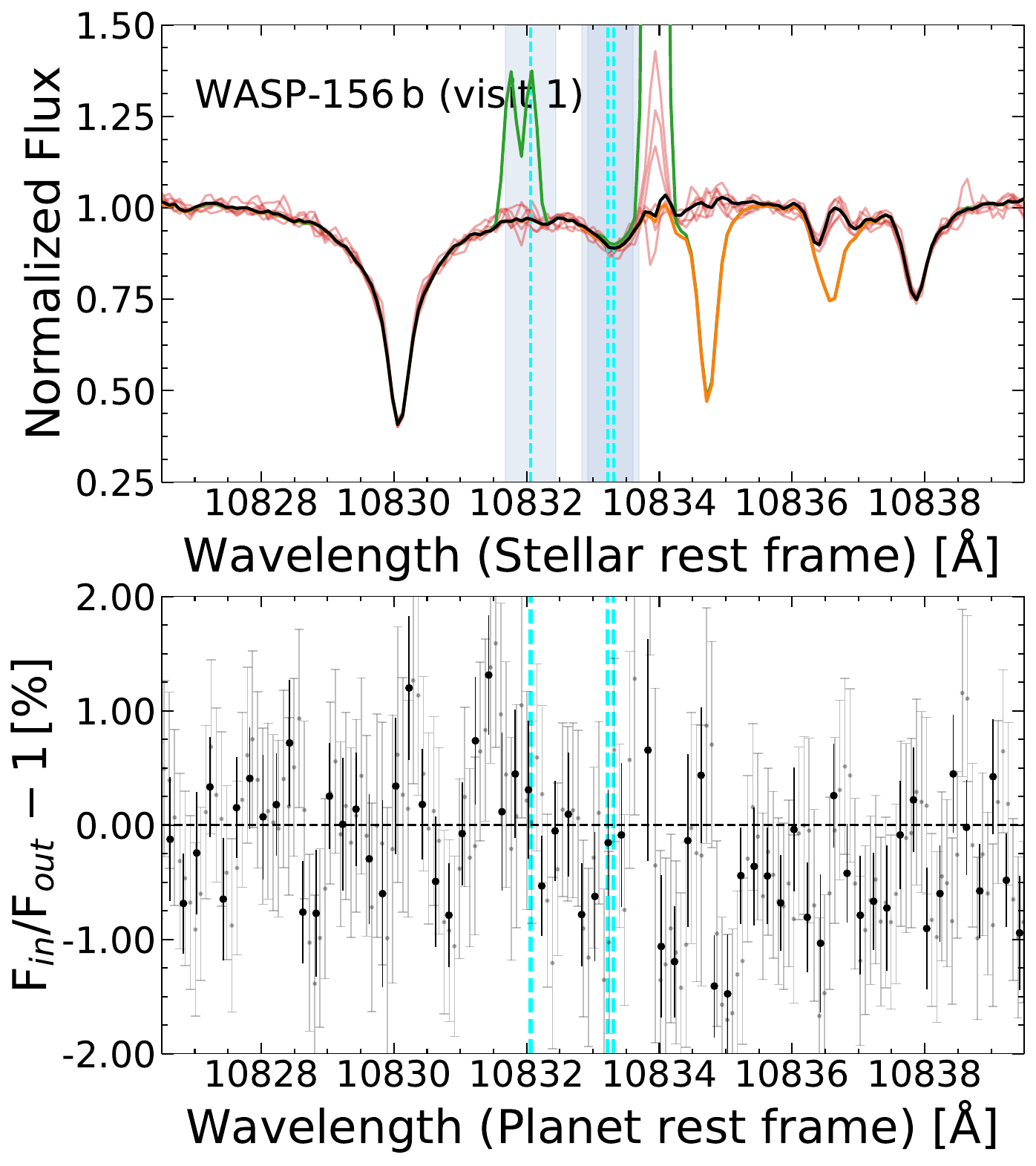}
   \includegraphics[width=0.32\linewidth]{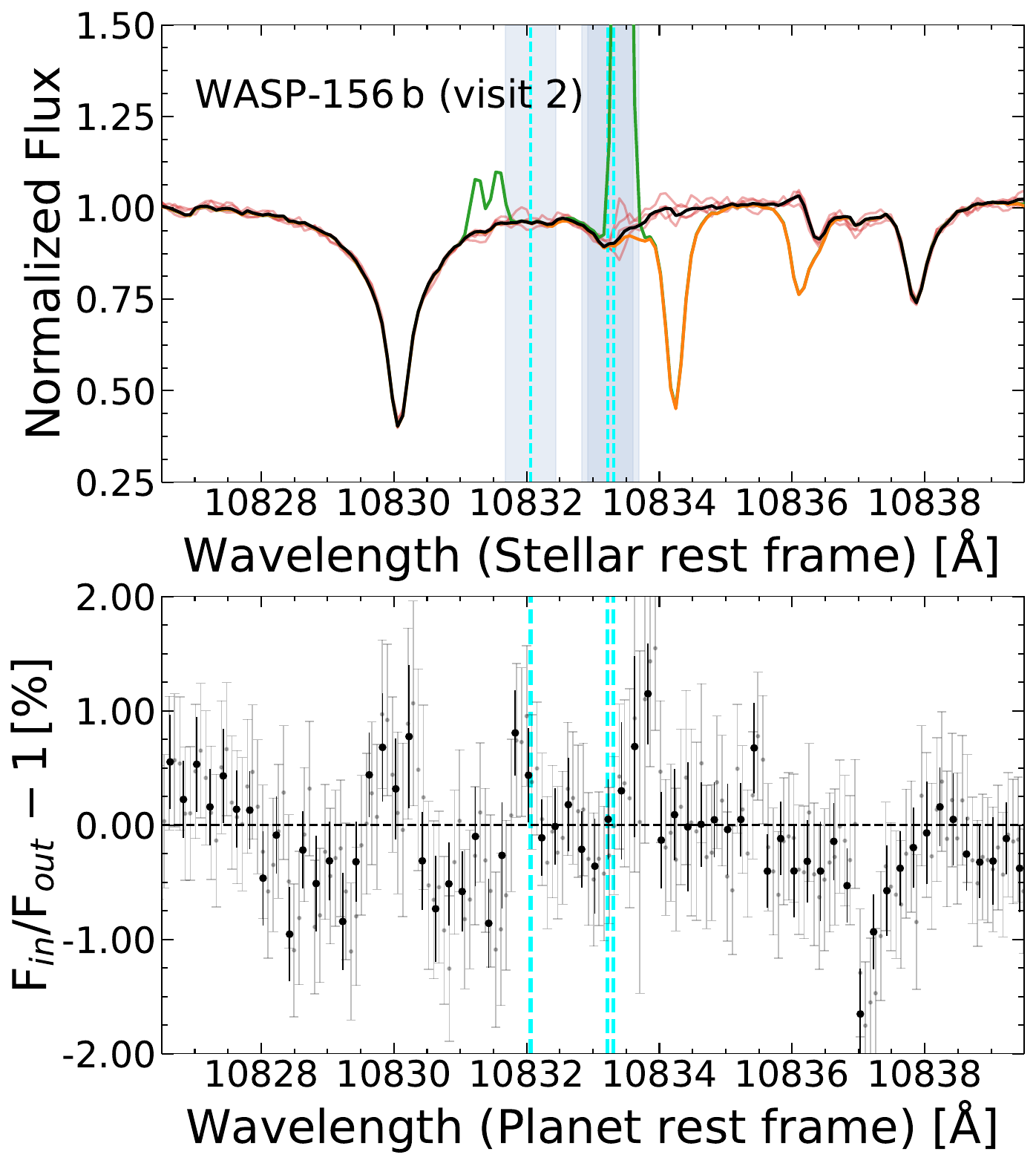}
   \caption{ \label{Fig: Visits plot 2}
   Same as Figure\,\ref{Fig: Visits plot 0} for the visits of WASP-69\,b, WASP-76\,b, WASP-80\,b, WASP-127\,b, and WASP-156\,b.
   }
\end{figure*}

\begin{figure*}[h!]
   \centering
   \includegraphics[width=0.32\linewidth]{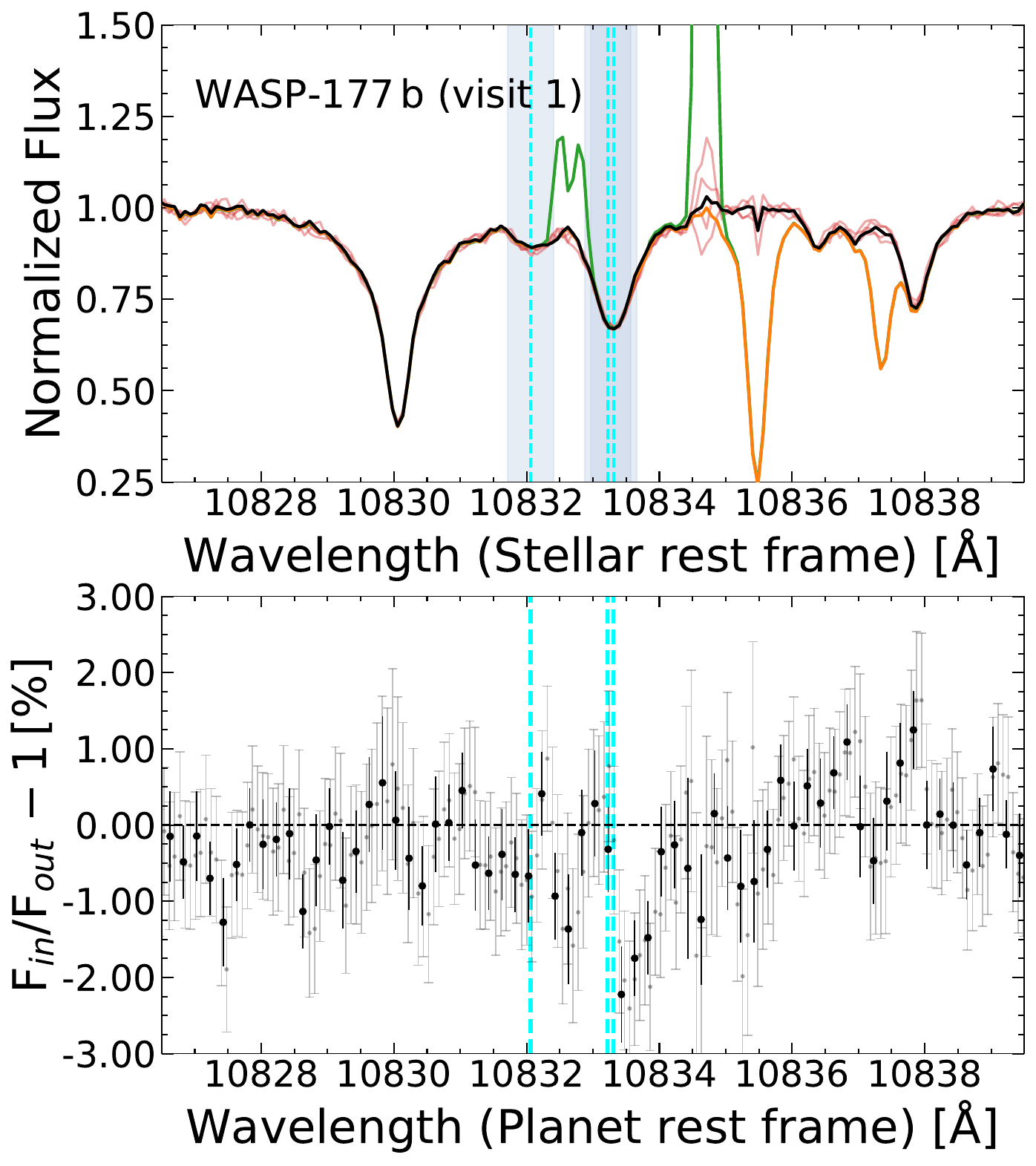}
   \includegraphics[width=0.32\linewidth]{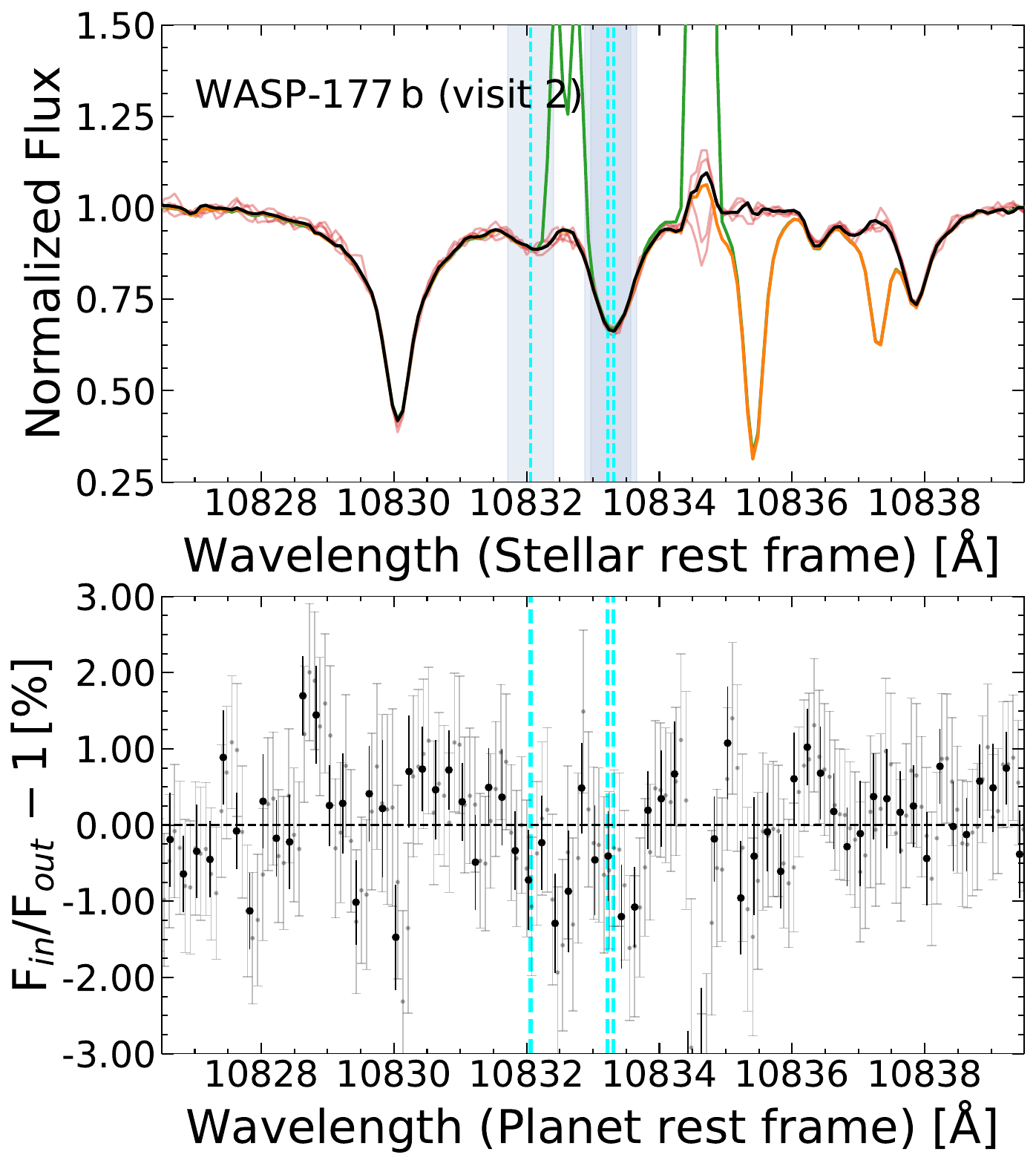}
   \caption{ \label{Fig: Visits plot 3}
   Same as Figure\,\ref{Fig: Visits plot 0} for WASP-177\,b.
   }
\end{figure*}


\bibliography{references}{}
\bibliographystyle{aasjournalv7}



\end{document}